\documentclass[amsmath,amssymb,aps,nofootinbib,prd,preprint,tightenlines]{revtex4-2}

\usepackage{bm}
\usepackage[colorlinks,citecolor=blue,urlcolor=blue,linkcolor=red]{hyperref}
\usepackage[mathscr]{euscript}
\usepackage{graphicx}
\usepackage{multirow,slashed}
\usepackage{xcolor}

\begin{document}
	
\baselineskip=17pt \parskip=5pt

\title{FCNC charmed-hadron decays with invisible singlet particles \\ in light of recent data}
\author{Geng Li$^1$\footnote[1]{ligeng@ucas.ac.cn} and Jusak Tandean$^2$\footnote[2]{jtandean@yahoo.com}\\}
\affiliation{$^1$School of Fundamental Physics and Mathematical Sciences, Hangzhou Institute for Advanced Study, UCAS, 310024 Hangzhou, China\\
	University of Chinese Academy of Sciences, 100190 Beijing, China \medskip \\
	$^2$Tsung-Dao Lee Institute, Shanghai Jiao Tong University, Shanghai 201210, China \\
	\vspace{0.6cm}}

\begin{abstract}
	
The flavor-changing neutral current (FCNC) decays of charmed hadrons with missing energy $(\slashed E)$ can serve as potentially promising hunting grounds for hints of new physics, as the standard-model backgrounds are very suppressed.
A few of such processes have been searched for in recent experiments, specifically $D^0\to\slashed E$ by Belle and $D^0\to\pi^0\slashed E$ and $\Lambda_c^+\to p\slashed E$ by BESIII, resulting in upper bounds on their branching fractions. 
We consider them to illuminate the possible contributions of the quark transition $c\to u\slashed E$ with a couple of invisible spinless bosons carrying away the missing energy, assuming that they are not charge conjugates of each other and hence can have unequal masses.
We find that these data are complementary in that they constrain different sets of the underlying operators and do not cover the same ranges of the bosons' masses, but there are regions not yet accessible.
From the allowed parameter space, we show that other $D$-meson decays, such as $D\to\rho\slashed E$, and the charmed-baryon ones $\Xi_c\to(\Sigma,\Lambda)\slashed E$ can have sizable branching fractions and therefore may offer further probes of the new-physics interactions.
We point out the importance of $D^0\to\gamma\slashed E$ which are not yet searched for but could access parts of the parameter space beyond the reach of the other modes.
In addition, we look at a scenario where the invisibles are instead fermionic, namely sterile neutrinos, and a scalar leptoquark mediates $c\to u\slashed E$.
We discuss the implications of the aforesaid bounds for this model.
The predictions we make for the various charmed-hadron decays in the different scenarios may be testable in the near future by BESIII and Belle II.
	
\end{abstract}

\maketitle

\newpage

\section{Introduction\label{intro}}

Flavor-changing neutral current (FCNC) hadron decays that alter the charm quantum number by one unit ($|\Delta C|$\,=\,1) and have missing energy ($\slashed E$) in the final states have received lots of theoretical attention over the years~\cite{Geng:2000if,Burdman:2001tf,Chen:2007cn,Chen:2007yn,Kamenik:2009kc,Badin:2010uh,Dorsner:2016wpm,Li:2020dpc,Fajfer:2021woc,Faisel:2020php,deBoer:2015boa,Bause:2020auq,Bause:2020xzj,Colangelo:2021myn,Alok:2021pdh,Golz:2022alh,Fajfer:2023nmz,Bhattacharya:2018msv,Geng:2022kmf,Boiarska:2019jym,MartinCamalich:2020dfe,Li:2018hgu,Gabrielli:2016cut,Fabbrichesi:2020wbt,Su:2020yze,Altmannshofer:2022hfs} because they are potentially valuable tools in the hunt for evidence of new physics (NP) beyond the standard model (SM).
In the SM such processes receive both short- and long-distance contributions.
The former comes from the quark transition $c\to u\nu\bar\nu$, with undetected neutrinos ($\nu\bar\nu$) being emitted, and is much suppressed because it arises from loop diagrams and is subject to highly efficient Glashow-Iliopoulos-Maiani cancellation.
Explicitly, the updated predictions for the branching fractions of a~number of charmed-hadron modes of interest due to the short-distance SM physics alone are
	\begin{align} \label{Bsmsd}
		{\cal B}(D^0\to\nu\bar\nu)_{\textsc{sm}}^{} & \,=\, 0 \,, & 
		{\cal B}(D^0\to\gamma\nu\bar\nu)_{\textsc{sm}}^{} & \,=\, 1.8\times10^{-19} \,, & \nonumber \\
		{\cal B}(D^0\to\pi^0\nu\bar\nu)_{\textsc{sm}}^{} & \,=\, 2.5\times10^{-17} \,, & 
		{\cal B}(D^0\to\rho^0\nu\bar\nu)_{\textsc{sm}}^{} & \,=\, 1.1\times10^{-17} \,, \nonumber \\
		{\cal B}(D^+\to\pi^+\nu\bar\nu)_{\textsc{sm}}^{} & \,=\, 1.3\times10^{-16} \,, & 
		{\cal B}(D^+\to\rho^+\nu\bar\nu)_{\textsc{sm}}^{} & \,=\, 5.9\times10^{-17} \,, \nonumber \\
		{\cal B}(D_s^+\to K^+\nu\bar\nu)_{\textsc{sm}}^{} & \,=\, 4.5\times10^{-17} \,, & 
		{\cal B}(D_s^+\to K^{*+}\nu\bar\nu)_{\textsc{sm}}^{} & \,=\, 3.3\times10^{-17} \,,  \nonumber \\
		{\cal B}(\Lambda_c^+\to p\nu\bar\nu)_{\textsc{sm}}^{} & \,=\, 7.3\times10^{-17} \,, & {\cal B}(\Xi_c^+\to\Sigma^+\nu\bar\nu)_{\textsc{sm}}^{} & \,=\, 1.1\times10^{-16} \,,
		\nonumber \\
		{\cal B}(\Xi_c^0\to\Sigma^0\nu\bar\nu)_{\textsc{sm}}^{} & \,=\, 1.8 \times10^{-17} \,, & {\cal B}(\Xi_c^0\to\Lambda\nu\bar\nu)_{\textsc{sm}}^{} & \,=\, 6.5\times10^{-18} \,,
	\end{align}
as evaluated later on in an appendix. 
The long-distance components have been estimated to be minuscule as well~\cite{Burdman:2001tf,Kamenik:2009kc,Colangelo:2021myn}.

In the presence of NP, the SM amplitudes might be modified and/or there might be extra channels involving one or more invisible nonstandard particles.
These are factors that could substantially enhance the rates.
Since the SM backgrounds are minimal, an observation of any of these decays at the present or near-future sensitivity level would likely be a sign of NP.

To date, there have been only a handful of attempts to seek FCNC $|\Delta C|$\,=\,1 transitions with missing energy, which came up empty, and the null outcomes translated into upper limits on their branching fractions~\cite{ParticleDataGroup:2022pth,Belle:2016qek,BESIII:2021slf,BESIII:2022vrr}.
The parent hadrons examined in these measurements, performed by the Belle and BESIII Collaborations, were the neutral pseudoscalar charmed-meson $D^0$ and the singly-charmed spin-1/2 baryon $\Lambda_c^+$.
Belle announced ${\cal B}(D^0\to\mbox{invisibles})<9.4\times10^{-5}$~\cite{Belle:2016qek}, while BESIII reported ${\cal B}(D^0\to\pi^0\nu\bar\nu)<2.1\times10^{-4}$~\cite{BESIII:2021slf} and ${\cal B}(\Lambda_c^+\to p\gamma')<8.0\times10^{-5}$~\cite{BESIII:2022vrr}, all at 90\% confidence level, with $\gamma'$ denoting a massless dark photon, which was unobservable.
In light of the smallness of the SM predictions above and the lack or scarcity of the corresponding experimental information, it is clear that the window of opportunity to discover NP in any one of these modes is wide open.

In view of their significance as beneficial probes of NP, it is hoped that more and more quests will be carried out for these kind of processes at already running operations, especially BESIII and Belle\,\,II.
At least it is anticipated that BESIII could better its aforementioned $D^0\to\pi^0\nu\bar\nu$ result~\cite{BESIII:2021slf} by a factor of \mbox{\footnotesize\,$\sim$\,}3 after its data sample gathered at center-of-mass energy \,$\sqrt s\simeq3.77$\,GeV  is increased from 2.93\,fb$^{-1}$ to 20 fb$^{-1}$ in a few years~\cite{BESIII:2021slf,BESIII:2020nme}.
In addition, Belle\,\,II is expected to improve on the Belle bound on \,$D^0\to\mbox{invisibles}$\, by a factor of seven~\cite{Belle-II:2018jsg}, and BESIII might push it down further to $10^{-6}$ with its final charm dataset~\cite{BESIII:2020nme}.
The foregoing suggests that for  $D^0\to\pi^0\nu\bar\nu$\, the ultimate reach of Belle II might be less than that of BESIII.
More distant in the future, searches for these decays with greater levels of sensitivity would presumably be feasible at the proposed Super Tau-Charm Facility (STCF), Circular Electron Positron Collider (CEPC), and Future Circular $e^+e^-$ Collider (FCC-ee).
The STCF~\cite{Achasov:2023gey} is designed to accumulate a data sample about 100 times that collected by BESIII and could therefore improve on the latter's reach for the preceding $D^0$ modes by a factor of 10 or better.
At the CEPC and FCC-ee, operating as $Z$-boson factories, the projected numbers of $D^0$ (and its antiparticle) from \,$Z\to c\bar c$\, are $1\times10^{11}$~\cite{CEPCStudyGroup:2018ghi} and~$6\times10^{11}$~\cite{Bause:2020xzj,FCC:2018byv}, respectively.
Since Belle II is anticipated to yield $8\times10^{10}$ of these mesons~\cite{Bause:2020xzj}, the CEPC and FCC-ee would expectedly be somewhat superior to Belle II for probing \,$D^0\to\slashed E,\pi^0\slashed E$\, if the three  have similar reconstruction efficiencies.
Given that the $D^0$ amount collected in each of these ongoing and proposed experiments \cite{Bause:2020xzj,Belle-II:2018jsg,BESIII:2020nme,CEPCStudyGroup:2018ghi} is bigger than those of the $D_{(s)}^+$ meson and charmed baryons, the sensitivities of these facilities to the other FCNC charmed-hadron transitions we will look at would probably be comparable or lower.

The prospect that a growing amount of fresh data on this subject is forthcoming has lately revived related theoretical efforts~\cite{deBoer:2015boa,Dorsner:2016wpm,Gabrielli:2016cut,Bhattacharya:2018msv,Li:2018hgu,Boiarska:2019jym,MartinCamalich:2020dfe,Fabbrichesi:2020wbt,Li:2020dpc,Su:2020yze,Bause:2020auq,Bause:2020xzj,Faisel:2020php,Fajfer:2021woc,Colangelo:2021myn,Alok:2021pdh,Golz:2022alh,Altmannshofer:2022hfs,Geng:2022kmf,Fajfer:2023nmz}.
	Various aspects of it have been explored to different extents, including the type of particles carrying away the missing energy and how many of them.
They might be a pair of ordinary neutrinos, and this often means that the restraints on the interactions of their charged-lepton partners would  at the same time squeeze the room for the NP affecting the dineutrino decays~\cite{deBoer:2015boa,Bause:2020auq,Bause:2020xzj,Faisel:2020php,Colangelo:2021myn,Alok:2021pdh,Golz:2022alh,Fajfer:2023nmz}.
Alternatively, the invisibles could be spin-1/2 fermions~\cite{Badin:2010uh,Dorsner:2016wpm,Li:2020dpc,Faisel:2020php,Fajfer:2021woc} or a pair of spin-0 bosons~\cite{Badin:2010uh,Li:2018hgu,Boiarska:2019jym} which hail from beyond the SM and are singlets under the SM gauge groups, implying that the restrictions pertaining to the charged leptons would likely have little, if any, bearing on the \,$c\to u\slashed E$\, sector.
Another possibility is that the missing energy is carried away instead by just one particle which is again a SM-gauge singlet and has to be a~boson.
It might be spinless~\cite{Boiarska:2019jym,MartinCamalich:2020dfe,Geng:2022kmf} or has spin 1, such as the massless dark photon~\cite{Gabrielli:2016cut,Fabbrichesi:2020wbt,Su:2020yze}.
It is worth remarking that similar transitions among down-type quarks with invisible bosons have also been much discussed in the past~\cite{Badin:2010uh,Li:2018hgu,Boiarska:2019jym,MartinCamalich:2020dfe,Gabrielli:2016cut,Fabbrichesi:2020wbt,Bird:2004ts,Bird:2006jd,Kim:2009qc,He:2010nt,Kamenik:2011vy,Su:2019ipw,Liao:2020boe,He:2020jly,Gori:2020xvq,Su:2020xwt,Li:2021sqe,He:2022ljo,Altmannshofer:2020pjb,Geng:2020seh,Hostert:2020xku,Li:2019cbk}\footnote{Corresponding processes with invisible new fermions have recently been analyzed in, {\it e.g.}, refs. \cite{Li:2019cbk,Hostert:2020xku,Dutta:2020scq,Su:2019tjn,Geng:2020seh,He:2021yoz,Li:2022tbh}.\vspace{2ex}} and the bosons might be dark-matter candidates.

Here we adopt a model-independent approach to investigate the decays of charmed hadrons brought about by effective \,$c\to u\slashed E$\, operators in which the invisibles comprise a couple of SM-gauge-singlet spin-0 bosons.\footnote{The two bosons could alternatively be of spin 1. As can be inferred from refs.\,\cite{Badin:2010uh,Kamenik:2011vy,He:2022ljo}, the situation would then be significantly more complicated than its spin-0 counterpart, with a much bigger number of effective operators, and, on top of that, their coefficients would have relatively far weaker experimental bounds~\cite{Kamenik:2011vy,He:2022ljo}. For these reasons, we opt not to deal with the spin-1 case here.\vspace{9ex}}
Unlike in most earlier papers, we assume that these particles are not charge conjugates of one another and hence may not have the same mass.
It turns out that whether or not their masses are equal could determine the feasibility of probing their couplings to the quarks.
Furthermore, taking into account the Belle and BESIII data quoted above and anticipating related upcoming measurements, we address a number of charmed-hadron processes, not only  $D^0\to\slashed E,\pi^0\slashed E$\, and \,$\Lambda_c^+\to p\slashed E$\, but also \,$D^0\to\gamma\slashed E$,\, more decays of pseudoscalar charmed-mesons into final states with a charmless meson, and analogous decays of the singly-charmed baryons $\Xi_c^+$ and $\Xi_c^0$.
After extracting the allowed couplings from the existing empirical constraints, we make predictions for these proposed modes which are potentially testable soon.
Moreover, we demonstrate that  $D^0\to\gamma\slashed E$\, besides \,$D^0\to\slashed E$\, would be especially advantageous, as it  could access parameter space that is outside the reach of the other decays.

We also examine the case where singlet spin-1/2 fermions instead act as the invisibles.
We suppose in particular that they are connected to the $u$ and $c$ quarks owing to their joint couplings to a scalar leptoquark.
This was first treated in detail in ref.\,\cite{Faisel:2020php}, where the singlet fermions' masses were taken to be negligible and consequently the already available limit on \,$D^0\to\slashed E$\, from Belle did not apply. 
The advent of the BESIII limits on \,$D^0\to\pi^0\slashed E$\, and \,$\Lambda_c^+\to p\slashed E$\, has opened up an opportunity to scrutinize the model more thoroughly, and with the invisible fermions' masses permitted to be nonzero the Belle result becomes relevant as well.
Thus, as in the bosonic scenario, we will explore the implications of these recent data for several analogous FCNC charmed-hadron decays with missing energy.
From all this exercise, we hope to learn some of the salient consequences of selecting different types of invisible particles and of doing a model-based study versus a model-independent one.

The structure of the remainder of the paper is the following.
In the next section, we first write down the operators for the effective \,$c\to u\slashed E$\, transition with the invisible light spinless bosons and subsequently derive the induced amplitudes for the hadron decays investigated here and the corresponding rates.
With them we perform the numerical analysis in section \ref{nums}.
In section \ref{fermions} we entertain the possibility that the invisibles produced in \,$c\to u\slashed E$\, are singlet spin-1/2 fermions.
In section \ref{concl} we give our conclusions.
In two appendices we specify the hadronic form factors needed in our computation and estimate the SM backgrounds to the various modes with missing energy.

\section{Interactions and hadron decays due to \boldmath$c\to u{\texttt S}\bar{\texttt S}{}'$\label{int}}

The invisible light spin-0 bosons are SM-gauge singlets described by complex fields \texttt S and $\texttt S'$.
They could be stable or sufficiently long-lived to escape detection.
We assume that they are charged under some dark-sector symmetry or odd under a $Z_2$ symmetry, \,$\texttt S{}^{(\prime)}\to-\texttt S{}^{(\prime)}$,\, which does not influence SM fields.
Accordingly,\, \texttt S and $\texttt S'$ do not interact singly with SM quarks.
The leading-order low-energy effective \,$|\Delta C|$\,=\,1\, operators containing these bosons are expressible as~\cite{Badin:2010uh,Kamenik:2011vy}
\begin{align} \label{Lss'}
	{\mathcal L}_{\texttt{SS}'} & \,=\, - \big( \kappa_{\texttt{SS}'}^{\textsc v} \overline u\gamma_\mu c + \kappa_{\texttt{SS}'}^{\textsc a} \overline u\gamma_\mu \gamma_5^{}c \big) i\big({\texttt S}^\dagger\partial^\mu{\texttt S}' - \partial^\mu{\texttt S}^\dagger {\texttt S}' \big) - \big( \kappa_{\texttt{SS}'}^{\textsc s} \overline u c + \kappa_{\texttt{SS}'}^{\textsc p} \overline u\gamma_5^{}c \big) m_c\, {\texttt S}^\dagger {\texttt S}'  \,+\, {\rm H.c.} \,,
\end{align}
where $\kappa_{\texttt{SS}'}^{\textsc x}$, $\textsc x=\textsc v,\textsc a,\textsc s,\textsc p$, are in general complex coefficients which have the dimensions of inverse squared mass and $m_c$ is the charm-quark mass.
These $\kappa$s are free parameters in our model-independent approach and will be treated phenomenologically in our numerical work.
We notice in ${\mathcal L}_{\texttt{SS}'}$ that $\kappa_{\texttt{SS}'}^{\textsc v}$ and $\kappa_{\texttt{SS}'}^{\textsc s}$ $(\kappa_{\texttt{SS}'}^{\textsc a}$ and $\kappa_{\texttt{SS}'}^{\textsc p}$) accompany quark bilinears which are parity even (odd).
We suppose that \,$\texttt S'\neq\texttt S$\, and hence their masses can be unequal.\footnote{Effective operators for quark FCNCs involving two invisible light new particles which may differ in mass have been considered before in the literature, such as refs. \cite{Altmannshofer:2020pjb,Hostert:2020xku} (\cite{Faisel:2020php,Geng:2020seh,Hostert:2020xku}) where the invisibles are dark spin-0 bosons (spin-1/2 fermions).\vspace{2ex}}

\begin{figure}[htbp]
	\includegraphics[width=0.27\textwidth]{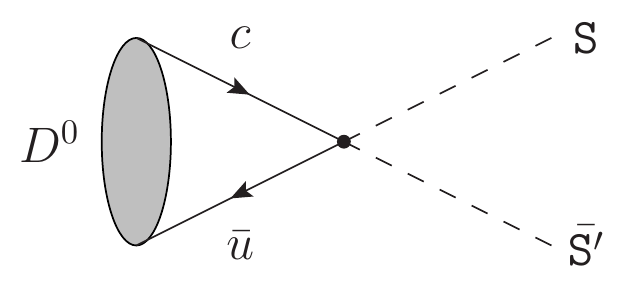} \hspace{7em}
	\includegraphics[width=0.27\textwidth]{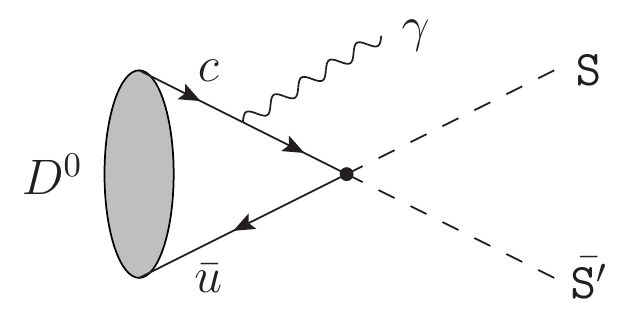}  
	
	{~ ~ ~ ~ \small(a) Fully invisible decay \hspace{7em} (b) Semi-invisible radiative decay} \vspace{5pt}
	
	~ ~ \includegraphics[width=0.35\textwidth]{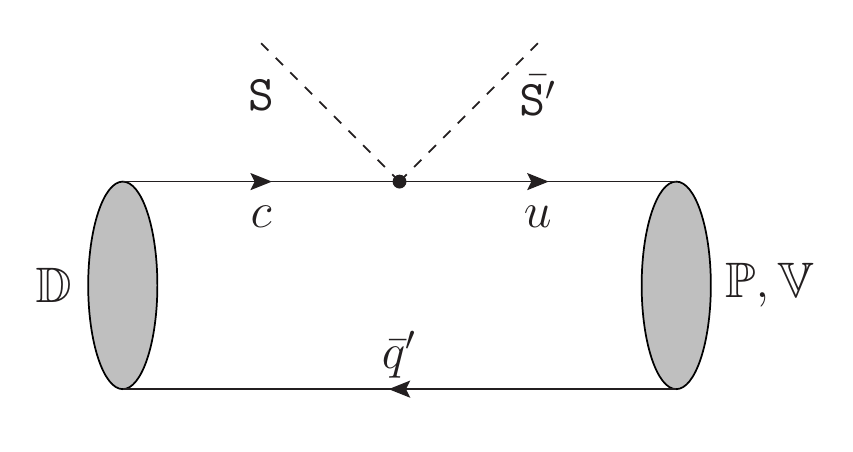} \hspace{4em}
	\includegraphics[width=0.35\textwidth]{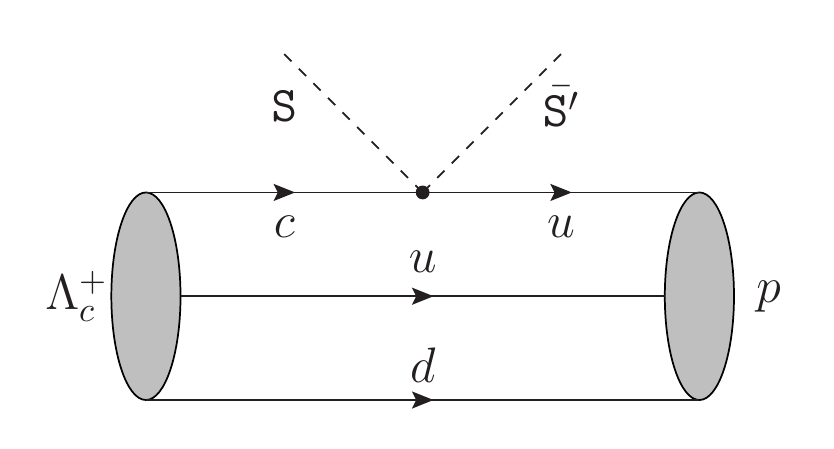} 
	
	{~ ~ ~ \small(c) Semi-invisible mesonic decay \hspace{6em} (d) Semi-invisible baryonic decay} \vspace{7pt}
	
	\caption{Diagrams of FCNC charmed-hadron decays with two invisible light spin-0 bosons. In (b) the photon can also be emitted from the $\bar u$-quark line.} \label{Feyn}
\end{figure}

It is interesting to comment that the emitted bosons not being charged conjugates of each other is beneficial because that helps avoid extra restrictions from the data on $D^0$-$\bar D^0$ mixing. 
For the latter gets contributions from four-quark operators \,$\overline u(1,\gamma^\alpha)\gamma_5^{}c\,\overline u(1,\gamma_\alpha)\gamma_5^{}c$\, that are generated by loop diagrams with \,\texttt S\, and \,$\texttt S'$ being in the loops and have coefficients proportional to linear combinations of $\kappa_{\texttt{SS}'}^{\textsc x}\kappa_{\texttt S'\texttt S}^{\textsc x}$, with $\textsc x=\textsc v,\textsc a,\textsc s,\textsc p$, which vanish if \,$\texttt S'\neq\texttt S$ and $\kappa_{\texttt S'\texttt S}^{\textsc x}=0$.\footnote{Similar situations occur in section \ref{fermions} and in the strangeness-changing (kaon and hyperon) sector~\cite{Su:2019tjn,Geng:2020seh}.\bigskip}
The operators in ${\mathcal L}_{\texttt{SS}'}$ give rise to many sorts of FCNC decays of charmed hadrons with missing energy.
Here we focus on the processes represented by the diagrams collected in figure \ref{Feyn}.
As already stated, the corresponding transitions in the SM, which have neutrinos in the final states, are highly suppressed.
Based on prior calculations~\cite{Geng:2000if,Burdman:2001tf,Badin:2010uh,Geng:2022kmf} and the updated estimates in appendix \ref{smpreds}, we can safely ignore the SM backgrounds to these hadron modes.
In the rest of this section, we discuss in detail the amplitudes for the latter and their rates. 

\subsection{Fully invisible decay\label{D0->SS'}}

The amplitude for the invisible channel $D^0\to{\texttt S}\bar{\texttt S}{}'$ can be expressed as
\begin{align}
	{\mathcal M}_{D^0\to{\texttt S}\bar{\texttt S}{}'} & \,=\, \kappa_{\texttt{SS}'}^{\textsc a}\, \langle0| \overline u\gamma^\mu\gamma_5^{}c|D^0\rangle ({\texttt k}-{\texttt k}')_\mu + \kappa_{\texttt{SS}'}^{\textsc p}\, m_c^{}\, \langle0|\overline u\gamma_5^{}c|D^0\rangle \,, &
\end{align}
with the mesonic matrix elements
\begin{align} \label{D->0}
	\langle0|\overline u\gamma^\mu\gamma_5^{}c|D^0\rangle & \,=\, -if_D^{}\, p_{D^0}^\mu \,, &
	\langle0|\overline u\gamma_5^{}c|D^0\rangle & \,=\, \frac{if_D^{}\, m_{D^0}^2}{m_u+m_c} \,, &
\end{align}
where $f_D^{}$ stands for the $D^0$ decay constant and $p_X^{}$ ($m_X$) is the momentum (mass) of $X$.
There are no contributions of $\kappa_{\texttt{SS}'}^{\textsc s}$ and $\kappa_{\texttt{SS}'}^{\textsc v}$ because
\,$\langle0|\overline u\gamma^\mu c|D^0\rangle = \langle0|\overline uc|D^0\rangle = 0$.\,
Neglecting $m_u$ compared to $m_c$ then leads to
\begin{align}
	{\mathcal M}_{D^0\to{\texttt S}\bar{\texttt S}{}'} & \,=\, i \big[ \kappa_{\texttt{SS}'}^{\textsc a} \big(m_{\texttt S'}^2-m_{\texttt S}^2\big) + \kappa_{\texttt{SS}'}^{\textsc p} m_{D^0}^2 \big] f_D^{} \,. &
\end{align}
From this follows the rate
\begin{align}
	\Gamma_{D^0\to{\texttt S}\bar{\texttt S}{}'} & \,=\, \frac{\lambda^{1/2}\big(m_{D^0}^2,m_{\texttt S}^2,m_{\texttt S'}^2\big)}{16\pi m_{D^0}^3} \big| \kappa_{\texttt{SS}'}^{\textsc a} \big(m_{\texttt S'}^2-m_{\texttt S}^2\big) + \kappa_{\texttt{SS}'}^{\textsc p} m_{D^0}^2 \big|^2 f_D^2 \,, &
	\label{D2SS'}
\end{align}
which contains the K$\ddot{\rm a}$ll$\acute{\rm e}$n function
\,$\lambda(x, y, z) = x^2 + y^2 +z^2 -2(xy+xz+yz)$.\, 
Evidently, \,$D^0\to{\texttt S}\bar{\texttt S}{}'$  can in general probe $\kappa_{\texttt{SS}'}^{\textsc p}$ and $\kappa_{\texttt{SS}'}^{\textsc a}$, which accompany the parity-odd quark bilinears in ${\mathcal L}_{\texttt{SS}'}$, but the sensitivity to $\kappa_{\texttt{SS}'}^{\textsc a}$ will be lost if \,$m_{\texttt S'}=m_{\texttt S}$.

\subsection{Semi-invisible radiative decay\label{D0->gSS'}}

Although $\kappa_{\texttt{SS}'}^{\textsc v}$ has no impact on the preceding mode, as does $\kappa_{\texttt{SS}'}^{\textsc a}$ if \,$\texttt S$\, and \,$\texttt S'$ are degenerate, these parameters can contribute together if an ordinary photon is radiated, namely in \,$D^0\to\gamma{\texttt S}\bar{\texttt S}{}'$.
The amplitude for it is
\begin{align} \label{MD2gSS'}
	{\mathcal M}_{D^0\to\gamma{\texttt S}\bar{\texttt S}{}'} & \,=\, \kappa_{\texttt{SS}'}^{\textsc v}\, \langle\gamma|\overline u\gamma_\mu c|D^0 \rangle ({\texttt k}-{\texttt k}')^\mu + \kappa_{\texttt{SS}'}^{\textsc a}\, \langle\gamma|\overline u\gamma^\mu\gamma_5^{}c|D^0 \rangle ({\texttt k}-{\texttt k}')_\mu \,, &
\end{align}
where \,$\texttt k^{(\prime)}$ designates the momentum of \,$\texttt S^{(\prime)}$ and
\begin{align} \label{D->g}
	\langle\gamma|\overline u\gamma_\mu c|D^0\rangle & \,=\, \frac{e F_V}{m_{D^0}}\epsilon_{\mu\zeta\eta\theta}^{}\, \varepsilon_\gamma^{\zeta*} p_{D^0}^\eta\, p_\gamma^\theta \,, &   
	\langle\gamma|\overline u\gamma^\mu\gamma_5^{}c|D^0\rangle & \,=\, \frac{i e F_A}{m_{D^0}} \big( p_\gamma^{}\!\cdot\!p_{D^0}^{}\, \varepsilon_\gamma^{\mu*} - \varepsilon_\gamma^*\!\cdot\!p_{D^0}^{}\, p_\gamma^\mu \big) \,,  
\end{align}
with $e$ being the proton charge, $\varepsilon_X^{}$ denoting the polarization vector of $X$, and $F_V$ and $F_A$ symbolizing form factors depending on the squared momentum-transfer $(p_{D^0}-p_\gamma)^2=({\texttt k}+{\texttt k}')^2\equiv\hat s$.
Since \,$\langle\gamma|\overline uc|D^0\rangle = \langle\gamma|\overline u\gamma_5^{}c|D^0\rangle=0$,\, there are no $\kappa_{\texttt{SS}'}^{\textsc s,\textsc p}$ terms in eq.\,(\ref{MD2gSS'}).
It is obvious that ${\mathcal M}_{D^0\to\gamma{\texttt S}\bar{\texttt S}{}'}$ satisfies the requirement of electromagnetic gauge-invariance.
Evaluating the absolute square of the amplitude times the three-body phase space, one obtains the differential rate
\begin{align} \label{G'D2gSS'}
	\frac{d\Gamma_{D^0\,\to\gamma{\texttt S}\bar{\texttt S}{}'}}{d\hat s} & \,=\, \frac{ \alpha_{\rm e}^{}\,
		\lambda^{3/2}\big(\hat s,m_{\texttt S}^2,m_{\texttt S'}^2\big)}{384 \pi^2\, m_{D^0}^5\, \hat s^2} \big(m_{D^0}^2-\hat s\big)^3\, \big(|\kappa_{\texttt{SS}'}^{\textsc v}|^2 F_V^2 + |\kappa_{\texttt{SS}'}^{\textsc a}|^2 F_A^2\big) \,, &
\end{align}
which is to be integrated over \,$(m_{\texttt S}+m_{\texttt S'})^2\le\hat s\le m_{D^0}^2$.\,
Thus, the invisible scalars' mass range covered by this mode is \,$0\le m_{\texttt S}+m_{\texttt S'}<m_{D^0}$,\, the same as that in the $D^0\,\to{\texttt S}\bar{\texttt S}{}'$ case.
All this illustrates the importance of \,$D^0\to\gamma\slashed E$\, as a valuable search tool for new physics, despite its rate having a suppression factor of \,$\alpha_{\rm e}=e^2/(4\pi)=1/137$.

\subsection{Semi-invisible mesonic decays\label{D->MSS'}}

The interactions in ${\mathcal L}_{\texttt{SS}'}$ can cause a pseudoscalar charmed-meson \,$\mathbb D$ to turn into a pseudoscalar or vector charmless-meson, \,$\mathbb P$ or $\mathbb V$, plus the \,$\texttt S\bar{\texttt S}{}'$ pair.
Specifically, we will look at the instances where \,${\mathbb D}=D^0,D^+,D_s^+$\, and the final mesons are \,${\mathbb P}=\pi^0,\pi^+,K^+$\, or \,${\mathbb V}=\rho^0,\rho^+,K^{*+}$,\, respectively.
The amplitudes for these channels are
\begin{align}
	{\mathcal M}_{{\mathbb D}\to{\mathbb P}{\texttt S}\bar{\texttt S}{}'} & \,=\, \kappa_{\texttt{SS}'}^{\textsc v}\, \langle {\mathbb P}|\overline u\gamma^\mu c|{\mathbb D}\rangle ({\texttt k}-{\texttt k}')_\mu + \kappa_{\texttt{SS}'}^{\textsc s}\, m_c^{}\, \langle{\mathbb P}|\overline uc|{\mathbb D}\rangle \,,
	\label{MD2PSS'} \\ \raisebox{3ex}{}
	{\mathcal M} _{{\mathbb D}\to{\mathbb V}{\texttt S}\bar{\texttt S}{}'} & \,=\, \kappa_{\texttt{SS}'}^{\textsc v}\, \langle{\mathbb V}|\overline u\gamma_\mu c|{\mathbb D}\rangle ({\texttt k}-{\texttt k}')^\mu + \kappa_{\texttt{SS}'}^{\textsc a}\, \langle{\mathbb V}|\overline u\gamma^\mu\gamma_5^{}c|{\mathbb D}\rangle ({\texttt k}-{\texttt k}')_\mu + \kappa_{\texttt{SS}'}^{\textsc p}\, m_c^{}\, \langle {\mathbb V}|\overline u\gamma_5^{}c|{\mathbb D}\rangle \,, ~~~
	\label{MD2VSS'}
\end{align}
which involve the momentum \,$\texttt k^{(\prime)}$ of \,$\texttt S^{(\prime)}$ and the mesonic matrix elements 
\begin{align} \label{D->P}
	\langle {\mathbb P}|\overline u\gamma^\mu c|{\mathbb D}\rangle & \,=\, \big( p_{\mathbb D}^\mu+p_{\mathbb P}^\mu\big) f_+^{} + \big(p_{\mathbb D}^\mu-p_{\mathbb P}^\mu\big)\big(f_0^{}-f_+^{}\big) \frac{m_{\mathbb D}^2-m_{\mathbb P}^2}{q_{\mathbb{DP}}^2} \,, 
	\nonumber \\
	\langle {\mathbb P}|\overline uc|{\mathbb D}\rangle & \,=\, \frac{m_{\mathbb D}^2-m_{\mathbb P}^2}{m_c-m_u}f_0^{} \,,
	\\ \raisebox{5ex}{}
	\langle {\mathbb V}|\overline u \gamma_\mu c|{\mathbb D}\rangle	& \,=\, \frac{2 V}{m_{\mathbb D}+m_{\mathbb V}}\, \epsilon_{\mu\beta\eta\theta}^{}\, \varepsilon_{\mathbb V}^{\beta*}p_{\mathbb V}^\eta\, p_{\mathbb D}^\theta \,,
	\nonumber \\
	\langle {\mathbb V}|\overline u\gamma^\mu\gamma_5^{}c|{\mathbb D}\rangle & \,=\, i(m_{\mathbb D}+m_{\mathbb V}) \varepsilon_{\mathbb V}^{\mu*} A_1^{} - \bigg[ \frac{p_{\mathbb D}^\mu+p_{\mathbb V}^\mu}{m_{\mathbb D}+m_{\mathbb V}} A_2 + \frac{p_{\mathbb D}^\mu-p_{\mathbb V}^\mu}{q_{\mathbb{DV}}^2} (A_3-A_0) 2m_{\mathbb V} \bigg] i \varepsilon_{\mathbb V}^*\cdot p_{\mathbb D}^{} \,,
	\nonumber \\
	\langle {\mathbb V}|\overline u\gamma_5^{}c|{\mathbb D}\rangle & \,=\, \frac{-2i A_0\, m_{\mathbb V}}{m_c+m_u}\, \varepsilon_{\mathbb V}^* \cdot p_{\mathbb D}^{} \,, &
	\label{D->V}
\end{align}
where $f_+^{}$ and $f_0^{}$ \,[$V$, $A_0$, $A_1$, and $A_2$] are form factors which are functions of the squared momentum-transfer \,$q_{\mathbb{DP}}^2=(p_{\mathbb D}-p_{\mathbb P})^2$\, $\big[q_{\mathbb{DV}}^2=(p_{\mathbb D}-p_{\mathbb V})^2\big]$ and 
\,$2 A_3 m_{\mathbb V} = (m_{\mathbb D}+m_{\mathbb V}) A_1-(m_{\mathbb D}-m_{\mathbb V}) A_2$.\,
Other $\kappa_{\texttt{SS}'}$ terms are absent from eqs.\,(\ref{MD2PSS'})-(\ref{MD2VSS'}) because \,$\langle{\mathbb P}|\overline u\gamma^\mu\gamma_5^{}c|{\mathbb D}\rangle = \langle {\mathbb P}|\overline u\gamma_5^{}c|{\mathbb D}\rangle = \langle {\mathbb V}|\overline uc|{\mathbb D}\rangle=0$.

Given that \,$m_u=0.002\,m_c$,\, henceforth we ignore $m_u$ relative to $m_c$ when calculating decay rates.
Accordingly, from the absolute squares of the amplitudes in eqs.\,(\ref{MD2PSS'})-(\ref{MD2VSS'}), we arrive at
\begin{align}
	\frac{d\Gamma_{{\mathbb D}\,\to{\mathbb P}{\texttt S}\bar{\texttt S}{}'}}{d\hat s} = \frac{ 2
		\tilde\lambda_{\mathbb{DP}}^{1/2} \tilde\lambda_{\texttt{SS}'}^{1/2} }{(8\pi m_{\mathbb D} \hat s)^3} &
	\bigg[ \frac{1}{3}|\kappa_{\texttt{SS}'}^{\textsc v}|^2 \tilde\lambda_{\mathbb{DP}}^{}
	\tilde\lambda_{\texttt{SS}'}^{} f_+^2 + \big| \kappa_{\texttt{SS}'}^{\textsc v}
	\big(m_{\texttt S}^2-m_{\texttt S'}^2\big) + \kappa_{\texttt{SS}'}^{\textsc s} \hat s \big|{}^2
	\big(m_{\mathbb D}^2-m_{\mathbb P}^2\big){}^2 f_0^2 \bigg] \,, &
	\nonumber \\ \label{G'D2VSS'} \raisebox{1cm}{}
	\frac{d\Gamma_{{\mathbb D}\,\to{\mathbb V}{\texttt S}\bar{\texttt S}{}'}}{d\hat s} = \frac{
		\tilde\lambda_{\mathbb{DV}}^{3/2} \tilde\lambda_{\texttt{SS}'}^{3/2} }{(8\pi m_{\mathbb D} \hat s)^3} &
	\! \begin{array}[t]{l} \displaystyle \Bigg\{ \frac{|\kappa_{\texttt{SS}'}^{\textsc a}|^2}
		{6\,m_{\mathbb V}^2} \Bigg[ 	\Bigg( 1 + \frac{12\,m_{\mathbb V}^2\hat s}{\tilde\lambda_{\mathbb{DV}}} \Bigg) A_1^2\,\widetilde{\texttt m}_+^2
		+ 2 (\hat s-\widetilde{\texttt m}_+\widetilde{\texttt m}_-) A_1^{} A_2^{}
		+ \frac{\tilde\lambda_{\mathbb{DV}}A_2^2}{\widetilde{\texttt m}_+^2} \Bigg]
		\\ \displaystyle
		\;+~ \frac{2 A_0^2}{\tilde\lambda_{\texttt{SS}'}} \big| \kappa_{\texttt{SS}'}^{\textsc a}
		\big(m_{\texttt S'}^2-m_{\texttt S}^2\big) + \kappa_{\texttt{SS}'}^{\textsc p}\,\hat s \big|\raisebox{2pt}{$^2$}
		+ \frac{4|\kappa_{\texttt{SS}'}^{\textsc v}|^2 \hat s V^2}{3\, \widetilde{\texttt m}_+^2} \Bigg\} \,, \end{array}
\end{align}
to be integrated over \,$(m_{\texttt S}+m_{\texttt S'})^2\le\hat s=({\texttt k}+{\texttt k}')^2\le(m_{\mathbb D}-m_{\mathbb P,\mathbb V})^2$,\, respectively, with
\begin{align}
	\tilde\lambda_{\texttt XY}^{} & = \lambda\big(m_{\texttt X}^2,m_{\texttt Y}^2,\hat s\big) \,, & \widetilde{\texttt m}_\pm^{} & = m_{\mathbb D}^{} \pm m_{\mathbb V}^{} \,. &
\end{align}

In eq.\,(\ref{G'D2VSS'}), we see that \,${\mathbb D}\,\to{\mathbb P}{\texttt S}\bar{\texttt S}{}'$ can probe not only $\kappa_{\texttt{SS}'}^{\textsc v}$ but also $\kappa_{\texttt{SS}'}^{\textsc s}$, which is inaccessible to $D^0\to{\texttt S}\bar{\texttt S}{}',\gamma{\texttt S}\bar{\texttt S}{}'$ as well as to \,${\mathbb D}\,\to{\mathbb V}{\texttt S}\bar{\texttt S}{}'$.
However, the latter is sensitive to the other three parameters, $\kappa_{\texttt{SS}'}^{\textsc v,\textsc p,\textsc a}$.

\subsection{Semi-invisible baryonic decays\label{Lc->pSS'}}

Given that at the moment the empirical information on FCNC $|\Delta C|$\,=\,1 decays with missing energy is still scarce, it is essential to investigate, in addition, this type of transitions among baryons.
As we demonstrate shortly, they can play a complementary role in the quest for hints of new physics in \,$c\to u\slashed E$.

Of interest here are \,$\Lambda_c^+\to p{\texttt S}\bar{\texttt S}{}'$\, and \,$\Xi_c^{+,0}\to\Sigma^{+,0}{\texttt S}\bar{\texttt S}{}'$\, plus \,$\Xi_c^0\to\Lambda{\texttt S}\bar{\texttt S}{}'$,\, but we explicitly treat only the amplitude for the first decay and its rate, as the corresponding quantities for the other three have analogous formulas.
Thus, we write
\begin{align}
	{\mathcal M}_{\Lambda_c^+\to p{\texttt S}\bar{\texttt S}{}'} & \,=\, \kappa_{\texttt{SS}'}^{\textsc v}\, \langle p|\overline u\gamma^\mu c|\Lambda_c^+\rangle ({\texttt k}-{\texttt k}')_\mu + \kappa_{\texttt{SS}'}^{\textsc a}\, \langle p|\overline u\gamma^\mu\gamma_5^{}c|\Lambda_c^+\rangle ({\texttt k}-{\texttt k}')_\mu
	\nonumber \\ & ~~~ +\, \kappa_{\texttt{SS}'}^{\textsc s}\, m_c^{}\, \langle p|\overline uc|\Lambda_c^+\rangle + \kappa_{\texttt{SS}'}^{\textsc p}\, m_c^{}\, \langle p|\overline u\gamma_5^{}c|\Lambda_c^+\rangle \,, &
	\label{MLc2pSS'}
\end{align}
where \,$\texttt k^{(\prime)}$ is again the momentum of \,$\texttt S^{(\prime)}$ and the baryonic matrix elements are expressible as
\begin{align} \label{<p|uc|Lc>}
	\langle p|\overline u\gamma^\mu c|\Lambda_c^+\rangle & \,=\, \bar u_p^{} \bigg\{ \bigg[ \gamma^\mu - \frac{{\texttt M}_+ \hat p^\mu - {\texttt M}_-^{} \hat q^\mu}{{\texttt M}_+^2-\hat q^2} \bigg] {\texttt F}_\perp^{} + \bigg[ \hat p^\mu - \frac{{\texttt M}_+ {\texttt M}_-^{} \hat q^\mu}{\hat q^2} \bigg] \frac{{\texttt M}_+\, {\texttt F}_+}{{\texttt M}_+^2\mbox{$-$}\hat q^2} + \frac{{\texttt M}_-^{} \hat q^\mu}{\hat q^2} {\texttt F}_0^{} \bigg\} u_{\Lambda_c}^{} \,,
	\nonumber \\  
	\langle p|\overline u\gamma^\mu\gamma_5^{}c|\Lambda_c^+\rangle & \,=\, \bar u_p^{} \bigg\{ \bigg[ \gamma^\mu + \frac{{\texttt M}_- \hat p^\mu - {\texttt M}_+ \hat q^\mu}{{\texttt M}_-^2-\hat q^2} \bigg] {\texttt G}_\perp^{} - \bigg[ \hat p^\mu - \frac{{\texttt M}_+ {\texttt M}_-^{}\hat q^\mu}{\hat q^2} \bigg] \frac{{\texttt M}_-\, {\texttt G}_+}{{\texttt M}_-^2\mbox{$-$}\hat q^2} - \frac{{\texttt M}_+^{}\hat q^\mu}{\hat q^2} {\texttt G}_0^{} \bigg\} \gamma_5^{} u_{\Lambda_c}^{} \,, ~~~
	\nonumber \\  
	\langle p|\overline uc|\Lambda_c^+\rangle & \,=\, \frac{{\texttt M}_-^{}\, {\texttt F}_0}{m_c-m_u}\, \bar u_p^{} u_{\Lambda_c}^{} \,, ~~~~ ~~~~~ 
	\langle p|\overline u\gamma_5^{}c|\Lambda_c^+\rangle \,=\, \frac{{\texttt M}_+\, {\texttt G}_0}{m_c+m_u}\, \bar u_p^{}\gamma_5^{}u_{\Lambda_c}^{} \,,  
\end{align}
where $u_p$ and $u_{\Lambda_c}$ designate the Dirac spinors of the baryons, ${\texttt F}_{\perp,+,0}$ and ${\texttt G}_{\perp,+,0}$ symbolize form factors which depend on $\hat s=\hat q^2$, 
\begin{align}  
	{\texttt M}_\pm & \,=\, m_{\Lambda_c} \pm m_p \,, & \hat p & \,=\, p_{\Lambda_c}^{} + p_p^{} \,, & \hat q & \,=\, p_{\Lambda_c}^{} - p_p^{} \,. &	
\end{align}
After averaging (summing) the absolute square of the amplitude over the initial (final) baryon polarizations and multiplying by the three-body phase space, we find the differential rate
\begin{align} \label{G'Lc2pSS'} 
	\frac{d\Gamma_{\Lambda_c^+\to p{\texttt S}\bar{\texttt S}{}'}}{d\hat s} \,=\, \frac{ 2 \tilde\lambda_{\Lambda_c^{}p}^{1/2}\, \tilde\lambda_{\texttt{SS}'}^{1/2} }{3(8\pi m_{\Lambda_c}\hat s)^3} \! & \begin{array}[t]{l} \Big\{
		\Big[ |\kappa_{\texttt{SS}'}^{\textsc v}|^2 \big(2 {\texttt F}_\perp^2\hat s+{\texttt F}_+^2 {\texttt M}_+^2\big) \hat\sigma_-^{} + |\kappa_{\texttt{SS}'}^{\textsc a}|^2 \big(2 {\texttt G}_\perp^2\hat s+{\texttt G}_+^2 {\texttt M}_-^2\big) \hat\sigma_+^{} \Big] \tilde\lambda_{\texttt{SS}'}
		\vspace{4pt} \\ 
		\;+~ 3 \big| \kappa_{\texttt{SS}'}^{\textsc v} \big(m_{\texttt S}^2-m_{\texttt S'}^2\big) + \kappa_{\texttt{SS}'}^{\textsc s} \hat s \big|{}^2\, \hat\sigma_+^{}\, {\texttt M}_-^2\, {\texttt F}_0^2 \end{array} ~~~
	\nonumber \\ &  
	\;+\, 3 \big| \kappa_{\texttt{SS}'}^{\textsc a} \big(m_{\texttt S'}^2-m_{\texttt S}^2\big)	+ \kappa_{\texttt{SS}'}^{\textsc p} \hat s \big|{}^2\, \hat\sigma_-^{}\, {\texttt M}_+^2\, {\texttt G}_0^2 \Big\} \,, 
\end{align}
where \,$\hat\sigma_\pm^{} = {\texttt M}_\pm^2 - \hat s$.\, 
It is to be integrated over \,$(m_{\texttt S}+m_{\texttt S'})^2\le\hat s\le(m_{\Lambda_c}-m_p)^2$.

It is clear from eq.\,(\ref{G'Lc2pSS'}) that all of the four coefficients, $\kappa_{\texttt{SS}'}^{\textsc{s,p,v,a}}$, can be probed with this channel,\footnote{The expression in eq.\,(\ref{G'Lc2pSS'}) for \,$m_{\texttt S}=m_{\texttt S'}$\, may be compared to the corresponding formula in ref.\,\cite{Li:2019cbk} for the rate of FCNC hyperon decay with invisible new bosons of equal mass in the final state.\medskip} unlike the mesonic cases of the previous subsections.
However, it is worth pointing out that the $m_{\texttt S}+m_{\texttt S'}$ ranges that can be covered in the aforesaid baryonic modes are less than those in $D^0\to(\gamma){\texttt S}\bar{\texttt S}{}'$\, and \,${\mathbb D}\,\to{\mathbb P}{\texttt S}\bar{\texttt S}{}'$.

\section{Numerical results for hadron decays induced by \boldmath$c\to u{\texttt S}\bar{\texttt S}{}'$\label{nums}}

\subsection{Constraints on effective couplings\label{constr}}

As mentioned in section \ref{intro}, so far there have been only three attempts to look for FCNC $|\Delta C|$\,=\,1 processes with missing energy and the null outcomes translated into caps on their branching fractions.
The first two are ${\cal B}(D^0\to{\rm invisibles})<9.4\times10^{-5}$ and ${\cal B}(D^0\to\pi^0\nu\bar\nu)<2.1\times10^{-4}$ both at 90\% CL~\cite{Belle:2016qek,BESIII:2021slf}.
Since the neutrinos in the second measurement were unobserved, we can apply these data to test the predictions for $D^0\to{\texttt S}\bar{\texttt S}{}'$ and $D^0\to\pi^0{\texttt S}\bar{\texttt S}{}'$, respectively.
The third finding, ${\cal B}(\Lambda_c^+\to p\gamma')<8.0\times10^{-5}$ at 90\% CL~\cite{BESIII:2022vrr}, concerns a two-body decay with the missing energy carried away solely by a massless dark photon ($\gamma'$) and therefore would not pertain directly to the $\Lambda_c^+$ three-body case under study.
Nevertheless, the fact that BESIII has only recently acquired this bound indicates that it might in the near future also report its three-body counterpart, which would perhaps be comparable in order of magnitude.\footnote{With the  data sample cited in ref.\,\cite{BESIII:2022vrr} the three-body bound would be relatively weaker due to a decreased detection efficiency, but fresh data to be collected in a few years might lead to a stronger bound not far from what we have adopted.} 
This implies that, for the following numerical exercise, it is reasonable to suppose that the $\Lambda_c^+$ result above is also the limit for the three-body mode, and consequently we may impose
\begin{align} \label{c2uss'lim}
	{\cal B}(D^0\to{\texttt S}\bar{\texttt S}{}') & < 9.4\times10^{-5} \,, & {\cal B}(D^0\to\pi^0{\texttt S}\bar{\texttt S}{}') & < 2.1\times10^{-4} \,, ~~~ ~~~~ \nonumber \\ {\cal B}(\Lambda_c^+\to p{\texttt S}\bar{\texttt S}{}') & < 8.0\times10^{-5} \,. & 
\end{align}  
For discussion purposes, we regard the third number on the same footing as the other two, while keeping in mind that it is only suggestive, being inspired by the $\Lambda_c^+\to p\gamma'$ data.

\begin{figure}[!t]
	\includegraphics[width=0.47\textwidth]{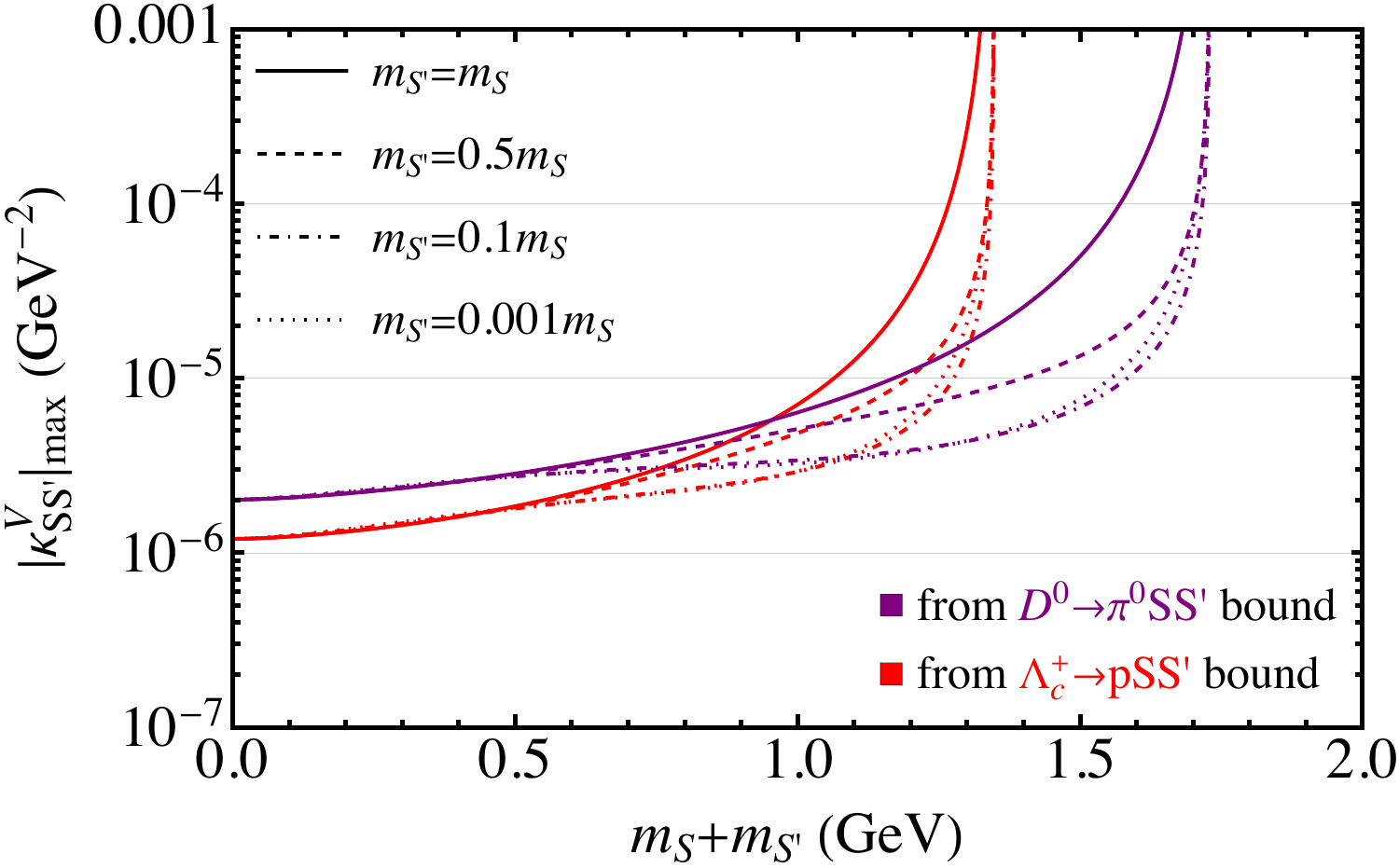} ~ ~ \includegraphics[width=0.47\textwidth]{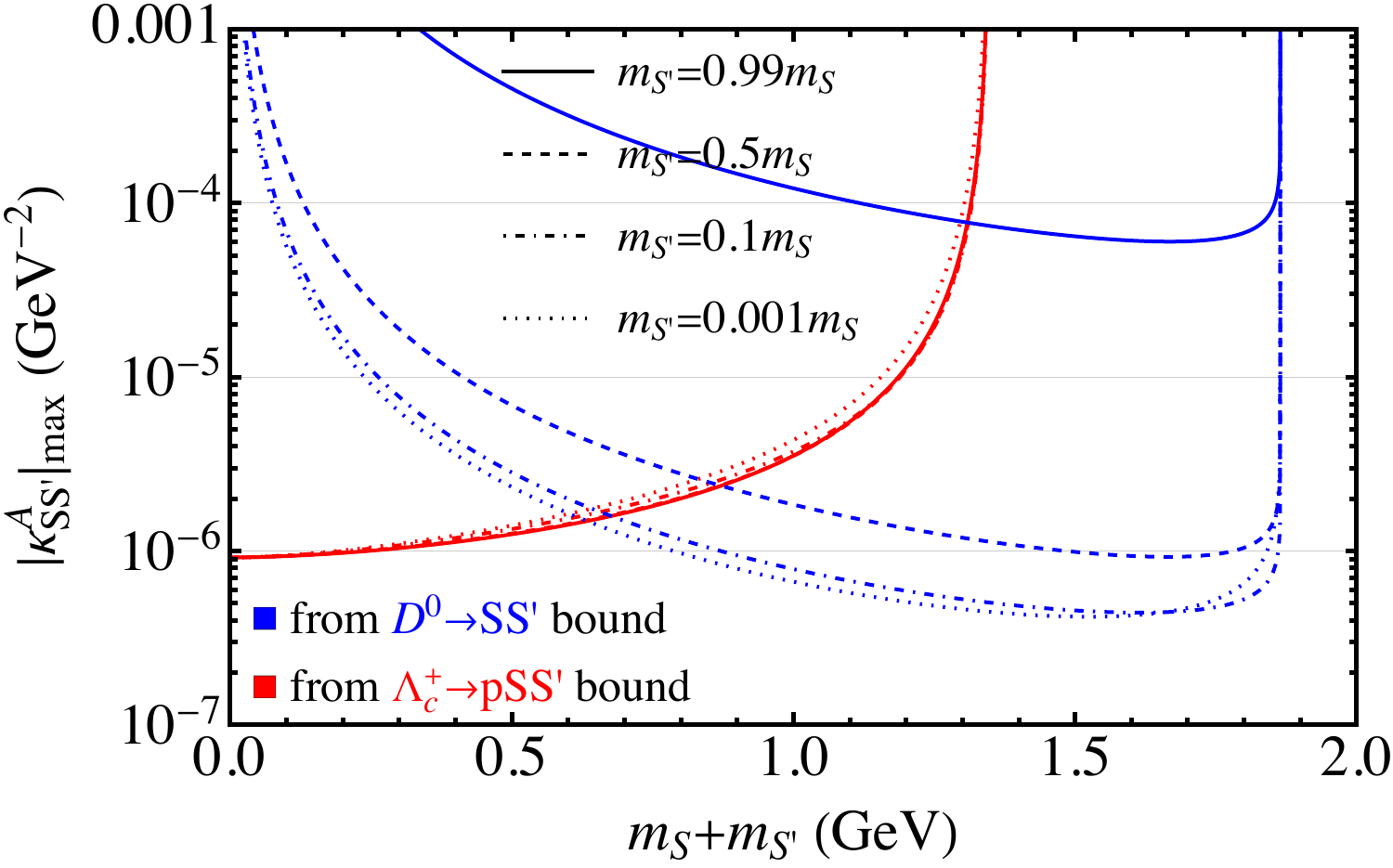}\vspace{7pt}\\
	\includegraphics[width=0.47\textwidth]{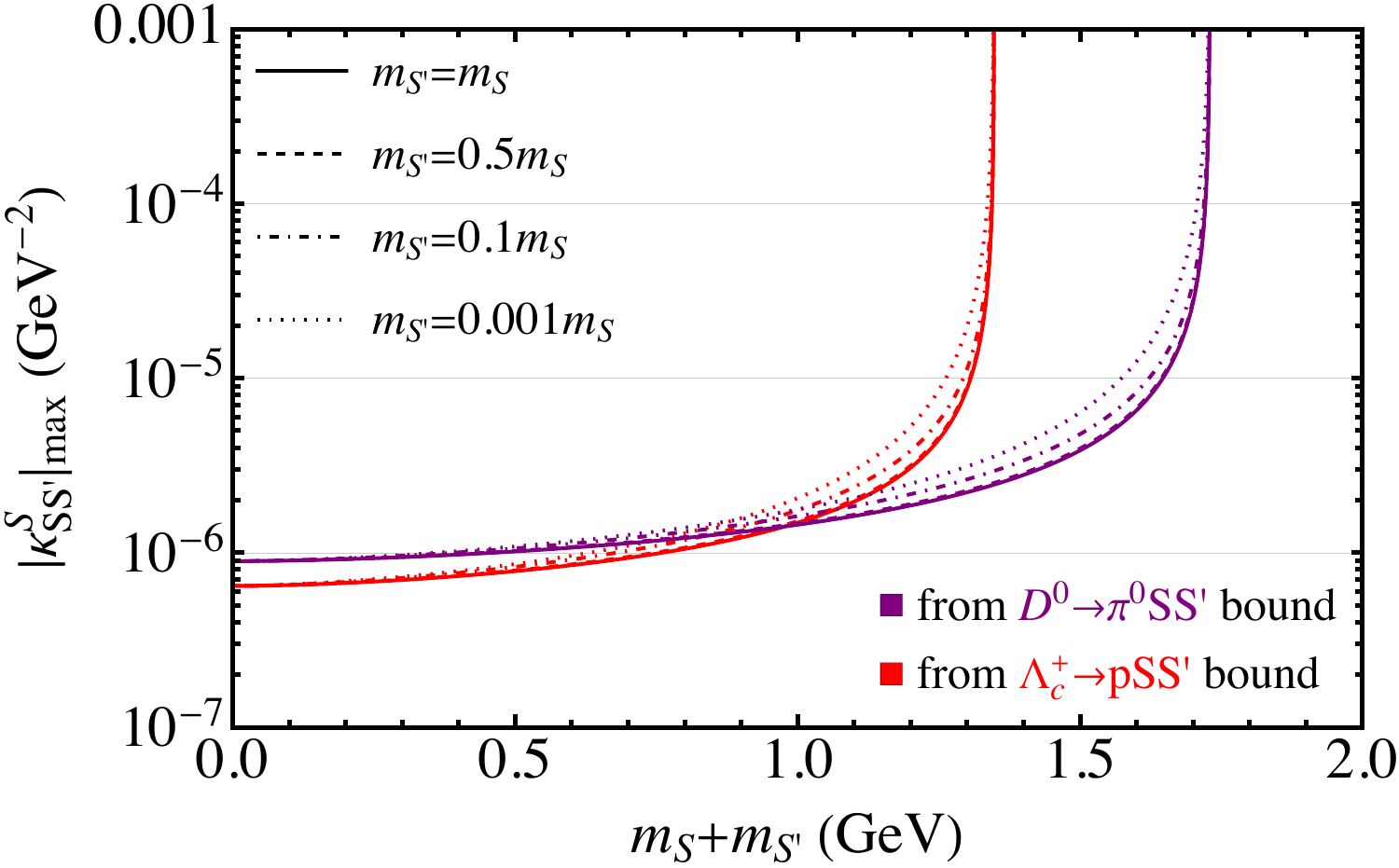} ~ ~ \includegraphics[width=0.47\textwidth]{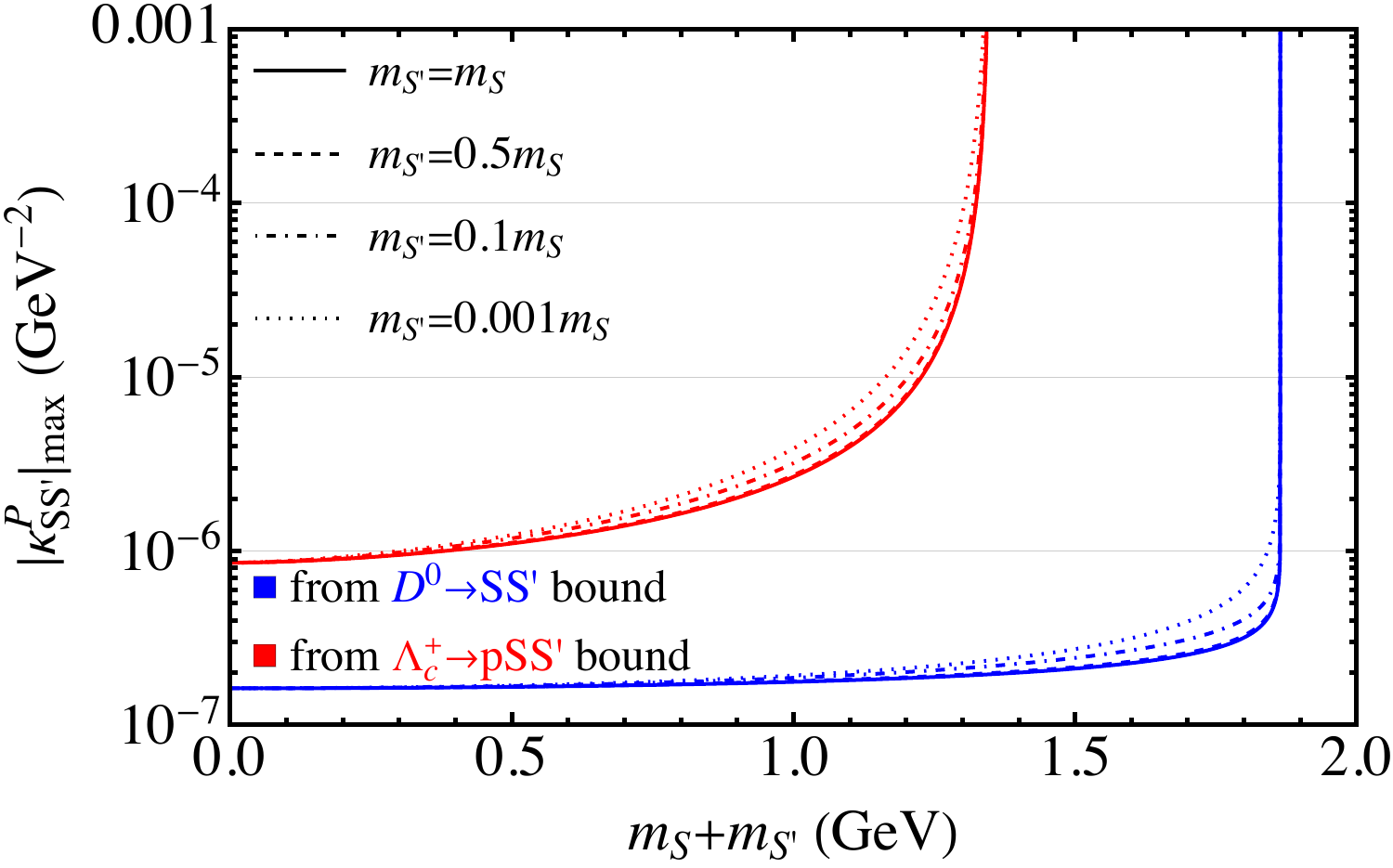}
	\caption{The upper limits on $|\kappa_{\texttt{SS}'}^{\textsc v}|$ (top left), $|\kappa_{\texttt{SS}'}^{\textsc a}|$ (top right), $|\kappa_{\texttt{SS}'}^{\textsc s}|$ (bottom left), and $|\kappa_{\texttt{SS}'}^{\textsc p}|$ (bottom right) versus \,$m_{\texttt S}+m_{\texttt S'}$\, obtained from the \,$D^0\to{\texttt S}\bar{\texttt S}{}'$ (blue), \,$D^0\to\pi^0{\texttt S}\bar{\texttt S}{}'$ (purple), and \,$\Lambda_c^+\to p{\texttt S}\bar{\texttt S}{}'$ (red) limits in eq.\,(\ref{c2uss'lim}) for various \,$m_{\texttt S'}/m_{\texttt S}$\, values if only one of $\kappa_{\texttt{SS}'}^{\textsc{v,a,s,p}}$ is nonzero at a time.}
	\label{k2}
\end{figure}

Hereafter, we entertain the possibility that merely one of the couplings $\kappa_{\texttt{SS}'}^{\textsc v,\textsc a,\textsc s,\textsc p}$ is nonvanishing at a time, which simplifies the analysis.
Moreover, accepting that $\texttt S$ and $\texttt S'$ can be nondegenerate, we include instances where \,$m_{\texttt S'}\neq m_{\texttt S}$.\,
In numerical calculations, we employ the central values of \,$f_D^{}=212.0(7)$\,MeV\, and the pertinent hadron lifetimes and masses from ref.\,\cite{ParticleDataGroup:2022pth} and the form factors specified in appendix \ref{FF}.

After implementing eq.\,(\ref{c2uss'lim}), we extract the maximal magnitudes of the individual couplings versus \,$m_{\texttt S}+m_{\texttt S'}$.\, 
The outcomes are depicted in figure \ref{k2}, where the blue, purple, and red curves correspond to the three limits in eq.\,(\ref{c2uss'lim}), respectively.
In each plot, the viable region for a particular \,$m_{\texttt S'}/m_{\texttt S}$\, case is below the lower of the purple or blue and red curves.

This figure makes plain, as alluded to earlier, that \,$\Lambda_c^+\to p{\texttt S}\bar{\texttt S}{}'$\, covers a narrower span of \,$m_{\texttt S}+m_{\texttt S'}$\, than \,$D^0\to\pi^0{\texttt S}\bar{\texttt S}{}'$\, can, and certainly more so than \,$D^0\to{\texttt S}\bar{\texttt S}{}'$.\,
Where the former two overlap in their mass coverage, we notice from the left portion of figure \ref{k2} that for \,$m_{\texttt S}+m_{\texttt S'} \;\mbox{\footnotesize$\lesssim$}\; 1$\,GeV  the values of $|\kappa_{\texttt{SS}'}^{\textsc v}|_{\rm max}^{}$ and $|\kappa_{\texttt{SS}'}^{\textsc s}|_{\rm max}^{}$ which are permitted by eq.\,(\ref{c2uss'lim}) are roughly comparable in order of magnitude.
By contrast, the top-right part of figure \ref{k2} reveals that $|\kappa_{\texttt{SS}'}^{\textsc a}|_{\rm max}^{}$ inferred from the \,$D^0\to{\texttt S}\bar{\texttt S}{}'$\, bound can be tremendously dissimilar to that from \,$\Lambda_c^+\to p{\texttt S}\bar{\texttt S}{}'$,\, depending on \,$m_{\texttt S}+m_{\texttt S'}$\, and \,$m_{\texttt S'}/m_{\texttt S}$,\, whereas the bottom-right graph shows that for $|\kappa_{\texttt{SS}'}^{\textsc p}|_{\rm max}^{}$ the \,$\Lambda_c^+\to p{\texttt S}\bar{\texttt S}{}'$\, limit is not competitive to the \,$D^0\to{\texttt S}\bar{\texttt S}{}'$\, one.
Furthermore, from the graphs in figure \ref{k2}, we learn that the $|\kappa_{\texttt{SS}'}^{\textsc s}|_{\rm max}^{}$ and $|\kappa_{\texttt{SS}'}^{\textsc p}|_{\rm max}^{}$ curves and the red $|\kappa_{\texttt{SS}'}^{\textsc a}|_{\rm max}^{}$ ones are not much affected by the choice of \,$m_{\texttt S'}/m_{\texttt S}$,\, the $|\kappa_{\texttt{SS}'}^{\textsc v}|_{\rm max}^{}$ curves are moderately dependent on this ratio, and the blue $|\kappa_{\texttt{SS}'}^{\textsc a}|_{\rm max}^{}$ ones manifest substantial variations with it.
The upward trend exhibited by the blue $|\kappa_{\texttt{SS}'}^{\textsc a}|_{\rm max}^{}$ curves in the top-right part of figure \ref{k2} as $m_{\texttt S'}$ approaches $m_{\texttt S}$ of course reflects the loosening of the \,$D^0\to{\texttt S}\bar{\texttt S}{}'$\, restraint, as dictated by eq.\,(\ref{D2SS'}).
Accordingly, it is interesting to observe that presently $\kappa_{\texttt{SS}'}^{\textsc a}$ is not subject to any empirical bound if \,$m_{\texttt S}=m_{\texttt S'}$\, and \,$m_{\texttt S}+m_{\texttt S'}>m_{\Lambda_c}-m_p\simeq1.36$\,\,GeV.  
On the other hand, there are still no experimental restrictions on $\kappa_{\texttt{SS}'}^{\textsc v}$ and $\kappa_{\texttt{SS}'}^{\textsc s}$ $\big(\kappa_{\texttt{SS}'}^{\textsc p}\big)$ if \,$m_{\texttt S}+m_{\texttt S'}$  exceeds \,$m_{D^0}-m_{\pi^0}\simeq1.73$\,\,GeV $(m_{D^0}\simeq1.86\rm\,GeV)$ regardless of $m_{\texttt S'}/m_{\texttt S}$.
Needles to say, the lack of constraints in the 1.36-1.86 GeV interval invites making the first effort to search for \,$D^0\to\gamma\slashed E$.

\subsection{Predictions\label{preds}}

\begin{figure}[b] \medskip
	\includegraphics[width=0.47\textwidth]{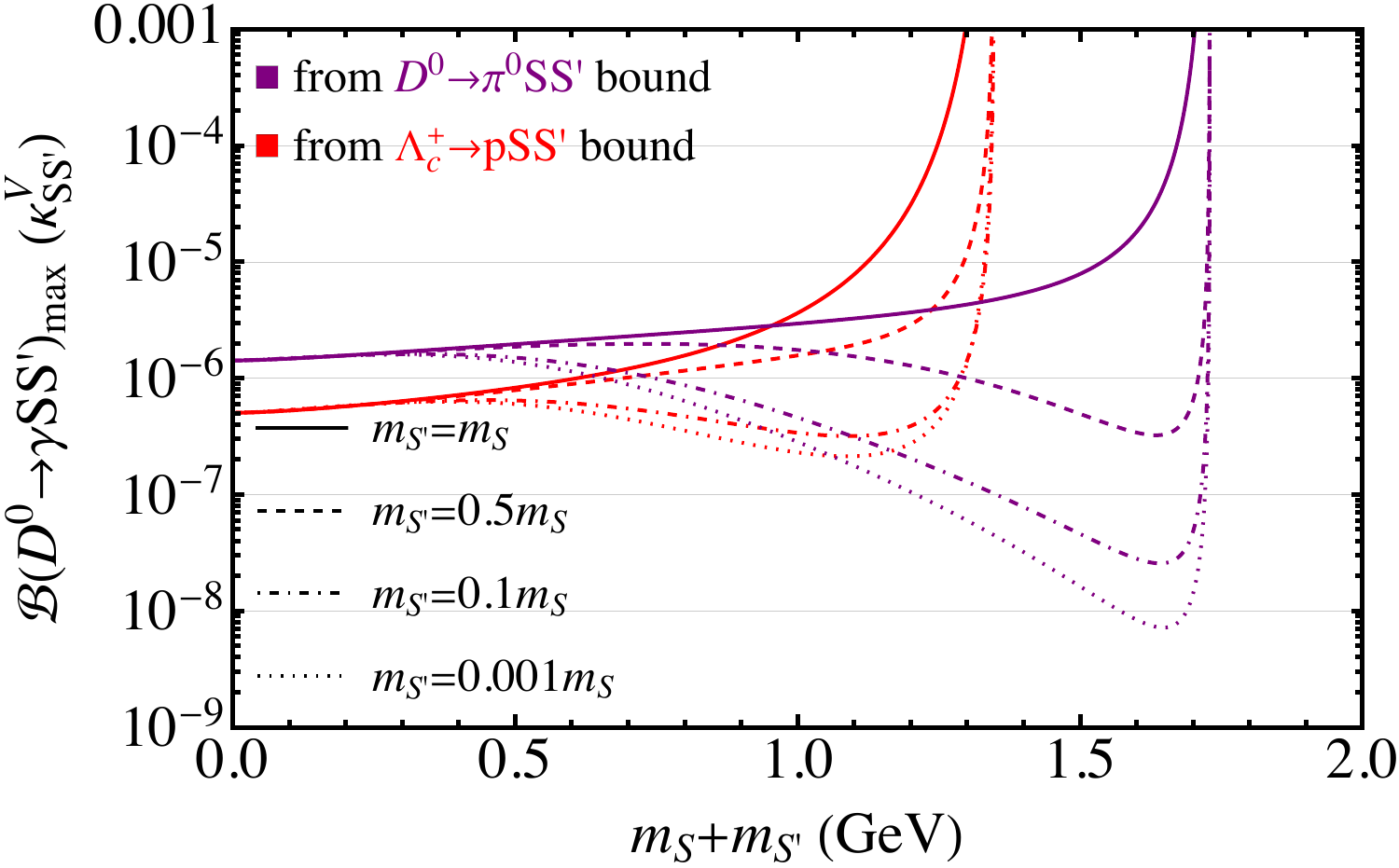} ~ ~ \includegraphics[width=0.47\textwidth]{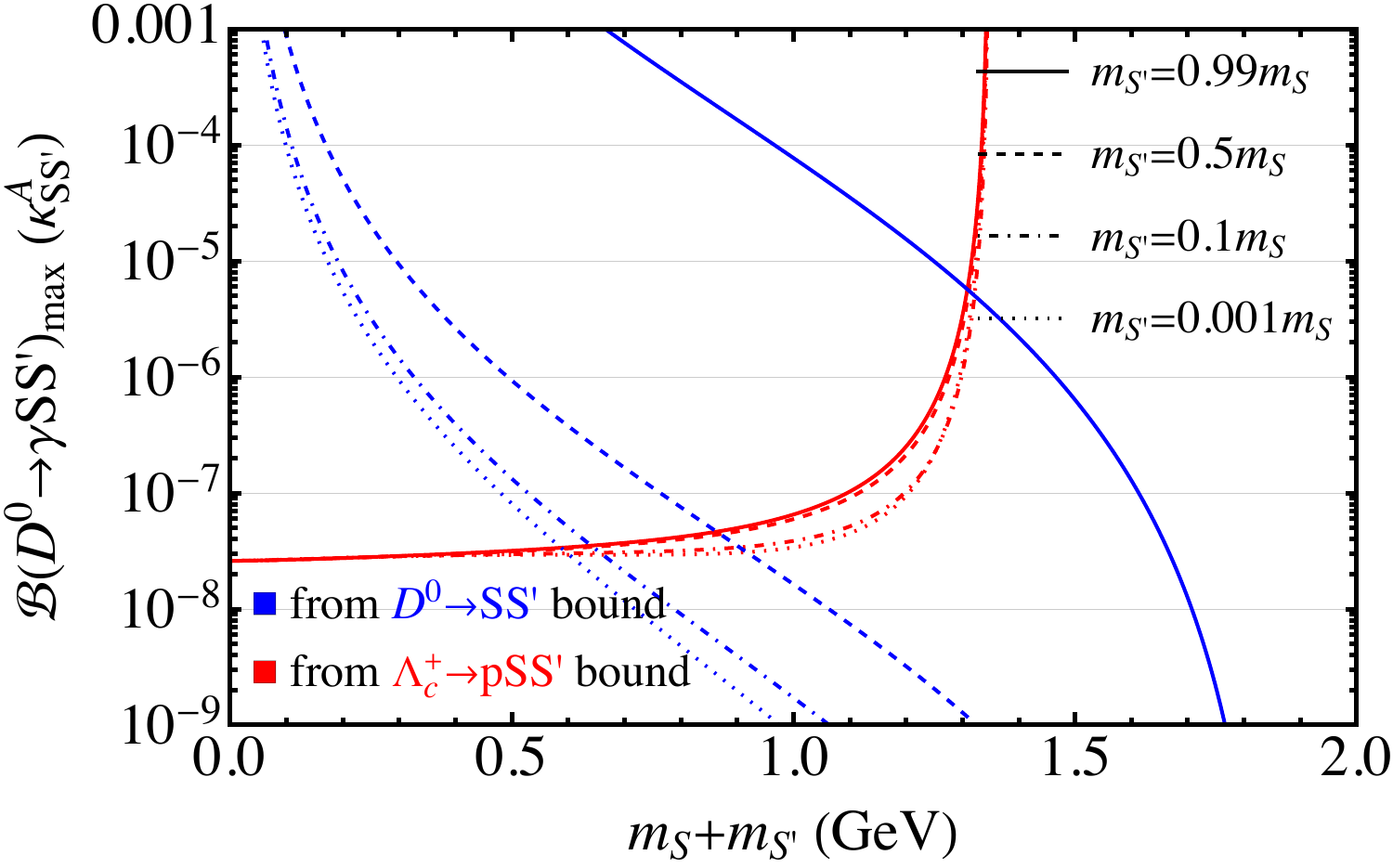} \vspace{-3pt}
	\caption{The maximal branching fraction of \,$D^0\to\gamma{\texttt S}\bar{\texttt S}{}'$\, due to $|\kappa_{\texttt{SS}'}^{\textsc v}|_{\rm max}^{}$ (left) or $|\kappa_{\texttt{SS}'}^{\textsc a}|_{\rm max}^{}$ (right) alone for various \,$m_{\texttt S'}/m_{\texttt S}$\, choices.}
	\label{D2gSS'}  
\end{figure}

\begin{figure}[t]  
	\includegraphics[width=0.47\textwidth]{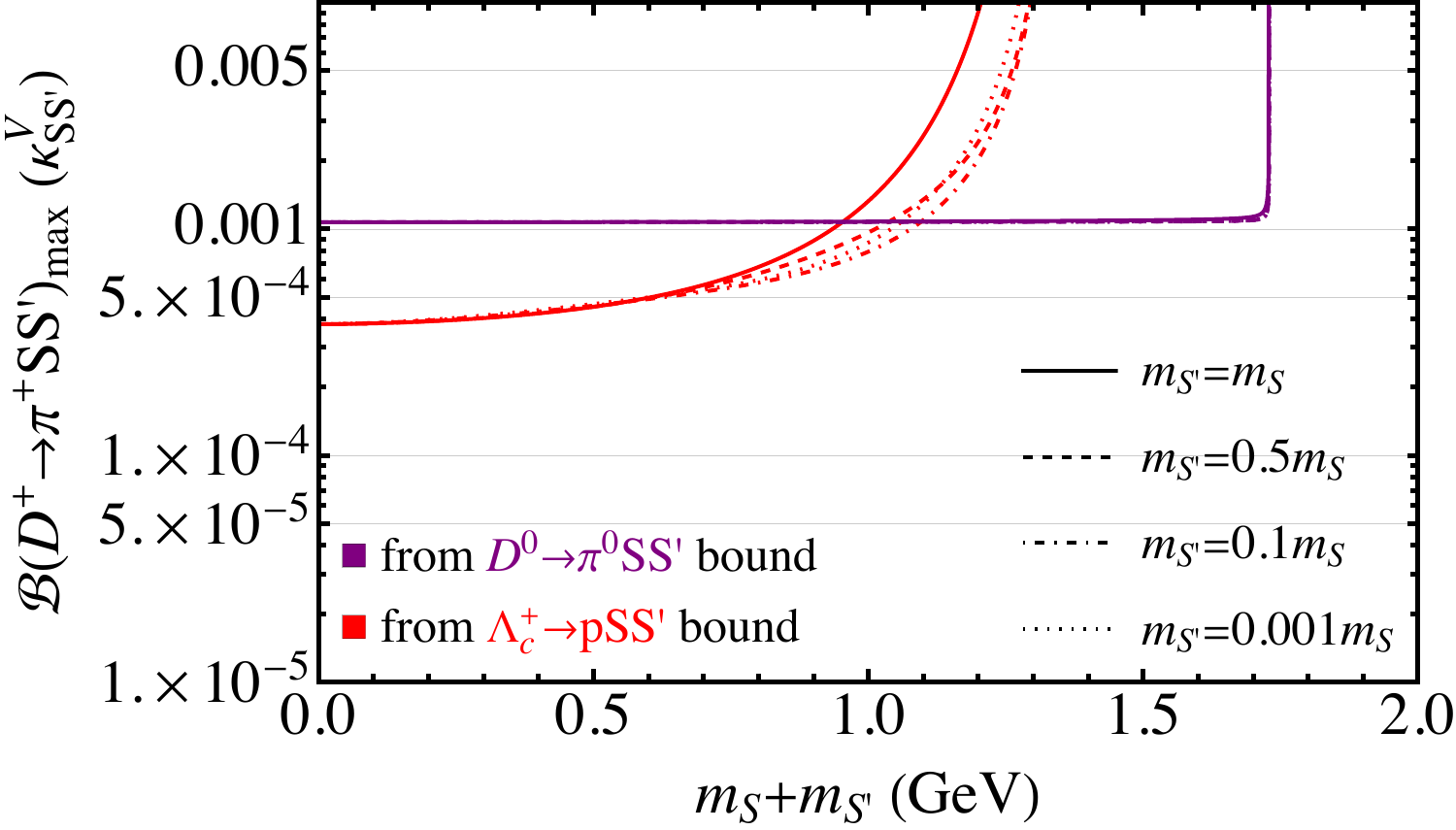} ~ ~ \includegraphics[width=0.47\textwidth]{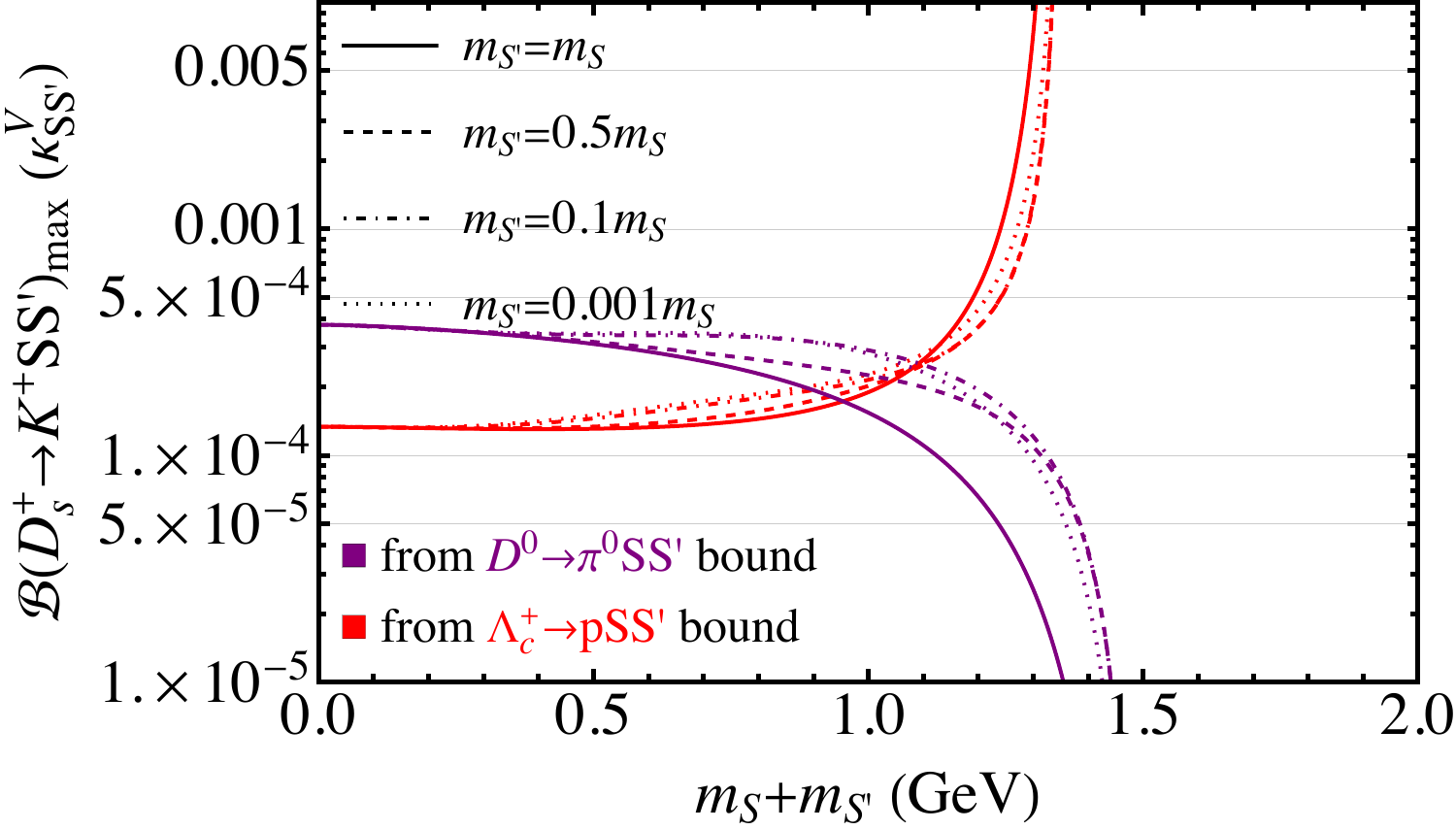}\vspace{9pt}\\
	\includegraphics[width=0.47\textwidth]{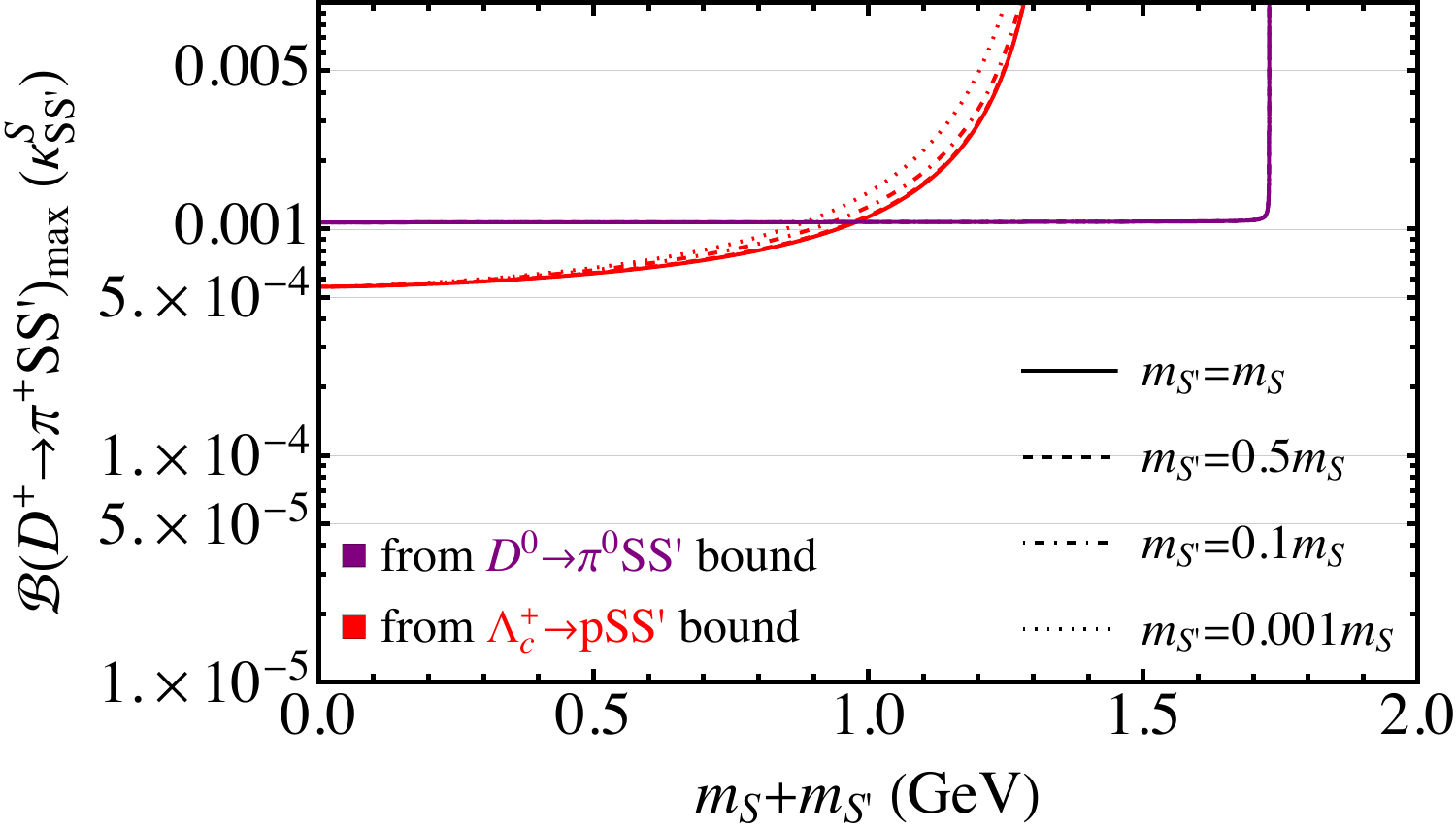} ~ ~ \includegraphics[width=0.47\textwidth]{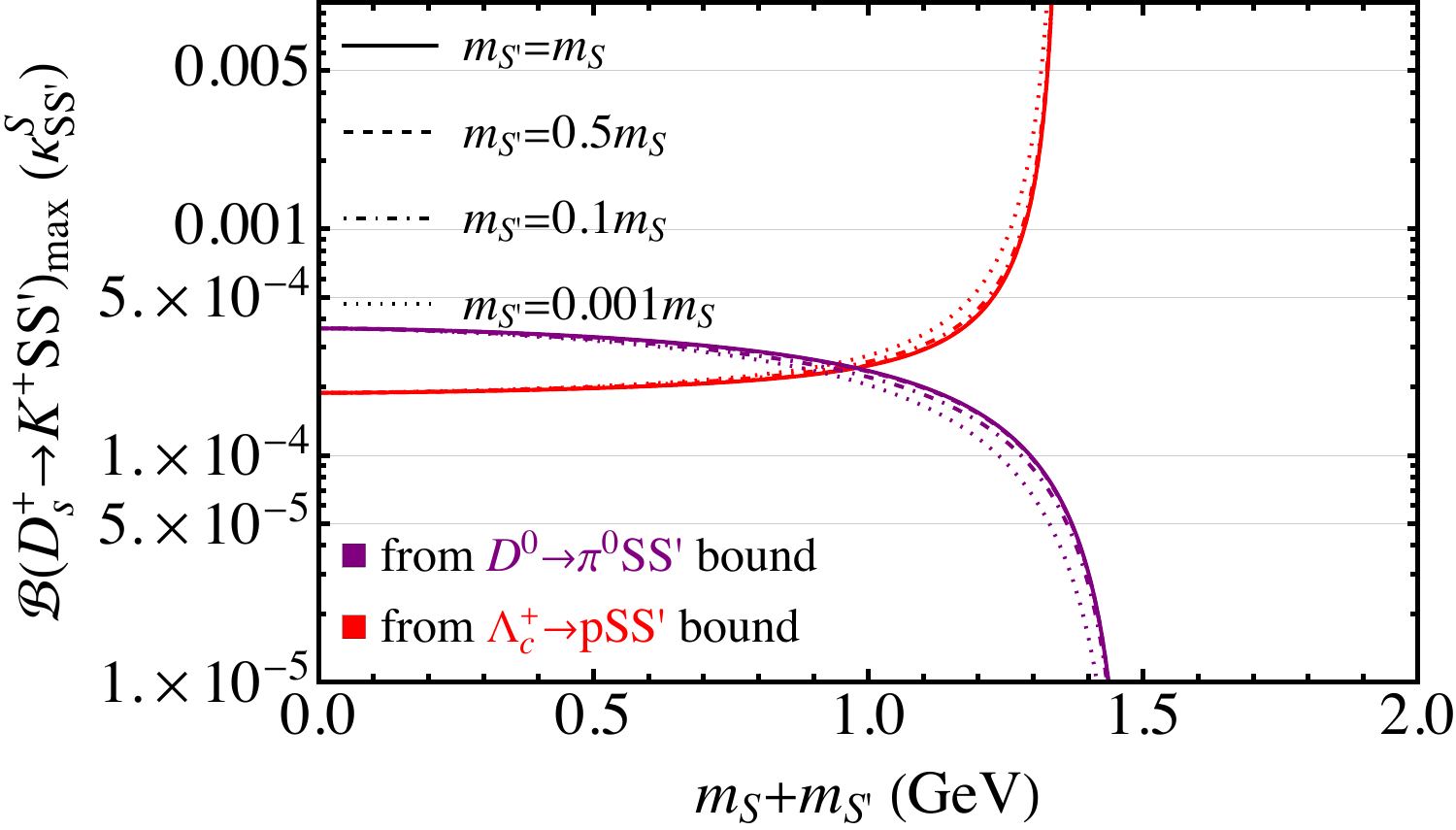}
	\vspace{-3pt}\\	
	\caption{The maximal branching fractions of \,$D^+\to\pi^+{\texttt S}\bar{\texttt S}{}'$ (left column) and \,$D_s^+\to K^+{\texttt S}\bar{\texttt S}{}'$ (right column) due to $|\kappa_{\texttt{SS}'}^{\textsc v}|_{\rm max}^{}$ (top row) or $|\kappa_{\texttt{SS}'}^{\textsc s}|_{\rm max}^{}$ (bottom row) alone for different \,$m_{\texttt S'}/m_{\texttt S}$\, values.}
	\label{D2PSS'}  
\end{figure}

\begin{figure}[t]
	\includegraphics[width=0.47\textwidth]{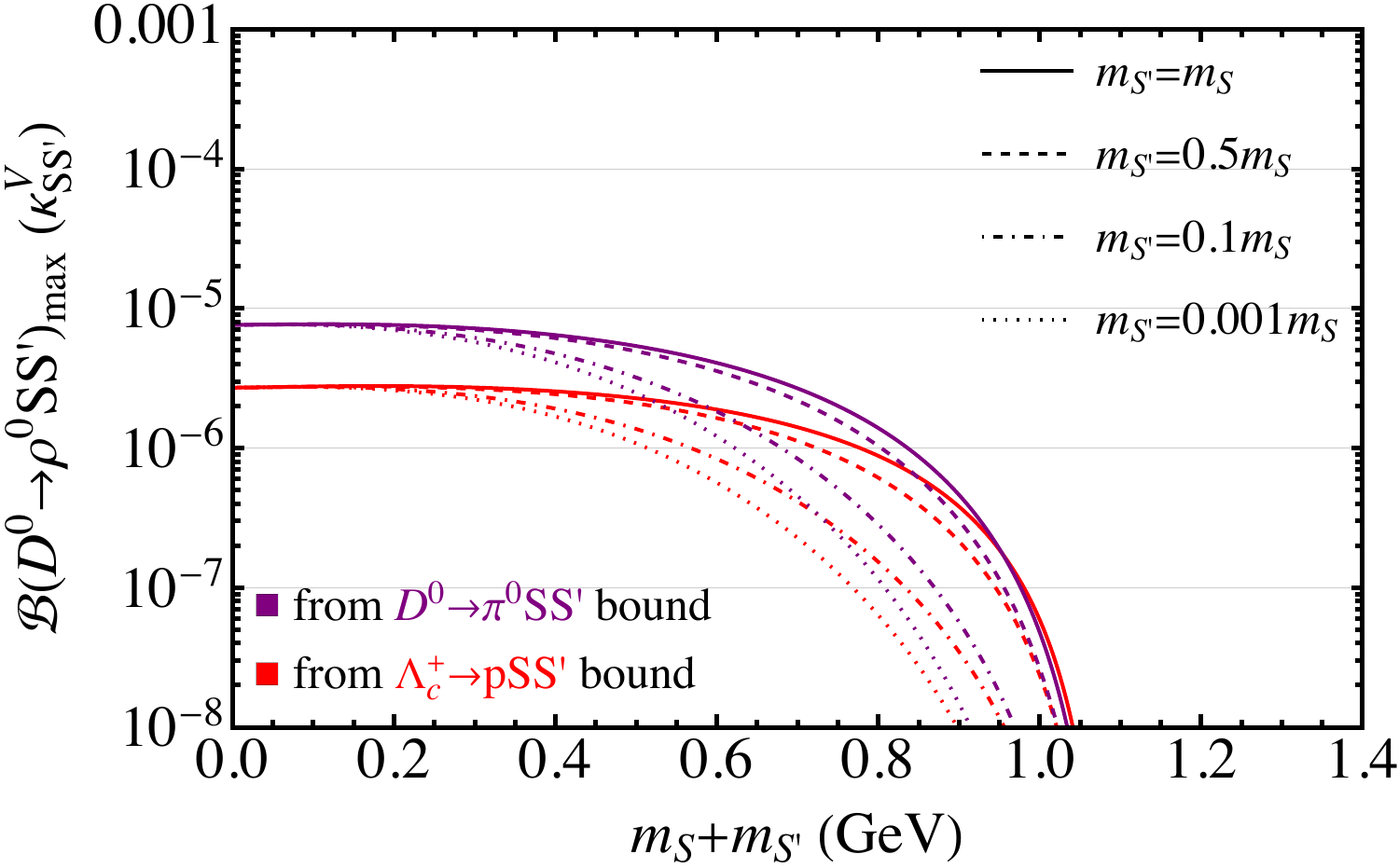} ~ ~ \includegraphics[width=0.47\textwidth]{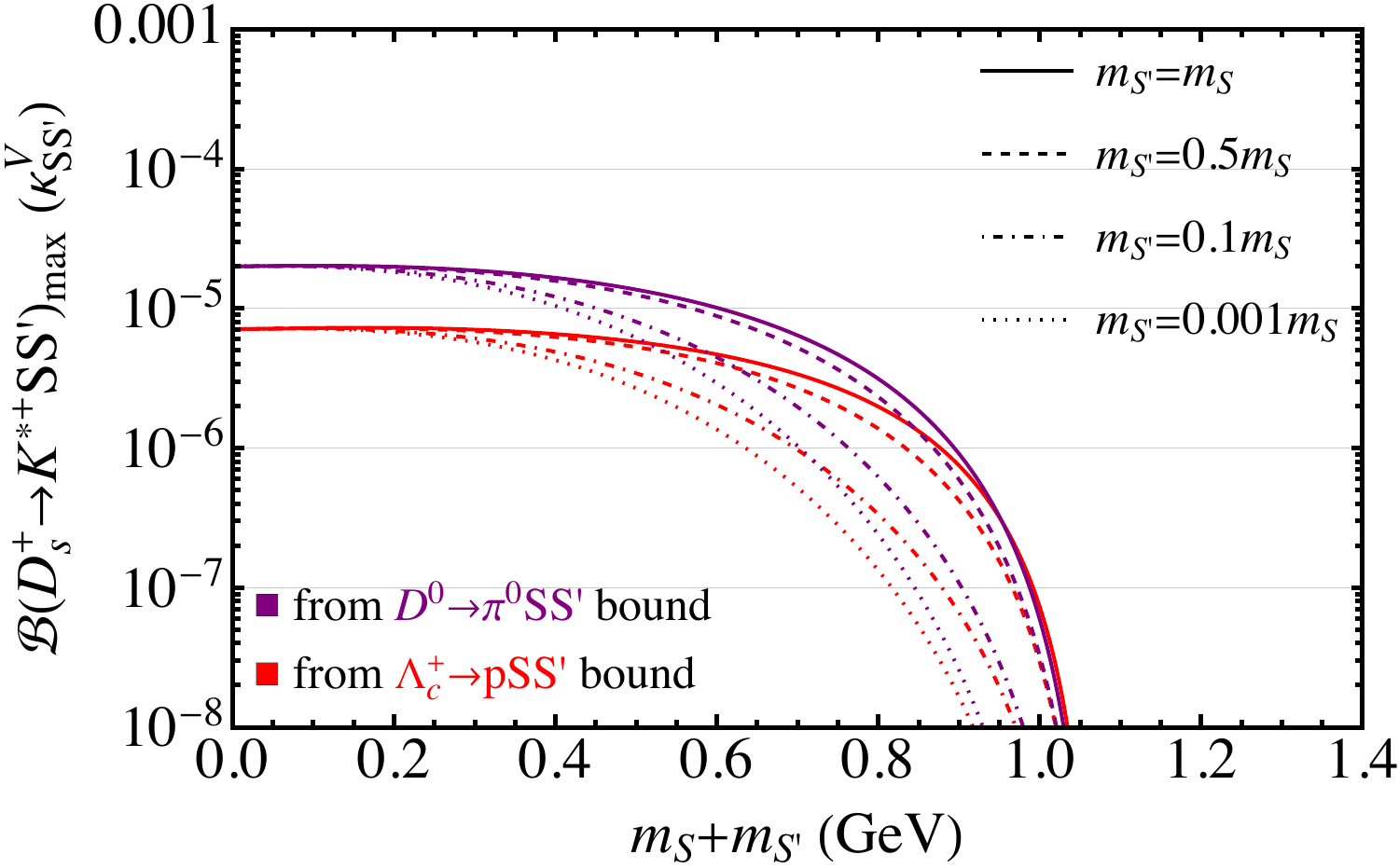}\vspace{7pt}\\	
	\includegraphics[width=0.47\textwidth]{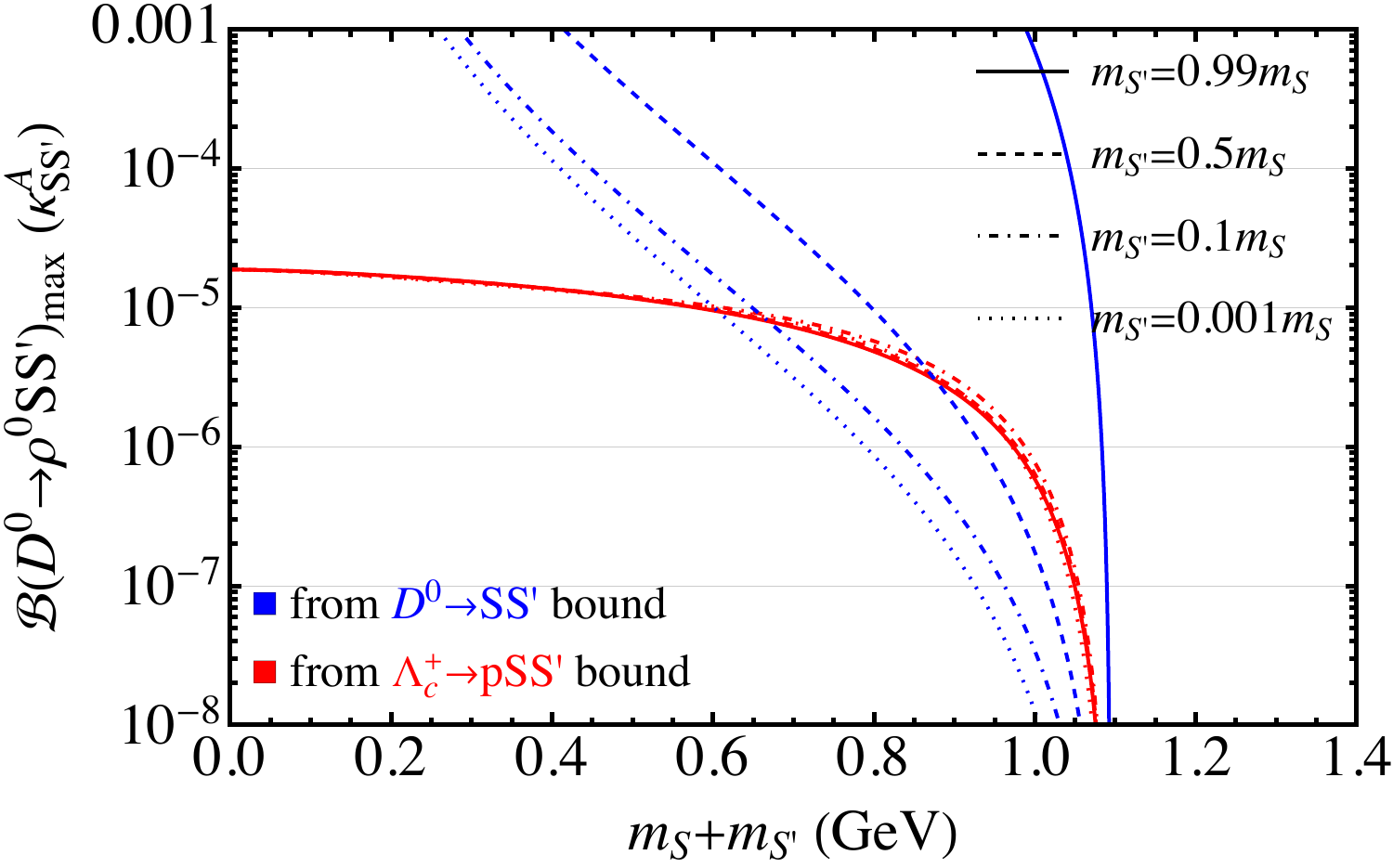} ~ ~ \includegraphics[width=0.47\textwidth]{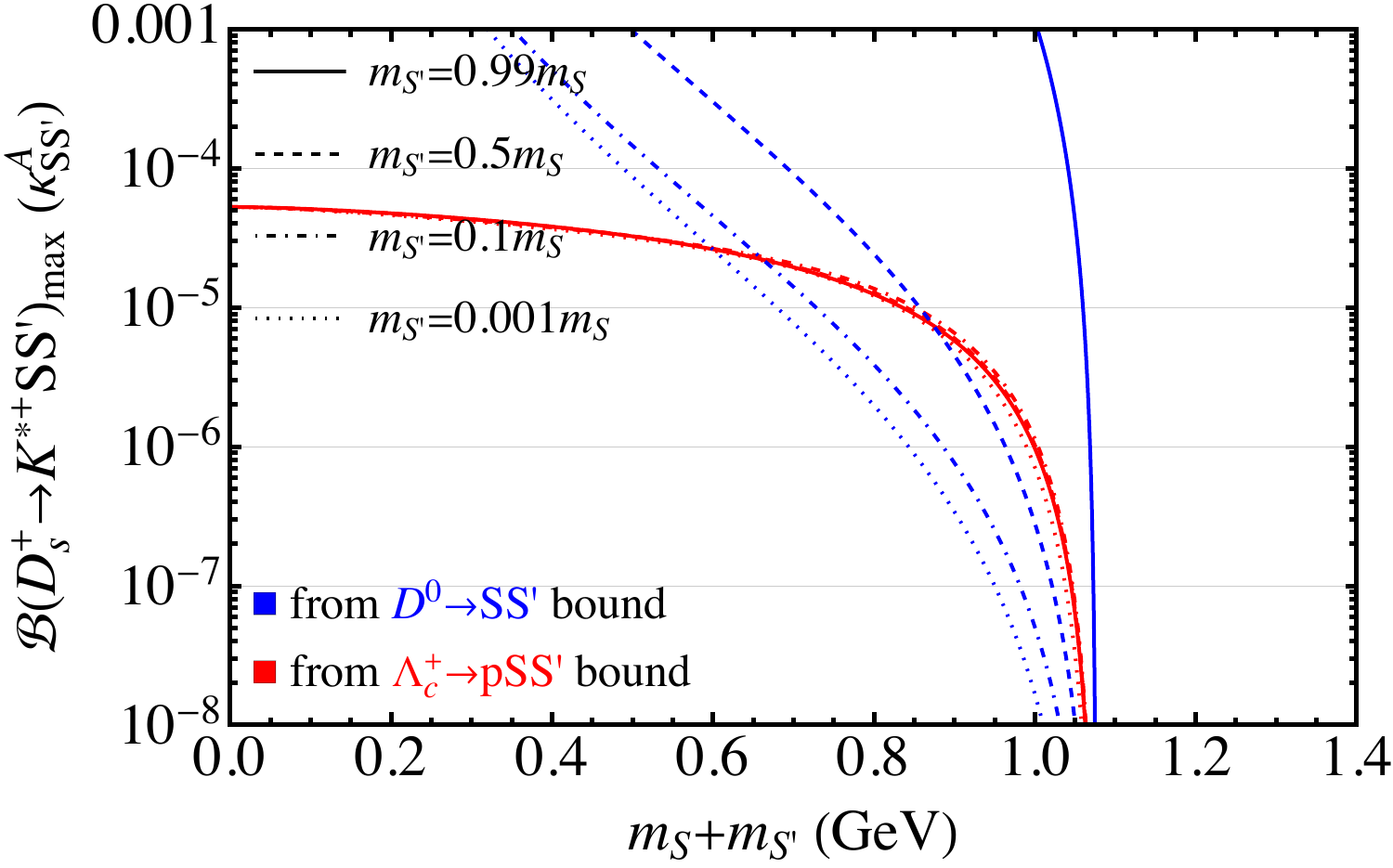}\vspace{7pt}\\	
	\includegraphics[width=0.47\textwidth]{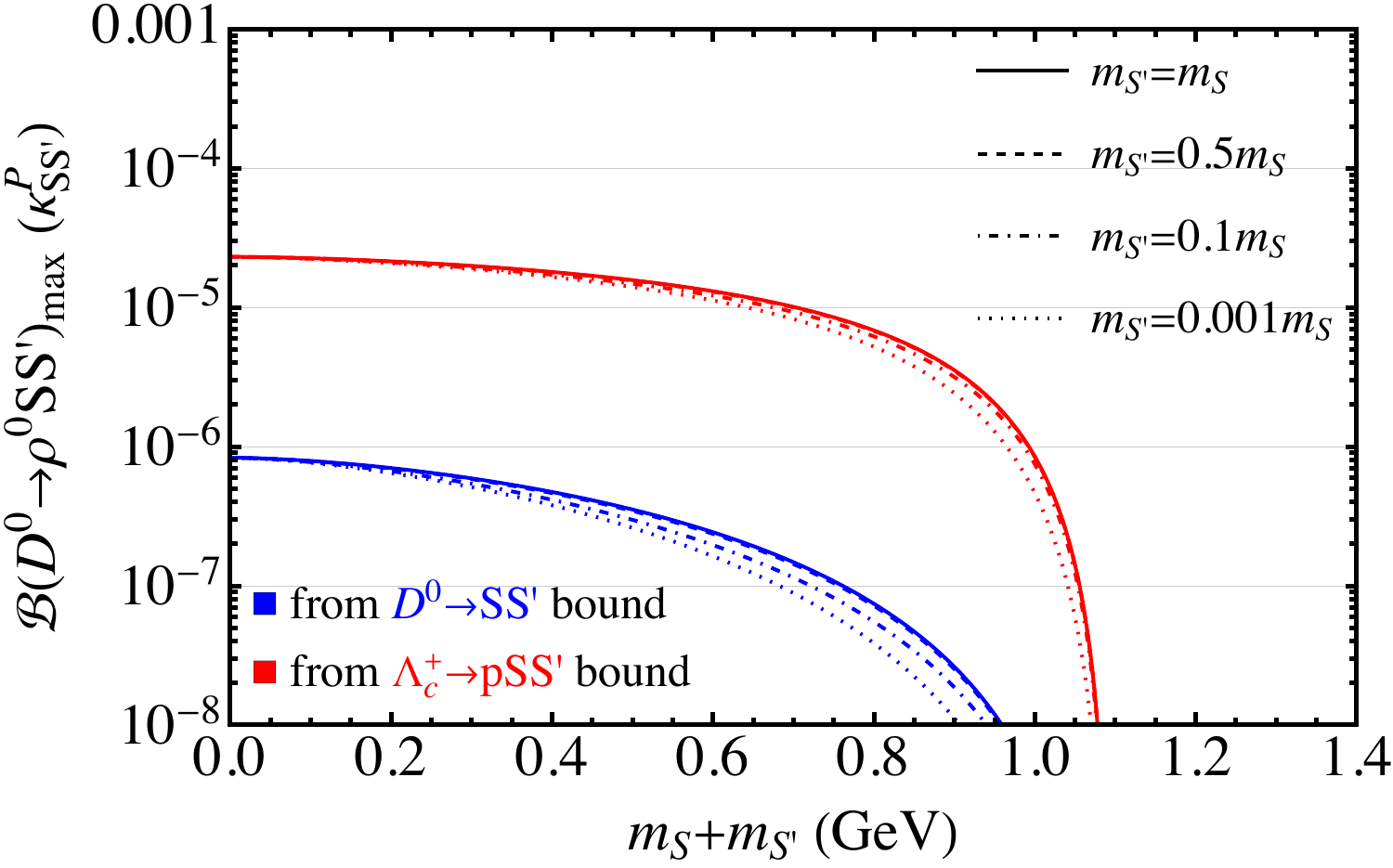} ~ ~ \includegraphics[width=0.47\textwidth]{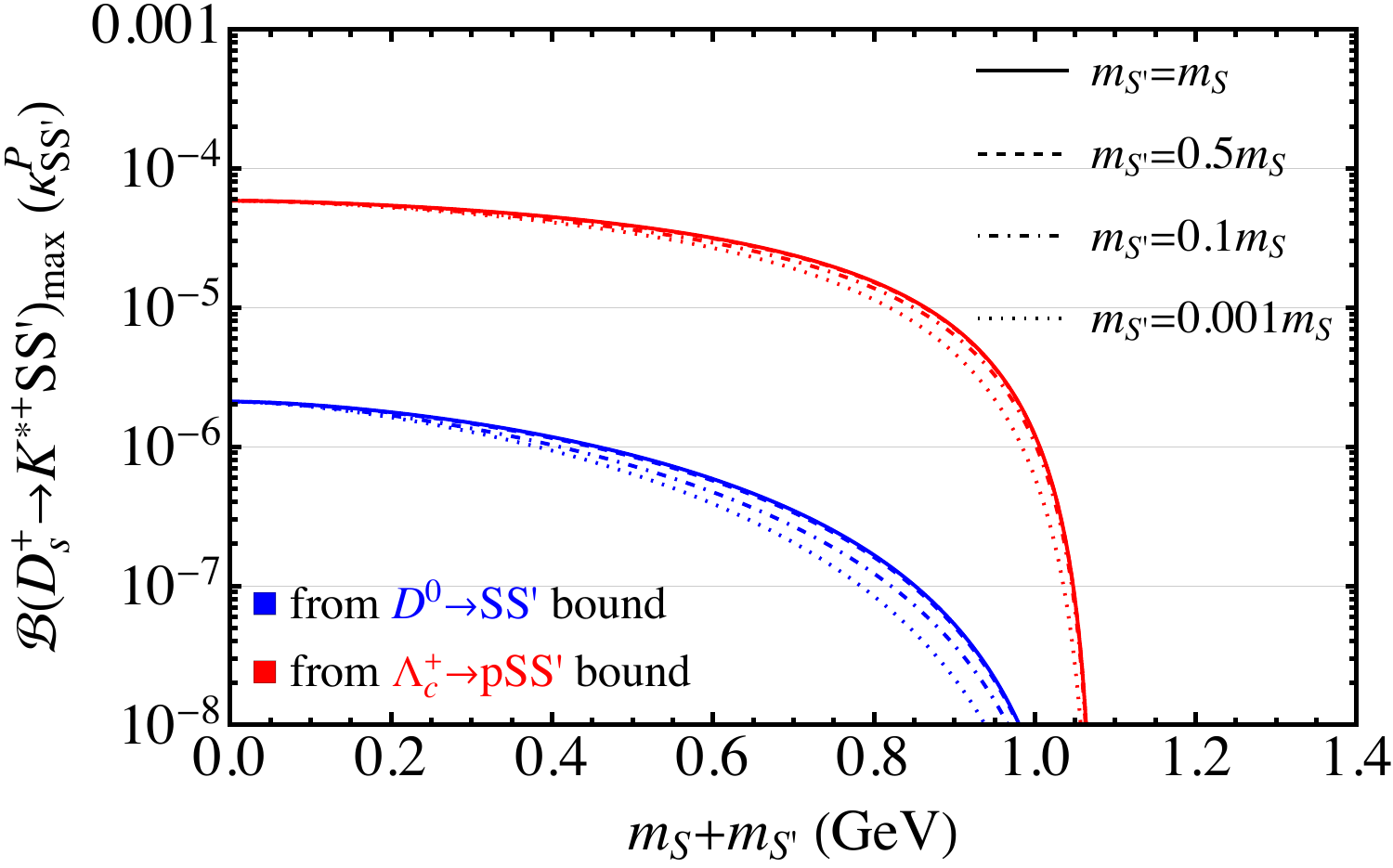}
	\vspace{-3pt}\\	
	\caption{The maximal branching fractions of \,$D^0\to\rho^0{\texttt S}\bar{\texttt S}{}'$ (left column) and \,$D_s^+\to K^{*+}{\texttt S}\bar{\texttt S}{}'$ (right column) due to $|\kappa_{\texttt{SS}'}^{\textsc v}|_{\rm max}^{}$ (top row) or $|\kappa_{\texttt{SS}'}^{\textsc a}|_{\rm max}^{}$ (middle row) or $|\kappa_{\texttt{SS}'}^{\textsc p}|_{\rm max}^{}$ (bottom row) alone.
		The $D^+\to\rho^+\texttt S\bar{\texttt S}{}'$ curves, not displayed, are approximately \,$2\tau_{D^+}/\tau_{D^0}\sim5$\, times their $D^0\to\rho^0{\texttt S}\bar{\texttt S}{}'$ counterparts.}
	\label{D2VSS'}  
\end{figure}

\begin{figure}[!t]
	\includegraphics[width=0.45\textwidth]{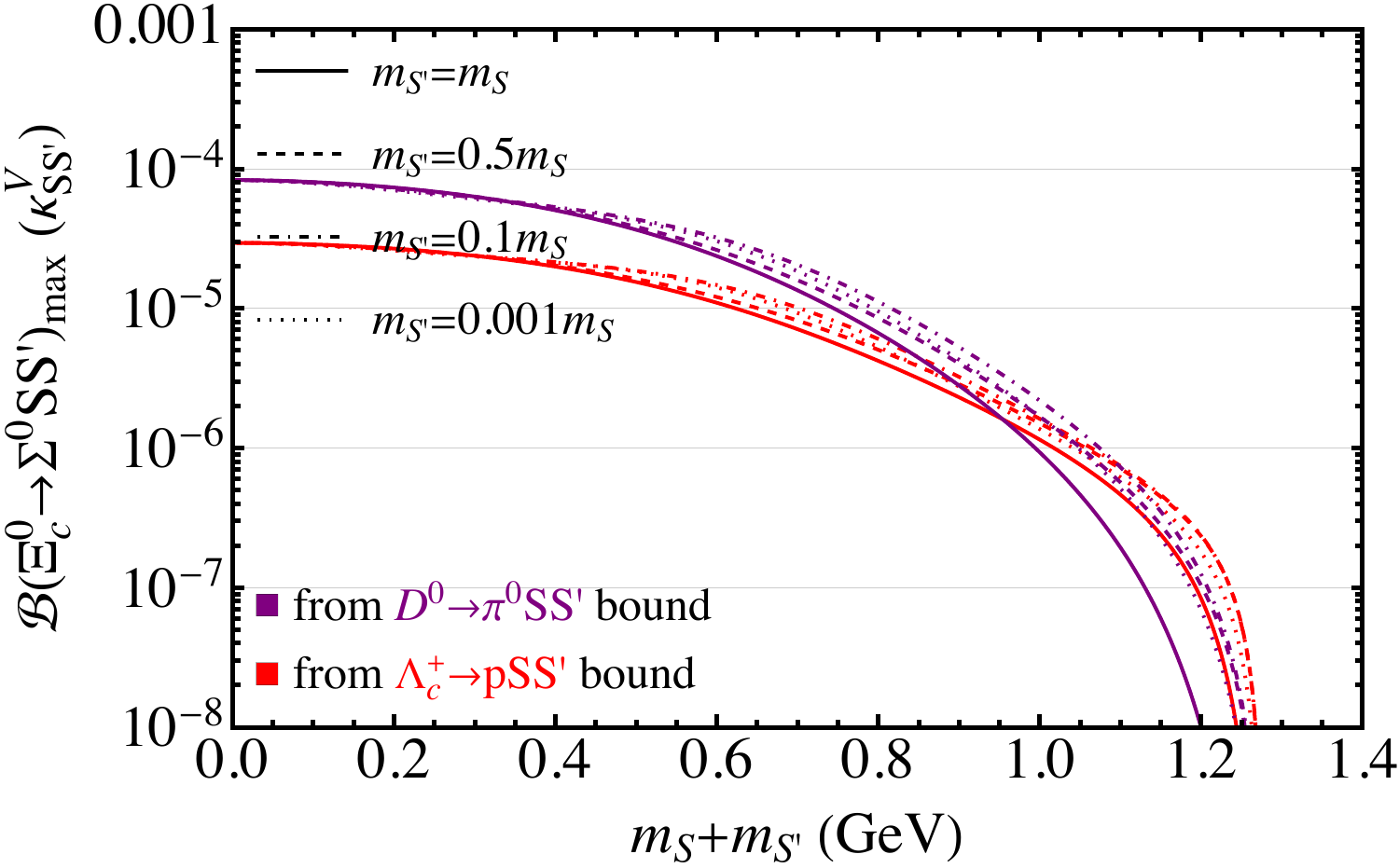} ~ ~ ~ \includegraphics[width=0.45\textwidth]{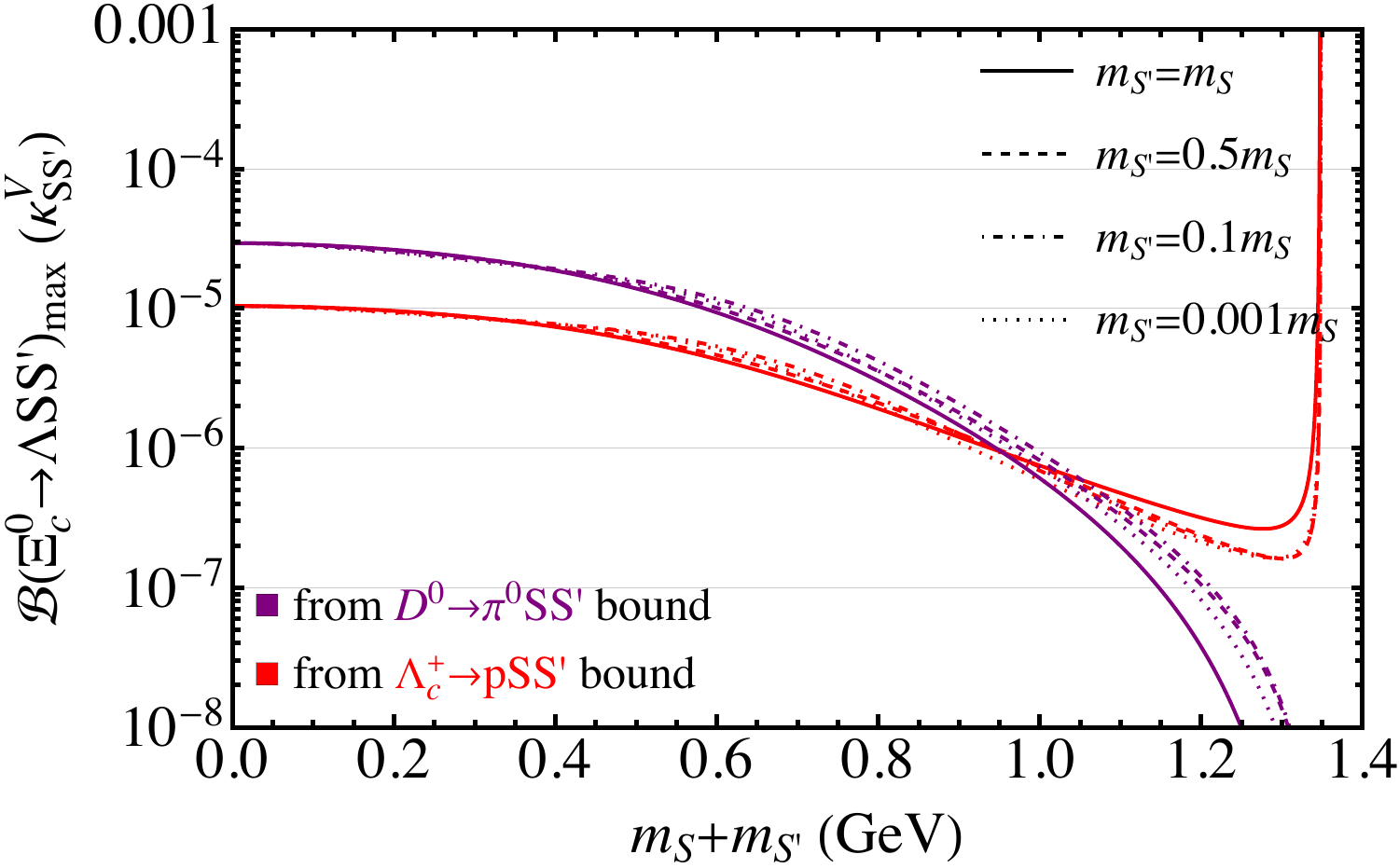}
	\vspace{2pt}\\
	\includegraphics[width=0.45\textwidth]{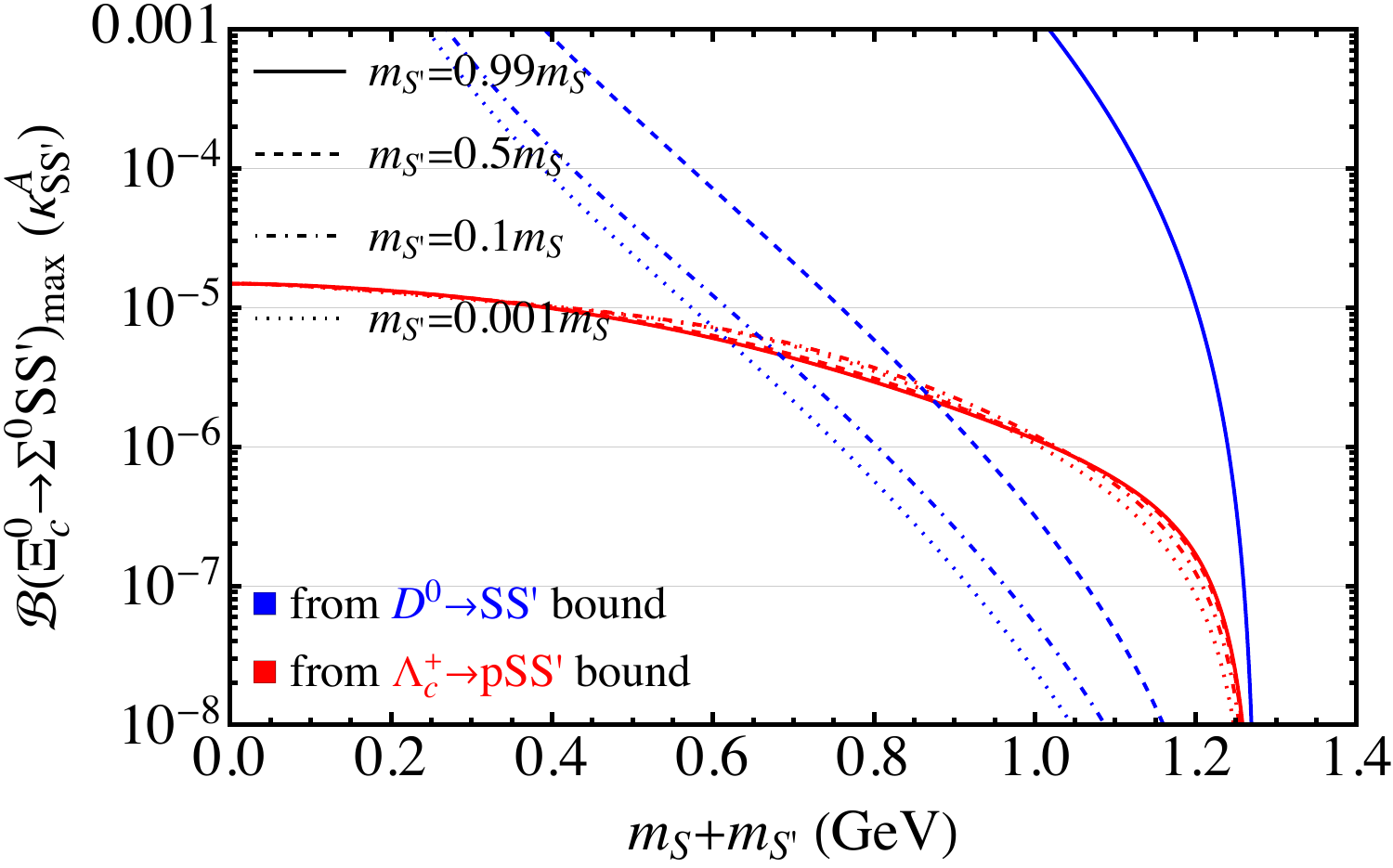} ~ ~ ~ \includegraphics[width=0.45\textwidth]{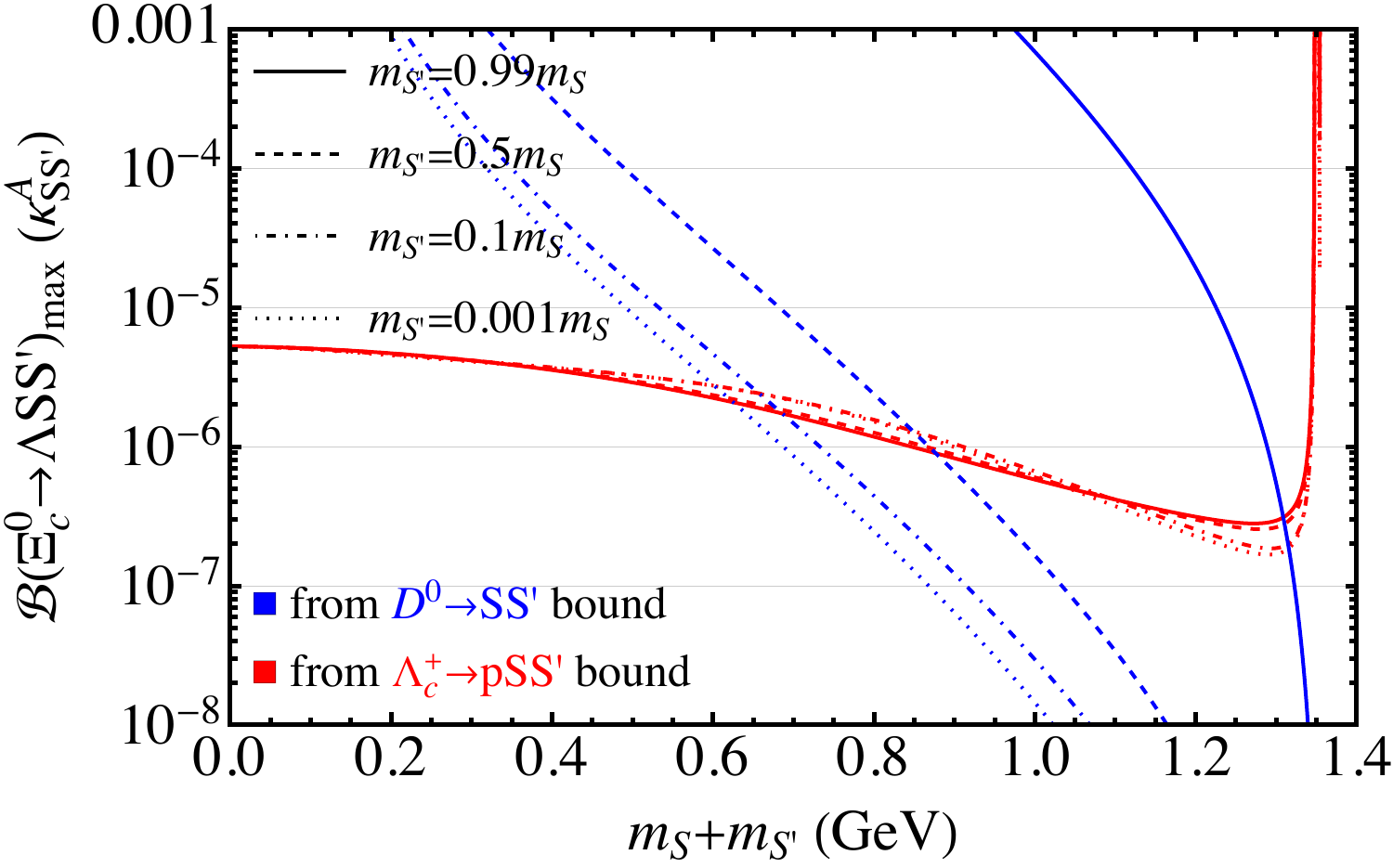}
	\vspace{2pt}\\
	\includegraphics[width=0.45\textwidth]{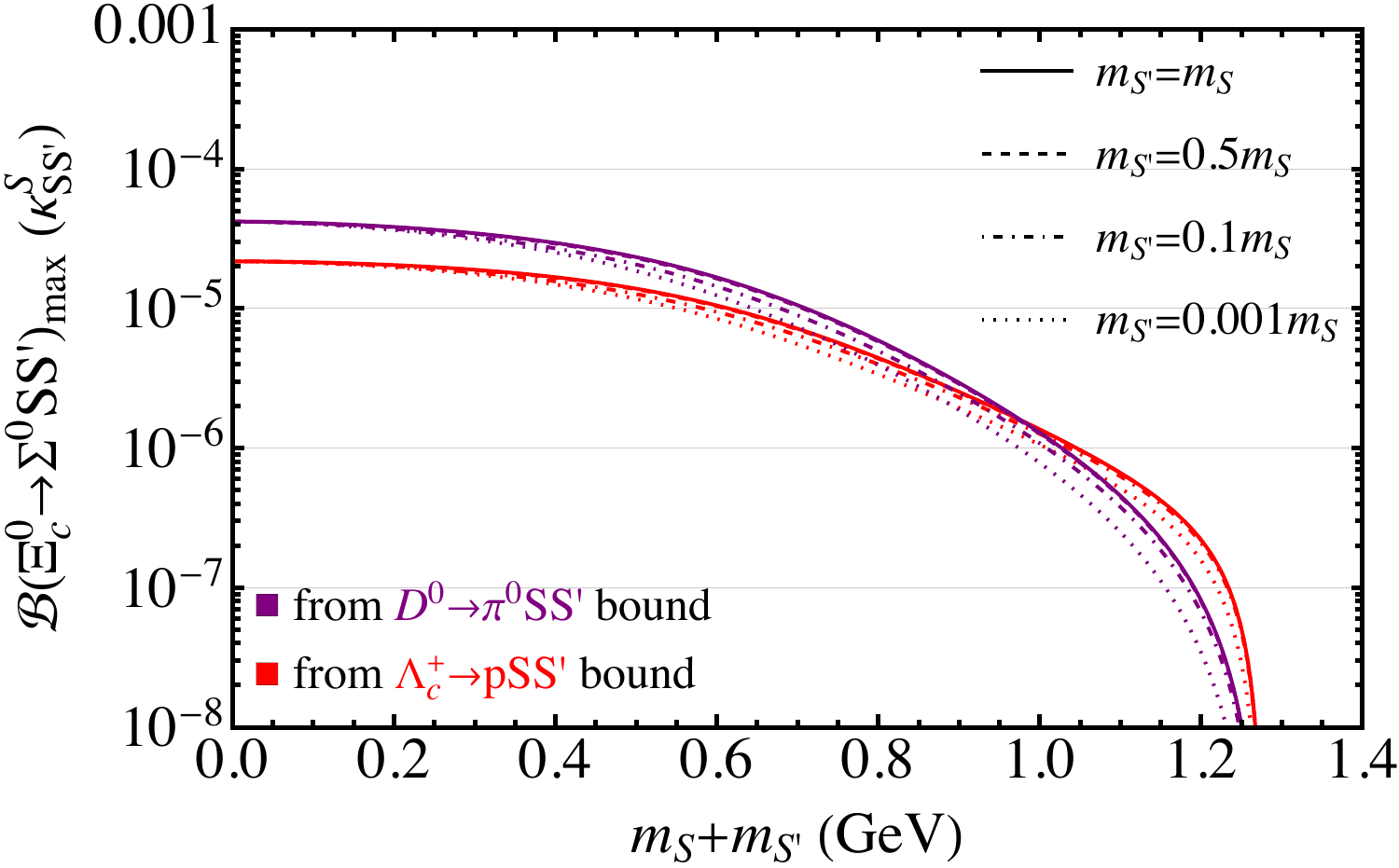} ~ ~ ~ \includegraphics[width=0.45\textwidth]{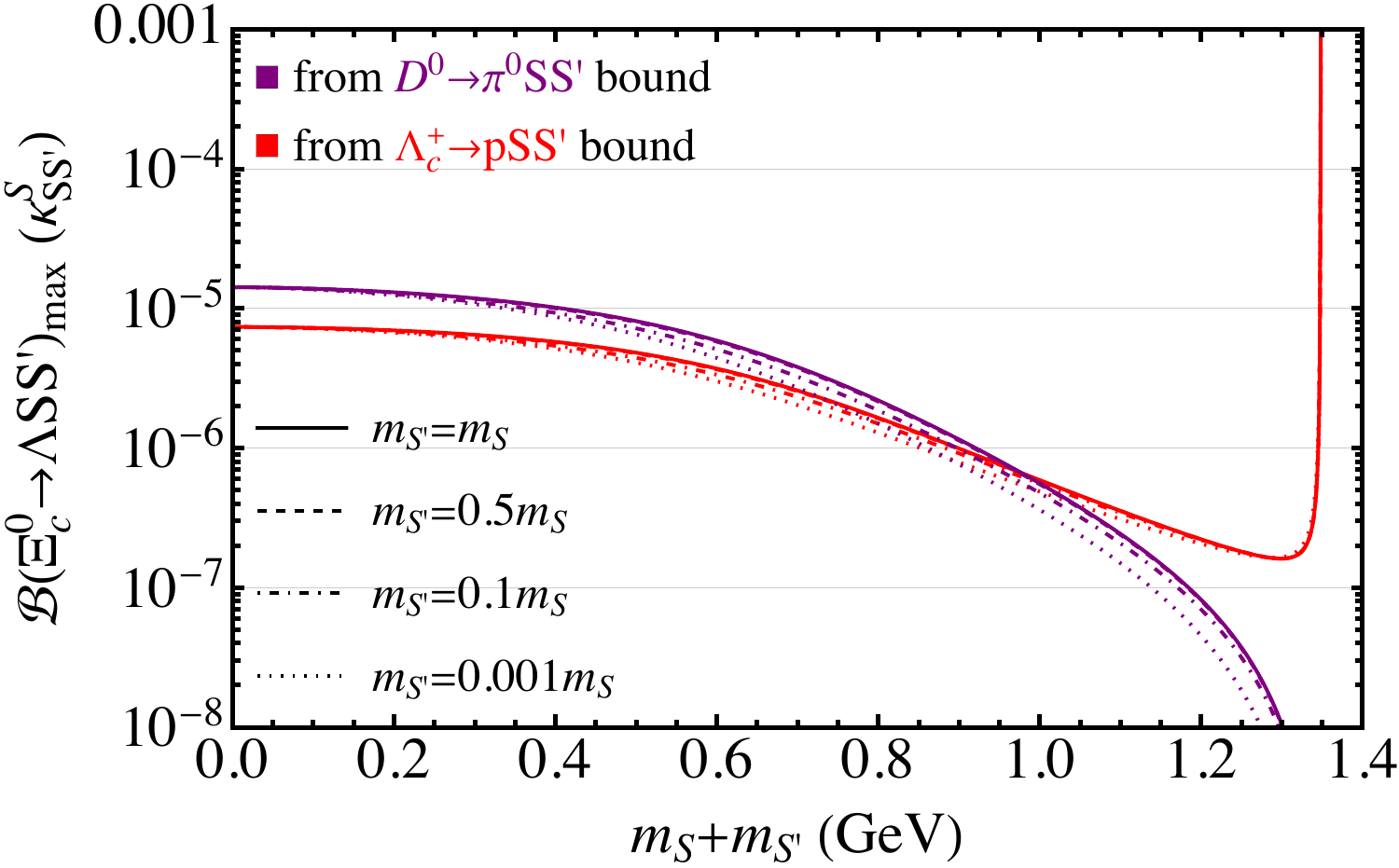}
	\vspace{2pt}\\
	\includegraphics[width=0.45\textwidth]{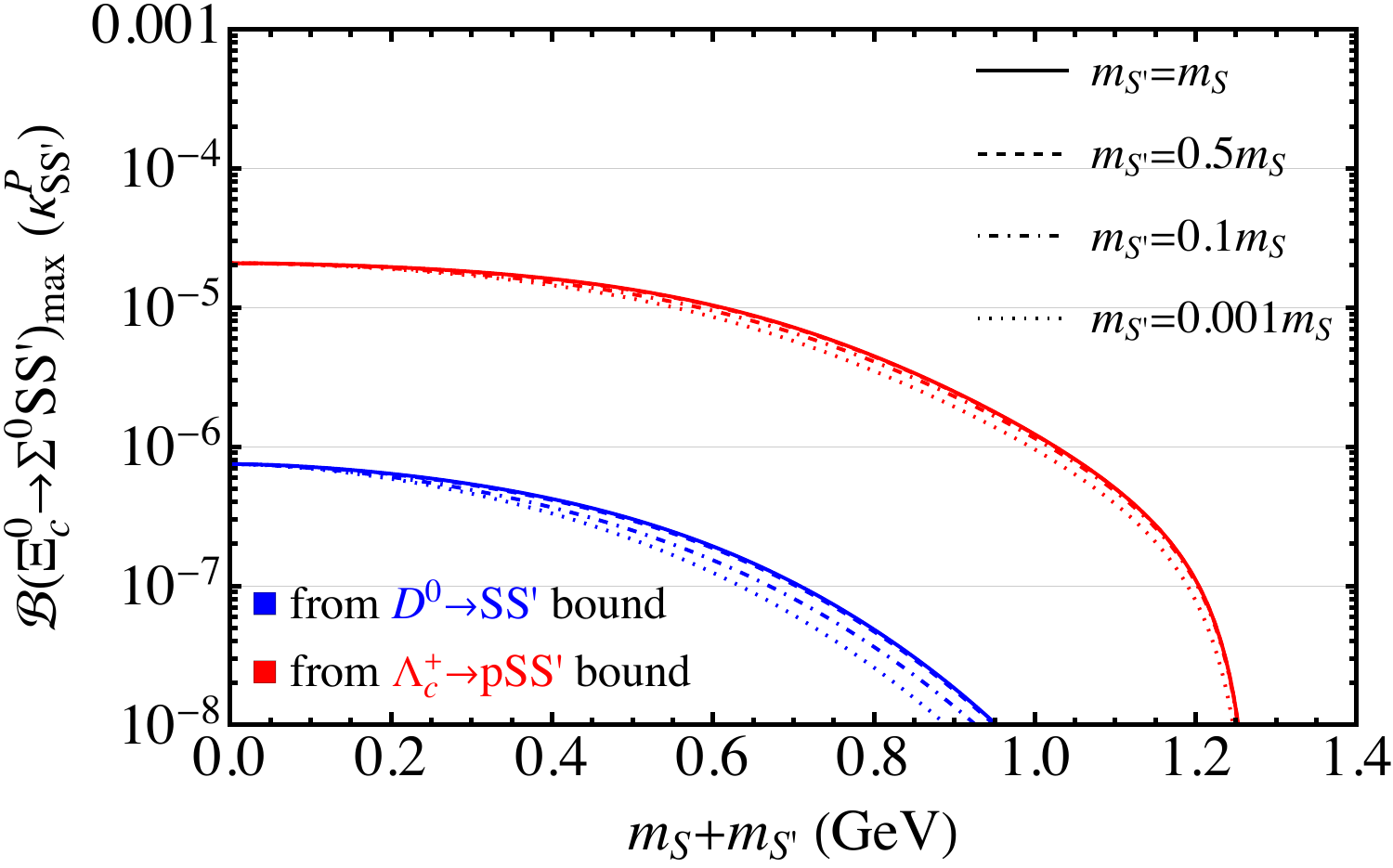} ~ ~ ~ \includegraphics[width=0.45\textwidth]{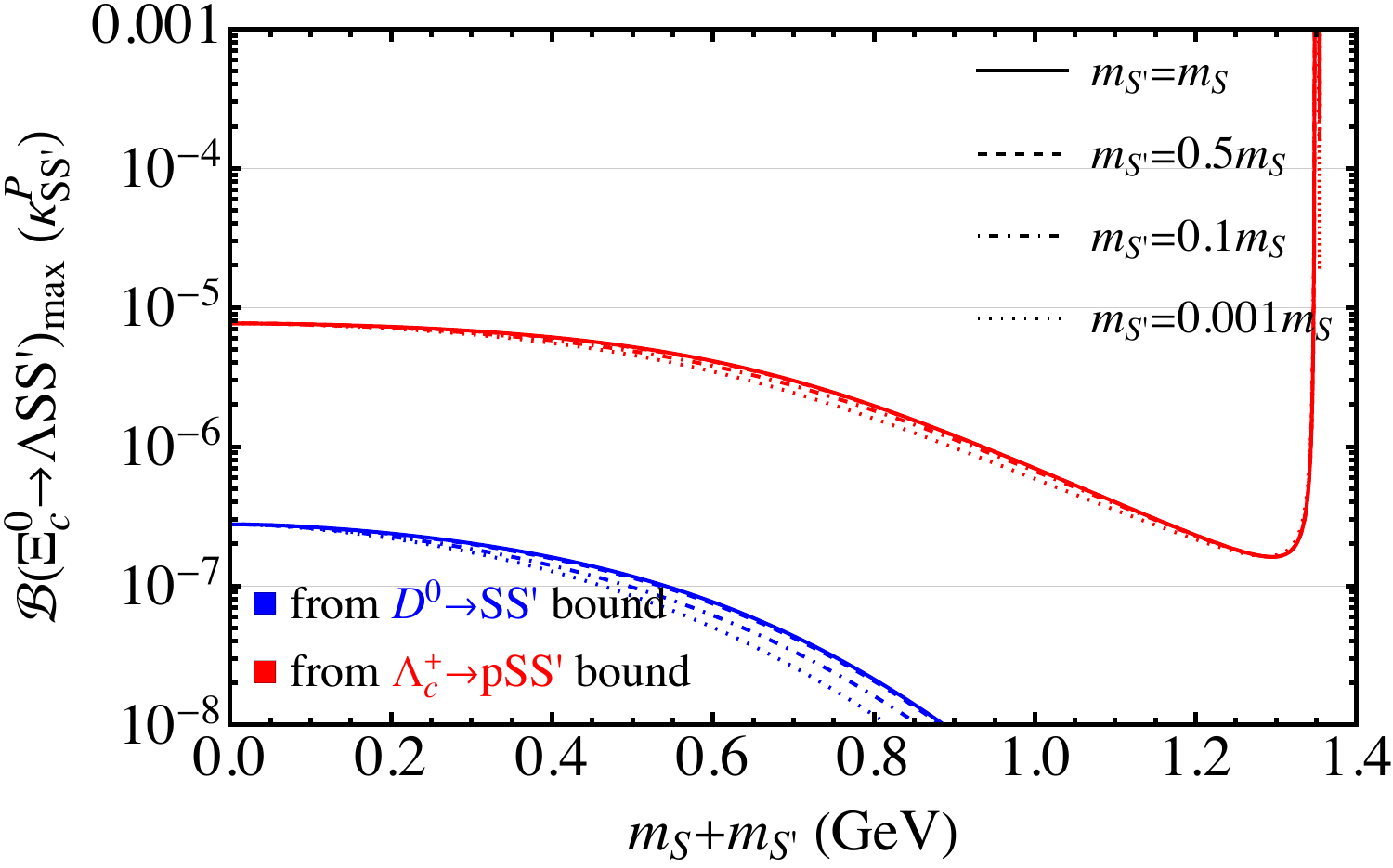}\vspace{-3pt}
	\caption{The maximal branching fractions of \,$\Xi_c^0\to\Sigma^0{\texttt S}\bar{\texttt S}{}'$ (left column) and \,$\Xi_c^0\to\Lambda{\texttt S}\bar{\texttt S}{}'$ (right column) due to $|\kappa_{\texttt{SS}'}^{\textsc v}|_{\rm max}^{}$ (top row) or $|\kappa_{\texttt{SS}'}^{\textsc a}|_{\rm max}^{}$ (second row) or $|\kappa_{\texttt{SS}'}^{\textsc s}|_{\rm max}^{}$ (third row) or $|\kappa_{\texttt{SS}'}^{\textsc p}|_{\rm max}^{}$ (bottom row) alone.
		The \,$\Xi_c^+\to\Sigma^+{\texttt S}\bar{\texttt S}{}'$\, curves, not shown, are approximately \,$2\tau_{\Xi_c^+}/\tau_{\Xi_c^0}\sim6$\, times their \,$\Xi_c^0\to\Sigma^0{\texttt S}\bar{\texttt S}{}'$\, counterparts.}
	\label{Xc->S}
\end{figure}

The caps on $|\kappa_{\texttt{SS}'}^{\textsc{v,a,s,p}}|$ can be turned into predictions for the maximal branching fractions of other FCNC charmed-hadron decays with ${\texttt S}\bar{\texttt S}{}'$ in the final states, again under the assumption that only one of the coefficients is nonvanishing at a time.
We have drawn the results with respect to \,$m_{\texttt S}+m_{\texttt S'}$\, in figures \ref{D2gSS'}-\ref{Xc->S}, where we have used the same curve styles for the constraints in eq.\,(\ref{c2uss'lim}) and the \,$m_{\texttt S'}/m_{\texttt S}$\, choices as in the corresponding $|\kappa_{\texttt{SS}'}^{\textsc{v,a,s,p}}|_{\rm max}^{}$ graphs in figure \ref{k2}.   
As will be illustrated in the following figures, whether or not $m_{\texttt S}$ and $m_{\texttt S'}$ are equal could significantly impact the decay rate, especially if $\kappa_{\texttt{SS}'}^{\textsc a}$ is the dominant coupling or sole one present.
In each of the branching-fraction plots, as before, the viable area for every \,$m_{\texttt S'}/m_{\texttt S}$\, case is below the lower of the purple or blue and red curves.

The two graphs in figure \ref{D2gSS'} reveal that $D^0\to\gamma{\texttt S}\bar{\texttt S}{}'$ currently has a branching fraction that is unconfined and hence could be quite sizable.
More precisely, it is less than $10^{-5}$ for total masses of up to 1.5 GeV or so, but there is no limitation on it if \,$m_{\texttt S}+m_{\texttt S'}>m_{D^0}-m_{\pi^0}$\, with at least $\kappa_{\texttt{SS}'}^{\textsc v}$ contributing or if both \,$m_{\texttt S}+m_{\texttt S'}>m_{\Lambda_c}-m_p$\, and \,$m_{\texttt S}=m_{\texttt S'}$\, with at least \,$\kappa_{\texttt{SS}'}^{\textsc a}\neq0$.\,
This condition will change if BESIII or Belle II pursues $D^0\to\gamma\slashed E$ and establishes a bound on it, if no discovery is made.
Any data on this channel would be greatly welcome.

From the left graphs in figure \ref{D2PSS'}, it is evident that \,${\cal B}(D^+\to\pi^+\texttt S\bar{\texttt S}{}')_{\rm max}\sim0.001$\, over the whole kinematical range. 
This number is about \,$2\tau_{D^+}/\tau_{D^0}\sim5$\, times the \,$D^0\to\pi^0\texttt S\bar{\texttt S}{}'$\, one in eq.\,(\ref{c2uss'lim}), as expected from approximate isospin symmetry. 
This relatively weak limit on the $D^+$ channel at the moment, especially for \,$m_{\Lambda_c}-m_p\;\mbox{\footnotesize$\lesssim$}\;m_{\texttt S}+m_{\texttt S'}<m_{D^+}-m_{\pi^+}$,\, encourages hunting \,$D^+\to\pi^+\slashed E$\, as well.\footnote{The charged modes \,$D_{(s)}^+\to\mathscr M_{(s)}^+\slashed E$\, with \,$\mathscr M_{(s)}=\pi,\rho\, (K,K^*)$\, have backgrounds from the sequential decays $D_{(s)}^+\to\tau^+\nu$\, and \,$\tau^+\to\mathscr M_{(s)}^+\bar \nu$ \cite{Burdman:2001tf,Kamenik:2009kc}, but we anticipate that they will be taken care of in the experimental searches.}
From the right column of figure \,\ref{D2PSS'}, we see that \,$D_s^+\to K^+\texttt S\bar{\texttt S}{}'$\, not only covers a shorter range of \,$m_{\texttt S}+m_{\texttt S'}$\, but also has a maximal branching-fraction which is comparatively smaller by several times or more.
The latter observation might continue to be the rough pattern formed by the limits on \,${\mathbb D}\to{\mathbb P}\texttt S\bar{\texttt S}{}'$\, from future quests. 
Nevertheless, if these decays are discovered, the acquired data can offer cross-checks on the effects of the responsible NP parametrized by $\kappa_{\texttt{SS}'}^{\textsc{v,s}}$.

From figure \ref{D2VSS'}, it may be inferred that the \,${\mathbb D}\to{\mathbb V}\texttt S\bar{\texttt S}{}'$\, channels have branching fractions which are further suppressed but can still reach roughly $1\times10^{-4}$.
The situation resembles that of the charmed-baryon decays \,$\Xi_c^0\to\Sigma^0\texttt S\bar{\texttt S}{}'$\, and \,$\Xi_c^0\to\Lambda\texttt S\bar{\texttt S}{}'$,\, illustrated in figure \ref{Xc->S}, as well as \,$\Xi_c^+\to\Sigma^+\texttt S\bar{\texttt S}{}'$,\, which has a branching fraction around \,$2\tau_{\Xi_c^+}/\tau_{\Xi_c^0}\sim6$\, times that of \,$\Xi_c^0\to\Sigma^0\texttt S\bar{\texttt S}{}'$\, but which is not included in the figure.

As a reminder, we remark that the red curves in our figures are only indicative for now, not being based on actual data on \,$\Lambda_c^+\to p\slashed E$\, with two invisibles being emitted.
Thus, empirical information on it would be highly desirable, as may also be concluded from the graphs we have produced. 
Importantly, this in addition means that the predicted upper-limits on branching fractions which we have discussed could be even bigger in the absence of the red curves. 
This is another incentive to look for these decay modes.

To demonstrate this more explicitly, as well as for completeness and reference, in Tables\,\,\ref{br1} and \ref{br2} we provide numerical examples of the maximal branching-fractions (the unbracketed entries in columns 2-4) for \,$m_{\texttt S}=m_{\texttt S'}=0$\, and \,$m_{\texttt S'}=0.1\,m_{\texttt S}=0.05\;$GeV,\,  respectively, if the \,$\Lambda_c^+\to p\texttt S\bar{\texttt S}{}'$\, bound is not present.
When it is taken into account and the stronger, we obtain the numbers placed in brackets.
It is worth noting that, as the third column of Table\,\,\ref{br1} makes clear, if  \,$m_{\texttt S}=m_{\texttt S'}=0$  and the \,$\Lambda_c^+\to p\texttt S\bar{\texttt S}{}'$\, bound is dropped, the branching fractions of modes which get a contribution from $\kappa_{\texttt{SS}'}^{\textsc a}$ is currently unconfined and hence could be substantial. 
\vspace{-1ex}

\begin{table}[!h]
	\setlength{\tabcolsep}{0.1cm}
	\caption{The upper limits on branching fractions, in units of $10^{-5}$, of various charmed-hadron decays induced by the \,$c\to u\texttt S\bar{\texttt S}{}'$ operators for \,$m_{\texttt S'}=m_{\texttt S}=0$\, if the \,$\Lambda_c^+\to p\texttt S\bar{\texttt S}{}'$\, bound is absent and, in brackets, if it is taken into account and the stronger. Only one of the coefficients $\kappa_{\texttt{SS}'}^{\textsc v,\textsc a,\textsc s,\textsc p}$ of the operators is taken to be nonzero at a time. A dash entry under \,$\kappa_{\texttt{SS}'}^{\textsc x}\neq0$\, means that $\kappa_{\texttt{SS}'}^{\textsc x}$ does not affect the decay.} \smallskip
	
	\begin{tabular*}{\textwidth}{@{}@{\extracolsep{\fill}}ccccc}
		\hline\hline
		Decay modes & $\kappa_{\texttt{SS}'}^{\textsc v}\neq0$ & $\kappa_{\texttt{SS}'}^{\textsc a}\neq0$ & $ \kappa_{\texttt{SS}'}^{\textsc s}\neq0$ & $ \kappa_{\texttt{SS}'}^{\textsc p}\neq0\vphantom{|_|^|}$ \\
		\hline
		$D^0\to\texttt S\bar{\texttt S}{}'\raisebox{12pt}{}$           & - & - & - & 9.4\raisebox{1pt}{\scriptsize~[Input]} \\
		$D^0\to\gamma\texttt S\bar{\texttt S}{}'$     & 0.14 \,(0.050) & (0.0026) & - & - \\
		$D^0\to\pi^0\texttt S\bar{\texttt S}{}'$      & 21\raisebox{1pt}{\scriptsize~[Input]} \,(7.5) & - & 21\raisebox{1pt}{\scriptsize~[Input]} \,(11) & - \\
		$D^+\to\pi^+\texttt S\bar{\texttt S}{}'$      & 107 \,(38)    & -     & 107 \,(55) & - \\
		$D_s^+\to K^+\texttt S\bar{\texttt S}{}'$     & 38 \,(13)     & -     & 36 \,(19) & - \\
		$D^0\to\rho^0\texttt S\bar{\texttt S}{}'$     & 0.74 \,(0.26) & (1.8) & - & 0.081 \\
		$D^+\to\rho^+\texttt S\bar{\texttt S}{}'$     & 3.8 \,(1.4)   & (9.4) & - & 0.42 \\
		$D_s^+\to K^{*+}\texttt S\bar{\texttt S}{}'$  & 2.0 \,(0.71)  & (5.3) & - & 0.21 \\
		$\Lambda_c^+\to p\texttt S\bar{\texttt S}{}'$ & 23 \,(8.0\raisebox{1pt}{\scriptsize~[Input]}) & (8.0\raisebox{1pt}{\scriptsize~[Input]}) & 15 \,(8.0\raisebox{1pt}{\scriptsize~[Input]}) & 0.29 \\
		$\Xi_c^+\to\Sigma^+\texttt S\bar{\texttt S}{}'$ & 49 \,(17)   & (8.7) & 25 \,(13) & 0.44 \\
		$\Xi_c^0\to\Sigma^0\texttt S\bar{\texttt S}{}'$ & 8.3 \,(2.9) & (1.5) & 4.2 \,(2.2) & 0.075 \\
		$\Xi_c^0\to\Lambda\texttt S\bar{\texttt S}{}'$  & 2.9 \,(1.0) & (0.52) & 1.4 \,(0.74) & 0.028 \\
		\hline\hline
		\label{br1}
	\end{tabular*} \vspace{-2em}
\end{table} 

\begin{table}[!h]
	\setlength{\tabcolsep}{0.1cm}
	\caption{The same as Table\,\,\ref{br1} but for \,$m_{\texttt S'}=0.1\,m_{\texttt S}=0.05\;$GeV.} 
	\begin{tabular*}{\textwidth}{@{}@{\extracolsep{\fill}}ccccc}
		\hline\hline
		Decay modes & $\kappa_{\texttt{SS}'}^{\textsc v}\neq0$ & $\kappa_{\texttt{SS}'}^{\textsc a}\neq0$ & $\kappa_{\texttt{SS}'}^{\textsc s}\neq0$ & $\kappa_{\texttt{SS}'}^{\textsc p}\neq0\vphantom{|_|^|}$ \\
		\hline
		$D^0\to\texttt S\bar{\texttt S}{}'\raisebox{12pt}{}$ & - & 9.4\raisebox{1pt}{\scriptsize~[Input]} \,(3.5) & - & 9.4\raisebox{1pt}{\scriptsize~[Input]} \\
		$D^0\to\gamma\texttt S\bar{\texttt S}{}'$ & 0.14 \,(0.063) & 0.0081 \,(0.0030) & - & - \\
		$D^0\to\pi^0\texttt S\bar{\texttt S}{}'$      & 21\raisebox{1pt}{\scriptsize~[Input]} \,(9.3) & - & 21\raisebox{1pt}{\scriptsize~[Input]} \,(13) & - \\
		$D^+\to\pi^+\texttt S\bar{\texttt S}{}'$ & 107 \,(47)      & - & 107 \,(68) & - \\
		$D_s^+\to K^+\texttt S\bar{\texttt S}{}'$ & 34 \,(15)      & - & 32 \,(20) & - \\
		$D^0\to\rho^0\texttt S\bar{\texttt S}{}'$ & 0.23 \,(0.10)  & 2.9 \,(1.1) & - & 0.024 \\
		$D^+\to\rho^+\texttt S\bar{\texttt S}{}'$ & 1.2 \,(0.55)   & 15 \,(5.6) & - & 0.12 \\
		$D_s^+\to K^{*+}\texttt S\bar{\texttt S}{}'$ & 0.62 \,(0.27) & 8.1 \,(3.0) & - & 0.060 \\
		$\Lambda_c^+\to p\texttt S\bar{\texttt S}{}'$ & 18 \,(8.0\raisebox{1pt}{\scriptsize~[Input]}) & 22 \,(8.0\raisebox{1pt}{\scriptsize~[Input]}) & 13 \,(8.0\raisebox{1pt}{\scriptsize~[Input]}) & 0.14 \\
		$\Xi_c^+\to\Sigma^+\texttt S\bar{\texttt S}{}'$ & 22 \,(9.9) & 13 \,(4.7) & 10 \,(6.5) & 0.12 \\
		$\Xi_c^0\to\Sigma^0\texttt S\bar{\texttt S}{}'$ & 3.8 \,(1.7) & 2.2 \,(0.80) & 1.7 \,(1.1) & 0.020 \\
		$\Xi_c^0\to\Lambda\texttt S\bar{\texttt S}{}'$ & 1.4 \,(0.61) & 0.82 \,(0.30) & 0.61 \,(0.39) & 0.0078 \\
		\hline\hline
		\label{br2}
	\end{tabular*}
\end{table}

\section{FCNC charm decay with invisible singlet fermions\label{fermions}}

The possibility that the missing energy in \,$c\to u\slashed E$\, is carried away by two SM-gauge-singlet spin-1/2 particles has been entertained in the past to varying extents~\cite{Badin:2010uh,Dorsner:2016wpm,Li:2020dpc,Faisel:2020php,Fajfer:2021woc}.
Instead of adopting a model-independent approach as in the last two sections, here we consider a specific new-physics scenario where a heavy scalar leptoquark (LQ) is responsible for linking three Dirac right-handed sterile neutrinos (referred to as \,${\texttt N}_1,{\texttt N}_2,{\texttt N}_3$), which are the singlet fermions, to up-type quarks which are also right-handed.
This is the least constrained of the LQ models investigated in ref.\,\cite{Faisel:2020php} in the \,$c\to u\slashed E$\, context and can now be scrutinized to a greater degree in light of the recent Belle and BESIII data.

In the nomenclature of ref.\,\cite{Dorsner:2016wpm}, the LQ is $\bar S_1$ which transforms as $(\bar 3,1,-2/3)$ under the SM gauge groups \,SU(3$)_{\rm color}\times{\rm SU}(2)_L\times{\rm U}(1)_Y$.
We write the Lagrangian for the renormalizable interaction of $\bar S_1$ with ${\texttt N}_{1,2,3}$ and the quarks as
\begin{align} \label{Llq}
	{\cal L}_{\textsc{lq}}^{} & \,=\, \bar{\texttt Y}_{jl}^{}\, \overline{{\cal U}_j^{\textsc c}} P_R^{}\,
	{\texttt N}_l^{}\, \bar S_1^{} \,+\, \rm H.c. \,,
\end{align}
where $\bar{\texttt Y}_{jl}$ are generally complex elements of the LQ Yukawa matrix $\bar{\texttt Y}$, summation over family indices \,$j,l=1,2,3$\, is implicit, \,${\cal U}_{1,2,3}=(u,c,t)$,\, and the superscript \textsc c indicates charge conjugation.
Assuming the LQ to be heavy, we can then derive an effective Lagrangian of the form in eq.\,(\ref{cuff'}) for
\,$c\to u\texttt N\bar{\texttt N}{}'$,\, with the coefficients being given by
\begin{align}
	{\texttt C}_{\texttt N_j\texttt N_l}^{\texttt V} & \,=\, {\texttt C}_{\texttt N_j\texttt N_l}^{\texttt A} \,=\,
	\tilde{\textsc c}_{\texttt N_j\texttt N_l}^{\textsc v} \,=\, \tilde{\textsc c}_{\texttt N_j\texttt N_l}^{\textsc a} \,=\, -\frac{\bar{\texttt Y}{}_{1j}^*\bar{\texttt Y}{}_{2l}^{}}{8 m_{\bar S_1}^2} \,. ~~~ ~~~
\end{align}

As explained in ref.\,\cite{Faisel:2020php}, the interactions in eq.\,(\ref{Llq}) also bring about one-loop contributions to $D^0$-$\bar D^0$ mixing which are proportional to the combination \,$\mbox{\footnotesize$\sum$}_j \bar{\texttt Y}{}_{1j}^*\bar{\texttt Y}{}_{2j}^{}/m_{\bar S_1}$\, and which therefore will vanish if the nonzero elements of the first and second rows of $\bar{\texttt Y}$ do not share same columns.
Hence the potentially stringent restrictions on the parameters of this model from $D^0$-$\bar D^0$ mixing could be completely evaded.
To realize this, for definiteness we choose, as one of the simplest examples,
\begin{align} \label{bY1}
	\bar{\texttt Y} & \,= \left(\begin{array}{ccc} 0 & \bar{\texttt y}_{u2} & 0 \vspace{1pt} \\ \bar{\texttt y}_{c1} & 0 & 0 \vspace{1pt} \\ 0 & 0 & 0 \end{array}\right) , ~~~ ~~~~
\end{align}
in which case only \,$c\to u{\texttt N}_2\bar{\texttt N}_1$\, can occur with
\begin{align} \label{KNN'}
	{\texttt C}_{\texttt N_2\texttt N_1}^{\texttt V} & \,=\, {\texttt C}_{\texttt N_2\texttt N_1}^{\texttt A} \,=\, \tilde{\textsc c}_{\texttt N_2\texttt N_1}^{\textsc v} \,=\, \tilde{\textsc c}_{\texttt N_2\texttt N_1}^{\textsc a} \,=\, -\frac{\bar{\texttt y}_{u2}^*\bar{\texttt y}_{c1}^{}}{8 m_{\bar S_1}^2} \,\equiv\, {\textsc k}_{\texttt{NN}'} , ~~~ ~~~~
\end{align}
the other coefficients vanishing.

Another empirical constraint, deduced from the latest LHC data, excludes at 95\% CL scalar LQs having masses up to 1.14\,\,TeV and decaying fully to a neutrino and a light-flavored quark~\cite{CMS:2019ybf}.
Since this is applicable to ${\cal L}_{\textsc{lq}}$, we select $m_{\bar S_1}>1.2$\,\,TeV.
There is additionally a theoretical requirement for the elements of $\bar{\texttt Y}$, namely that their size not exceed $\sqrt{4\pi}$ to ensure perturbativity.
It follows that $|{\textsc k}_{\texttt{NN}'}| < 1.1\rm\;TeV^{-2}$.

\begin{figure}[t]
	\includegraphics[width=0.52\textwidth]{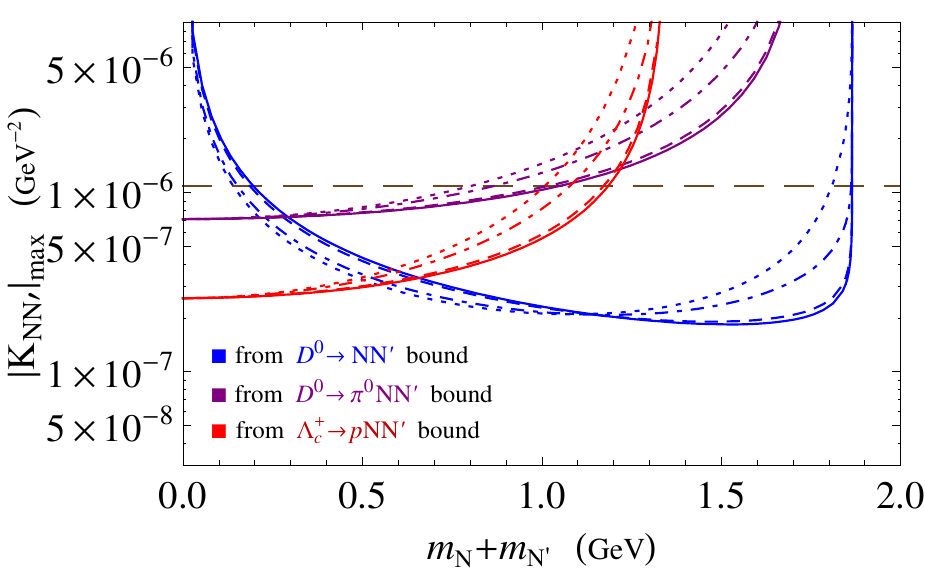} \vspace{-5pt}
	\caption{The upper limits on $|{\textsc k}_{\texttt N\texttt N'}|$ versus \,$m_{\texttt N}+m_{\texttt N'}$, with \,$\texttt N=\texttt N_2$\, and \,$\texttt N'=\texttt N_1$,\, implied by the \,$D^0\to\texttt N\bar{\texttt N}{}'$ (blue), \,$D^0\to\pi^0\texttt N\bar{\texttt N}{}'$ (purple), and \,$\Lambda_c^+\to p\texttt N\bar{\texttt N}{}'$ (red) bounds in eq.\,(\ref{c2unn'lim}) for \,$m_{\texttt N'}/m_{\texttt N}=0.001$ (dotted curves), 0.1 (dash-dotted curves), 0.5 (dashed curves), 1 (solid curves).
		The horizontal brown dashed line marks \,$|{\textsc k}_{\texttt{NN}'}|<1.1\rm\;TeV^{-2}$\, inferred from collider and perturbativity restrictions.\label{KNN'max}}  
\end{figure}

We now examine how the aforementioned Belle and BESIII measurements may test this particular case, with ${\textsc k}_{\texttt{NN}'}$ and the masses of \,$\texttt N=\texttt N_2$ and \,$\texttt N'=\texttt N_1$ being the only free parameters.
To begin, analogously to eq.\,(\ref{c2uss'lim}) we impose
\begin{align} \label{c2unn'lim}
	{\cal B}(D^0\to\texttt N\bar{\texttt N}{}') & \,<\, 9.4\times10^{-5} \,, & {\cal B}(D^0\to\pi^0\texttt N\bar{\texttt N}{}') & \,<\, 2.1\times10^{-4} \,, ~~~ ~~~~ \nonumber \\
	{\cal B}(\Lambda_c^+\to p\texttt N\bar{\texttt N}{}') & \,<\, 8.0\times10^{-5} \,.
\end{align}
Subsequently, after incorporating eq.\,(\ref{KNN'}) into the relevant rate formulas from eqs.\,\,(\ref{BD2ff'})-(\ref{D2Vff'}) with the appropriate form factors from appendix \ref{FF}, we extract the maximal values of $|{\textsc k}_{\texttt{NN}'}|$ over the permitted range of \,$m_{\texttt N}+m_{\texttt N'}$\, for a few choices of $m_{\texttt N'}/m_{\texttt N}$.
We display the results in figure\,\,\ref{KNN'max}, where the blue, purple, and red curves correspond, respectively, to the three limits in eq.\,(\ref{c2unn'lim}).
The allowed $|{\textsc k}_{\texttt{NN}'}|$ range for each $(m_{\texttt N},m_{\texttt N'})$ pair is below the lowest curve. 

We learn from this figure that the restraints on ${\textsc k}_{\texttt{NN}'}$ from \,$D^0\to\texttt N\bar{\texttt N}{}'$\, and \,$\Lambda_c^+\to p\texttt N\bar{\texttt N}{}'$\, in tandem are more stringent than the one from $D^0\to\pi^0\texttt N\bar{\texttt N}{}'$ and 3-4 times stronger than the one implied by collider data and perturbativity.
We also notice that none of the sets of curves of the same color exhibit substantial variations with \,$m_{\texttt N'}/m_{\texttt N}$,\, and so there is no drastic weakening of the constraints when \,$m_{\texttt N'}\to m_{\texttt N}$,\, unlike the situation depicted by figure \ref{k2} in the invisible-boson case. 
This is mostly because of the difference in dependence on the invisibles' masses between the \,$D^0\to\texttt N\bar{\texttt N}{}'$\, rate and the $\kappa_{\texttt{SS}'}^{\textsc a}$ part of $\Gamma_{D^0\to\texttt S\bar{\texttt S}{}'}$, as can be viewed in eqs.\,\,(\ref{BD2ff'}) and (\ref{D2SS'}).  
What we see in figure\,\,\ref{KNN'max} again illustrates the importance of the mesonic and baryonic modes as complementary tools in the quest for new-physics signals.

A further comparison of figures \ref{k2} and \ref{KNN'max} highlights one of the main differences between a model-independent analysis and a model-based one.
In figure \ref{k2} the coefficients of the operators contributing to \,$c\to u\slashed E$ are taken to be independent of one another and consequently each have to respect only a subset of the pertinent data. 
By contrast, the coefficients described in figure \ref{KNN'max}, associated with the operators listed in eq.\,(\ref{cuff'}), are connected via eq.\,(\ref{KNN'}), and the same ${\textsc k}_{\texttt{NN}'}$ must satisfy all of the requisites in eq.\,(\ref{c2unn'lim}), resulting by and large in a stronger restriction on it.

\begin{figure}[!t]
	\includegraphics[width=0.48\textwidth]{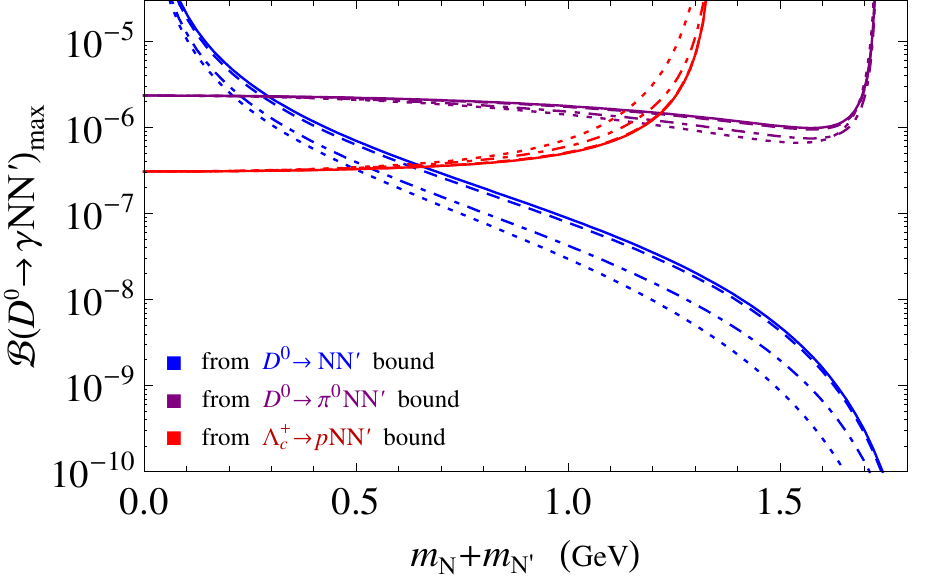} ~ ~ \includegraphics[width=0.48\textwidth]{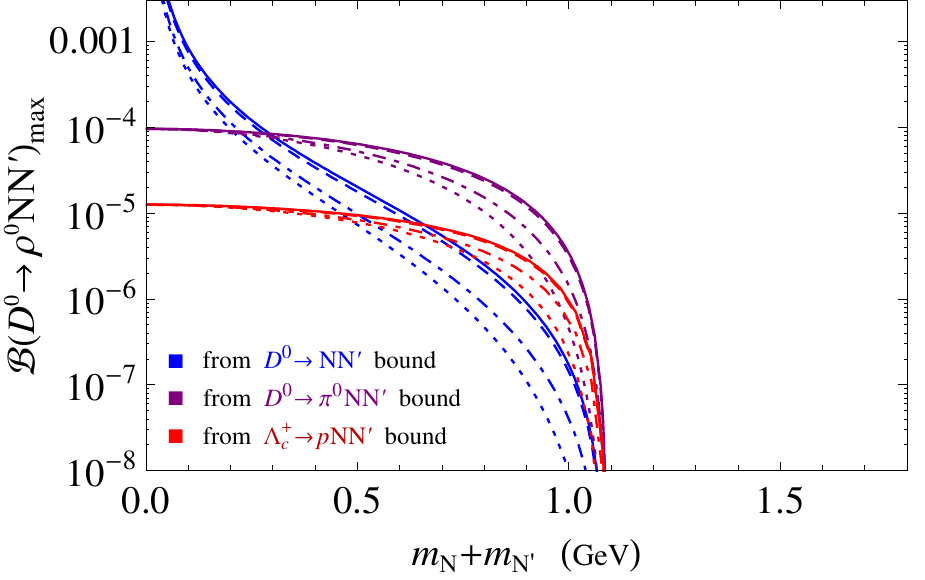} \vspace{5pt}\\
	\includegraphics[width=0.48\textwidth]{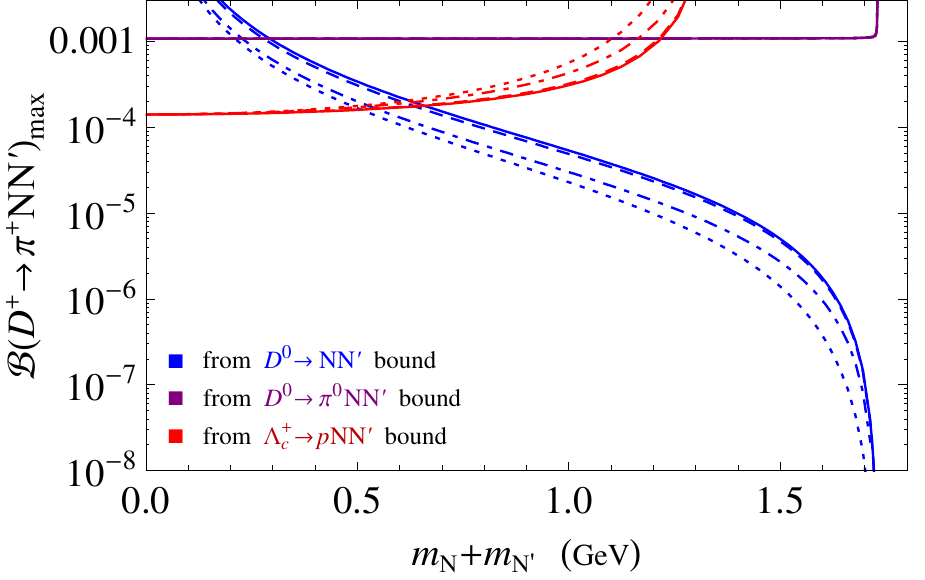} ~ ~ \includegraphics[width=0.48\textwidth]{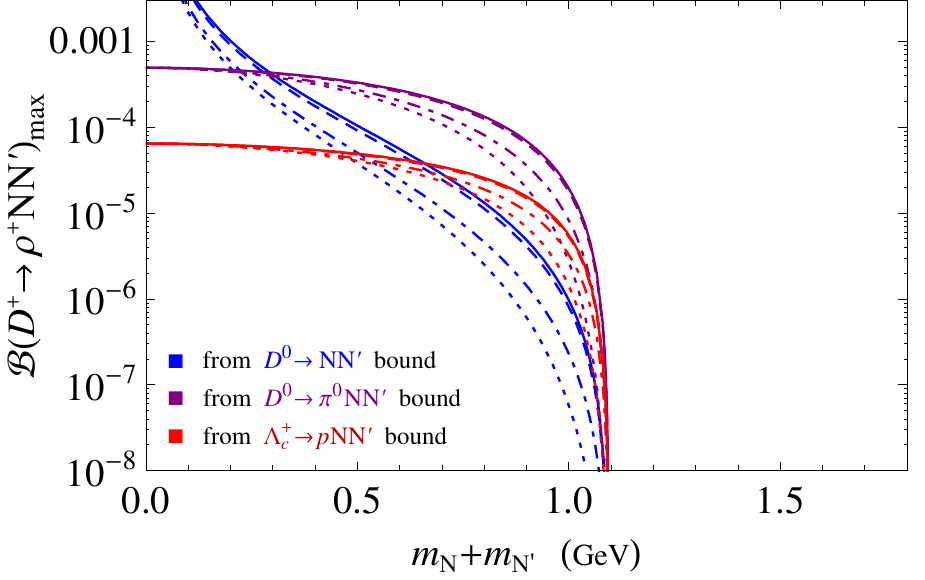}\vspace{5pt}\\
	\includegraphics[width=0.48\textwidth]{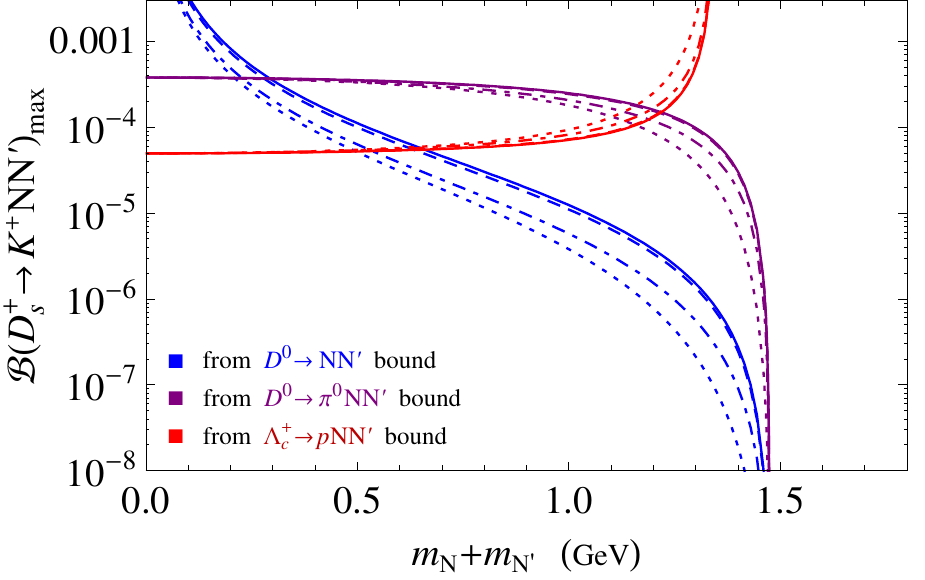} ~ ~ \includegraphics[width=0.48\textwidth]{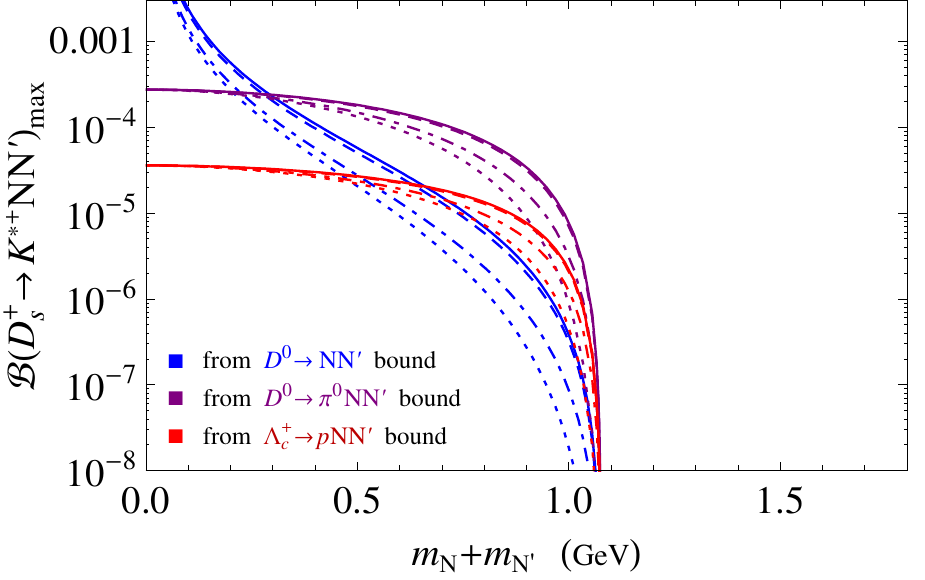}\vspace{-5pt}
	\caption{Maximal branching fractions of \,$D^0\to(\gamma,\rho^0)\texttt N\bar{\texttt N}{}'$,\, $D^+\to(\pi^+,\rho^+)\texttt N\bar{\texttt N}{}'$,\, and \,$D_s^+\to(K^+,K^{*+})\texttt N\bar{\texttt N}{}'$\, translated from the $|{\textsc k}_{\texttt{NN}'}|_{\rm max}$ values in figure \ref{KNN'max} inferred from the \,$D^0\to\texttt N\bar{\texttt N}{}'$ (blue), $D^0\to\pi^0\texttt N\bar{\texttt N}{}'$ (purple), and \,$\Lambda_c^+\to p\texttt N\bar{\texttt N}{}'$ (red) bounds in eq.\,(\ref{c2unn'lim}).\label{B-D2gNN'}}
\end{figure}

\begin{figure}[t]
	\includegraphics[width=0.48\textwidth]{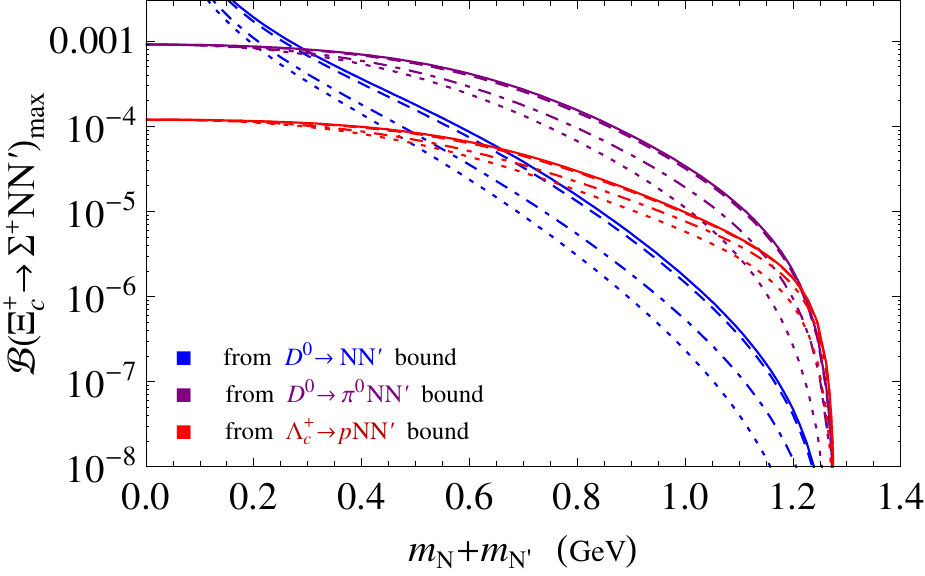} ~ ~ \includegraphics[width=0.48\textwidth]{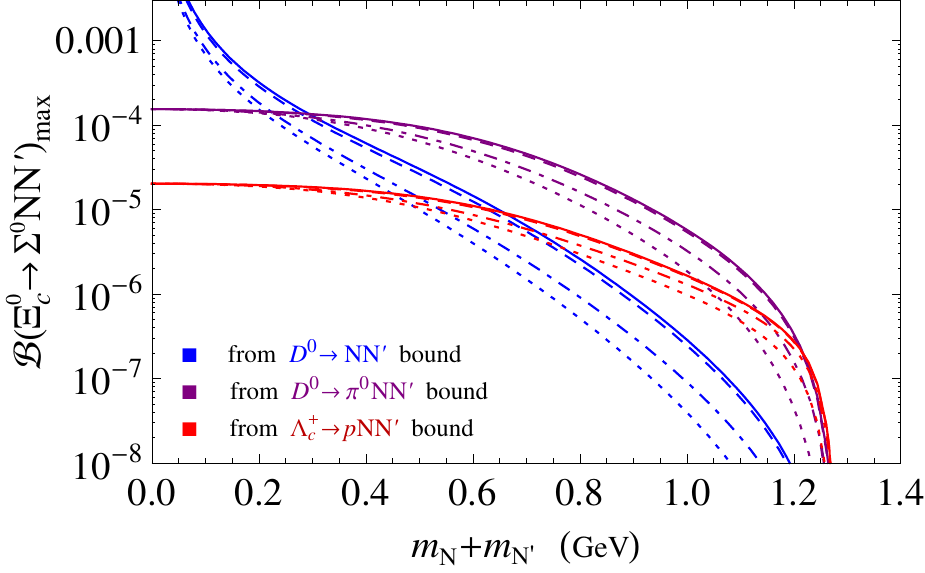} \vspace{7pt}\\ \includegraphics[width=0.48\textwidth]{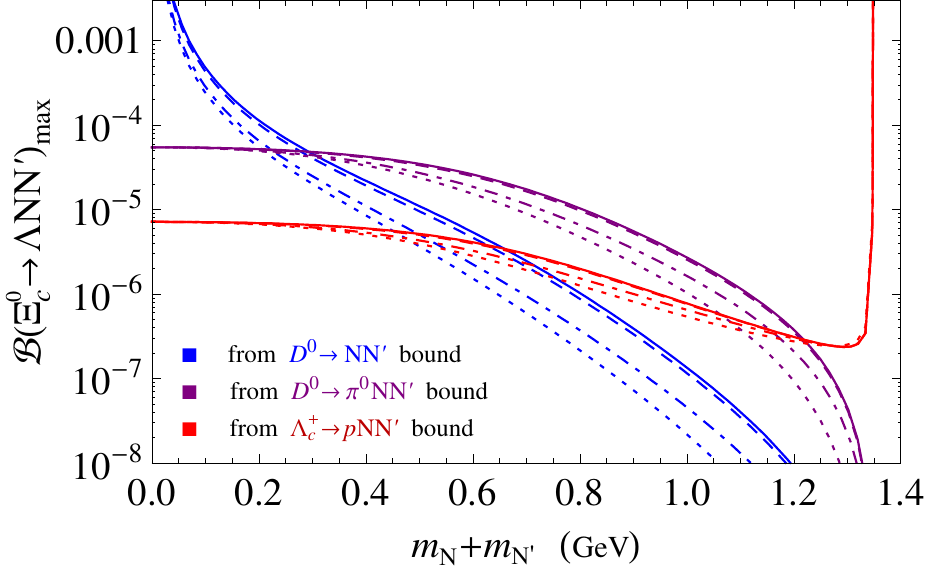} \vspace{-3pt}
	\caption{The same as figure \ref{B-D2gNN'} but for \,$\Xi_c^{+,0}\to\Sigma^{+,0}\texttt N\bar{\texttt N}{}'$\, and \,$\Xi_c^0\to\Lambda\texttt N\bar{\texttt N}{}'$.\label{B-X2SNN'}}
\end{figure}

From the $|{\textsc k}_{\texttt{NN}'}|_{\rm max}$ values, we can predict the maximal branching-fractions of a number of hadron decays arising from the \,$c\to u\texttt N\bar{\texttt N}{}'$\, operators. 
The results, plotted in figures \ref{B-D2gNN'} and \ref{B-X2SNN'}, are on the whole lower than the corresponding ones in the scalar case, in figures \ref{D2gSS'}-\ref{Xc->S}, considering that the effects of the scalar coefficient $\kappa_{\texttt{SS}'}^{\textsc a}$ if \,$m_{\texttt S}\simeq m_{\texttt S'}$\, are presently unknown and consequently could be sizable. 
Nevertheless, as figures \ref{B-D2gNN'} and \ref{B-X2SNN'} reveal, the predictions can still be significant, especially if  $m_{\texttt N}+m_{\texttt N'}<300$\,\,MeV\, and the red curves are ignored. 
This is shown explicitly by the numerical examples quoted in table \ref{br3}, which may be compared with tables \ref{br1} and \ref{br2}. 
We can then conclude that this specific new-physics scenario remains alive and attractive, notably because it accommodates both a leptoquark and sterile neutrinos, and will be probed more thoroughly by future data.

\begin{table}[!h]
	\setlength{\tabcolsep}{0.1cm}
	\caption{The upper limits on the branching fractions, in units of $10^{-5}$, of various charmed-hadron decays induced by the \,$c\to u\texttt N\bar{\texttt N}{}'$\, interactions and evaluated with the lowest $|{\textsc k}_{\texttt{NN}'}|_{\rm max}$ for \,$m_{\texttt N'}=m_{\texttt N}=0$\, and \,$m_{\texttt N'}=0.1\,m_{\texttt N}=0.05\;$GeV\, if the \,$\Lambda_c^+\to p\texttt N\bar{\texttt N}{}'$\, bound is absent and, in brackets, if it is included and the strongest.} \smallskip
	
	\begin{tabular*}{\textwidth}{@{}@{\extracolsep{\fill}}ccc}
		\hline\hline
		Decay modes & $m_{\texttt N'}=m_{\texttt N}=0\vphantom{|^|}$ & $m_{\texttt N'}=0.1\,m_{\texttt N}=0.05\;$GeV \\
		\hline
		$D^0\to\texttt N\bar{\texttt N}{}'\raisebox{12pt}{}$ & - & 9.4\,\footnotesize[Input] \\
		$D^0\to\gamma\texttt N\bar{\texttt N}{}'$       & 0.15 \,(0.020) & 0.021 \\
		$D^0\to\pi^0\texttt N\bar{\texttt N}{}'$        & 21\,{\footnotesize[Input]} \,(2.8) & 3.1 \\
		$D^+\to\pi^+\texttt N\bar{\texttt N}{}'$        & 107 \,(14)     & 16 \\
		$D_s^+\to K^+\texttt N\bar{\texttt N}{}'$       & 38 \,(4.9)     & 4.9 \\
		$D^0\to\rho^0\texttt N\bar{\texttt N}{}'$       & 9.6 \,(1.3)    & 0.68 \\
		$D^+\to\rho^+\texttt N\bar{\texttt N}{}'$       & 49 \,(6.4)     & 3.5 \\
		$D_s^+\to K^{*+}\texttt N\bar{\texttt N}{}'$    & 27 \,(3.6)     & 1.9 \\
		$\Lambda_c^+\to p\texttt N\bar{\texttt N}{}'$ & 61 \,(8.0\,{\footnotesize[Input]}) & 7.2 \\
		$\Xi_c^+\to\Sigma^+\texttt N\bar{\texttt N}{}'$ & 91 \,(12)    & 5.4 \\
		$\Xi_c^0\to\Sigma^0\texttt N\bar{\texttt N}{}'$ & 15 \,(2.0)   & 0.91 \\
		$\Xi_c^0\to\Lambda\texttt N\bar{\texttt N}{}'$  & 5.5 \,(0.71) & 0.34 \\
		\hline\hline
		\label{br3}
	\end{tabular*}  
\end{table}

\section{Conclusion\label{concl}}

We have explored the possibility that the FCNC decays of charmed hadrons with missing energy are enhanced by new physics affecting the \,$c\to u\slashed E$\, transition, where the missing energy is carried away by either a couple of spinless bosons or a pair of spin-1/2 Dirac fermions, all of which are singlets under the SM gauge groups. 
We study how the outcomes of the latest hunts for \,$D^0\to\slashed E$\, by Belle and \,$D^0\to\pi^0\slashed E$\, and \,$\Lambda_c^+\to p\slashed E$\, by BESIII can lead to constraints on the underlying operators describing \,$c\to u\slashed E$.\,
We demonstrate in our numerical work that these mesonic and baryonic modes already play valuable complementary roles in probing the operators.
Yet, additional data on these decays are needed to improve on the existing empirical bounds, which are not yet very stringent. 
Moreover, other channels are also important to search for because they could provide extra means to restrain the potential new physics.  
Of great interest among them is \,$D^0\to\gamma\slashed E$,\, which could still have a substantial branching fraction and covers the mass ranges of the invisible particles more than most of the other modes can. 
However, the latter channels, such as \,$D\to\rho\slashed E$\, and \,$\Xi_c\to\Sigma\slashed E$,\, are of consequence as well, with branching fractions that are not very small.
Many of the predictions we have made are expectedly testable by BESIII and Belle II in the near future

\section{Acknowledgments}
We thank Cong Geng and Yu Liu for information on experimental matters.
JT thanks the Tsung-Dao Lee Institute, Shanghai Jiao Tong University, and the Hangzhou Institute for Advanced Study, University of Chinese Academy of Sciences, for their hospitality during the course of this work.
It was supported in part by the National Key Research and Development Program of China under Grant No.~2020YFC2201501 and the National Natural Science Foundation of China (NSFC) under Grant No.~12147103 and No.~12247160.

\appendix

\section{Form factors in matrix elements of \boldmath$c\to u$ currents\label{FF}}

The form factors $F_V$ and $F_A$ for the $D^0\to\gamma$ transition in eq.\,(\ref{D->g}) have been addressed in the literature~\cite{Geng:2000if,Badin:2010uh,Hazard:2017udp}.
We employ the formulas from ref.\,\cite{Hazard:2017udp}
\begin{align}
	F_V & \,=\, \frac{(2/3)(-0.49)}{1-\hat q^2/(2.0\rm~GeV)^2} \,, &
	F_A & \,=\, \frac{(2/3)(-0.17)}{1-\hat q^2/(2.3\rm~GeV)^2} \,, &
\end{align}
which are functions of the squared momentum-transfer $\hat q^2$.

In the remainder of this appendix, we rely on isospin symmetry to relate the hadronic matrix elements of the $c\to u$ bilinears to those of $c\to d$ in the references cited.
One of the implications is that the form factors of $D^0\to\pi^0(\rho^0)$ are $1/\sqrt2$ times the corresponding ones of $D^+\to\pi^+(\rho^+)$ and those of $\Xi_c^0\to\Sigma^0$ are $1/\sqrt2$ times their $\Xi_c^+\to\Sigma^+$ counterparts.

For $f_+^{}$ and $f_0^{}$ in the \,${\mathbb D}\to\mathbb P$ matrix elements defined by eq.\,(\ref{D->P}), we adopt the lattice-QCD results of ref.\,\cite{FermilabLattice:2022gku} for $D\to\pi$ and $D_s\to K$ decays.
The dependence of $f_{+,0}^{}$ on $\hat q^2$ is given by~\cite{FermilabLattice:2022gku}
\begin{align}
	f_+^{} & = \frac{1}{1-\hat q^2/(2.00685\rm\,GeV)^2}\, \mbox{\footnotesize$\displaystyle\sum_{n=0}^3$}\, a_n^{} \Bigg[ z^n-\frac{n\, z^4}{4 (-1)^{4-n}} \Bigg] \,, ~~~ ~~~~
	f_0^{} = \frac{1}{1-\hat q^2/(2.3\rm\,GeV)^2}\, \mbox{\footnotesize$\displaystyle\sum_{n=0}^3$}\, b_n^{} z^n \,,  
	\nonumber \\
	z & = \frac{\sqrt{(M_D+M_\pi)^2-\hat q^2}-M_D-M_\pi}{\sqrt{(M_D+M_\pi)^2-\hat q^2}+M_D+M_\pi} \,, ~~~~ M_D = 1864.83 {\rm~MeV} , ~~~~ M_\pi = 134.9768 {\rm~MeV} , ~~~
\end{align}
where for $D^+\to\pi^+$
\begin{equation}
	a_0 = b_0 = 0.63 , ~~ a_1 = -0.61 , ~~ a_2 = -0.2 , ~~ a_3 = 0.3 , ~~ b_1 = 0.33 , ~~ b_2 = -0.31 , ~~ b_3 = -1.9
\end{equation}
and for $D_s^+\to K^+$
\begin{equation}
	a_0 = b_0 = 0.6307 , ~ a_1 = -0.562 , ~ a_2 = -0.19 , ~ a_3 = 0.33 , ~ b_1 = 0.347 , ~ b_2 = 0.44 , ~ b_3 = -0.21 \,.
\end{equation}

Concerning the \,${\mathbb D}\to\mathbb V$ form factors $A_{0,1,2}$ and $V$ in eq.\,(\ref{D->V}), to our knowledge there are as yet no lattice computations of them.
Therefore, we opt for the outcomes of a so-called symmetry-preserving formulation of a vector-vector contact interaction quite recently implemented in ref.\,\cite{Xing:2022sor}, which have the form
\begin{align}
	{\mathbb F}(\hat{\textsc f}_0,\hat{\textsc a},\hat{\textsc b}) & \,=\, \frac{\hat{\textsc f}_0}{1-\hat{\textsc a}\,\hat q^2/m_P^2+\hat{\textsc b}\,\big(\hat q^2/m_P^2\big)^2} \,, &
\end{align}
where $\hat{\textsc f}_0$, $\hat{\textsc a}$, $\hat{\textsc b}$, and $m_P$ are numbers obtained therein.
Thus, for $D^+\to\rho^+$
\begin{align}
	A_0 & = \,{\mathbb F}(0.61,1.29,0.27) \,, & A_1 & = \,{\mathbb F}(0.52,0.15,-0.14) \,, & A_2 & = \,{\mathbb F}(0.36,0.6,-0.042) \,, ~~~ \nonumber \\ V & = \,{\mathbb F}(0.83,0.87,0.0009) \,, & m_P & = 1.87{\rm~GeV}
\end{align}
and for $D_s^+\to K^{*+}$
\begin{align}
	A_0 & = \,{\mathbb F}(0.62,1.4,0.27) \,, & A_1 & = \,{\mathbb F}(0.56,0.22,-0.2) \,, & A_2 & = \,{\mathbb F}(0.4,0.72,-0.047) \,, ~~~ \nonumber \\ V & = \,{\mathbb F}(0.94,0.98,-0.0011) \,, & m_P & = 1.96{\rm~GeV} \,.
\end{align}

For \,${\tt F}_{\perp,+,0}$ and \,${\tt G}_{\perp,+,0}$ in the $\Lambda_c^+\to p$ matrix elements given by eq.\,(\ref{<p|uc|Lc>}), we use the lattice-QCD results of ref.\,\cite{Meinel:2017ggx} which are parametrized as
\begin{align}
	\tilde{\mathbb F}(\textsc a_0,\textsc a_1,\textsc a_2) & \,=\, \frac{\textsc a_0+\textsc a_1\tilde z+\textsc a_2\tilde z^2}{1-\hat q^2/m_{\rm pole}^2} \,, &
	\tilde z & \,=\, \frac{\sqrt{\tilde{\tt t}_+-\hat q^2}-\sqrt{\tilde{\tt t}_+-(m_{\Lambda_c}-m_N)^2}}{\sqrt{\tilde{\tt t}_+-\hat q^2}+\sqrt{\tilde{\tt t}_+-(m_{\Lambda_c}-m_N)^2}} \,, ~  
\end{align}
where \,$\tilde{\tt t}_+ = (1.87+0.135)^2\rm\;GeV^2$ and $m_N$ is the average nucleon mass. 
Accordingly, we have~\cite{Meinel:2017ggx}
\begin{align}
	{\tt F}_\perp & \,=\, \tilde{\mathbb F}(1.36,-1.7,0.71) \,, & {\tt F}_+ & \,=\, \tilde{\mathbb F}(0.83,-2.33,8.41) \,, & {\tt F}_0 & \,=\, \tilde{\mathbb F}(0.84,-2.57,9.87) \,, & \nonumber \\
	{\tt G}_\perp & \,=\, \tilde{\mathbb F}(0.69,-0.68,0.7) \,, & {\tt G}_+ & \,=\, \tilde{\mathbb F}(0.69,-0.9,2.25) \,, & {\tt G}_0 & \,=\, \tilde{\mathbb F}(0.73,-0.97,0.83) \,,
\end{align}
with $m_{\rm pole}=2.01\;$GeV for \,${\tt F}_{\perp,+}$,\, 2.351$\;$GeV for \,${\tt F}_0$,\, 2.423$\;$GeV for \,${\tt G}_{\perp,+}$,\, 1.87$\;$GeV for \,${\tt G}_0$.
We note that, instead of eq.\,(\ref{<p|uc|Lc>}), one can alternatively write 
\begin{align} \label{Lc-to-p}  
	\langle p|\overline u\gamma^\mu c|\Lambda_c^+\rangle & \,=\, \bar u_p^{} \bigg( \gamma^\mu f_1^{} + \frac{[\gamma^\mu,\gamma^\omega] \hat q_\omega^{}}{2 m_{\Lambda_c}} f_2^{} + \frac{\hat q^\mu}{m_{\Lambda_c}} f_3^{} \bigg) u_{\Lambda_c}^{} \,, 
	\nonumber \\
	\langle p|\overline u\gamma^\mu\gamma_5^{}c|\Lambda_c^+\rangle & \,=\, \bar u_p^{} \bigg( \gamma^\mu g_1^{} + \frac{[\gamma^\mu,\gamma^\omega] \hat q_\omega^{}}{2 m_{\Lambda_c}}\, g_2^{} + \frac{\hat q^\mu}{m_{\Lambda_c}}\, g_3^{} \bigg) \gamma_5^{} u_{\Lambda_c}^{} \,, & 
\end{align}
where $f_{1,2,3}^{}$ and $g_{1,2,3}^{}$ are connected to ${\tt F}_{\perp,+,0}$ and ${\tt G}_{\perp,+,0}$ by
\begin{align}
	{\tt F}_\perp^{} & \,=\, f_1^{} + \frac{\texttt M_+\, f_2^{}}{m_{\Lambda_c}} \,, & 
	{\tt F}_+^{} & \,=\, f_1^{} + \frac{\hat q^2 f_2^{}}{m_{\Lambda_c}\, \texttt M_+} \,, & 
	{\tt F}_0^{} & \,=\, f_1^{} + \frac{\hat q^2 f_3^{}}{m_{\Lambda_c}\, \texttt M_-} \,, 
	\nonumber \\
	{\tt G}_\perp^{} & \,=\, g_1^{} - \frac{\texttt M_-\, g_2^{}}{m_{\Lambda_c}} \,, & 
	{\tt G}_+^{} & \,=\, g_1^{} - \frac{\hat q^2 g_2^{}}{m_{\Lambda_c}\, \texttt M_-} \,, & 
	{\tt G}_0^{} & \,=\, g_1^{} - \frac{\hat q^2 g_3^{}}{m_{\Lambda_c}\, \texttt M_+} \,.
\end{align}

With regard to $\Xi_c\to\Sigma,\Lambda$, there is still no lattice analysis on their form factors as far as we can tell.
Hence we adopt those estimated in ref.\,\cite{Geng:2020gjh} in the light-front constituent quark model, which are expressible as $f_{1,2,3}^{}$ and $g_{1,2,3}^{}$, defined analogously to their $\Lambda_c^+\to p$ counterparts in eq.\,(\ref{Lc-to-p}) and having the form $\tilde F(\kappa_0,\kappa_1,\kappa_2) = \kappa_0/(1-\kappa_1\,\hat q^2+\kappa_2\,\hat q^4)$, with $\kappa_{0,1,2}$ being constants calculated therein. 
Thus, for $\Xi_c^+\to\Sigma^+$  
\begin{align}
	f_1^{} & = \tilde F(0.73, 1.49, 2.35) \,, & f_2^{} & = \tilde F(0.99, 1.43, 2.38) \,, & \nonumber \\
	g_1^{} & = \tilde F(0.63, 1.18, 1.79) \,, & g_2^{} & = \tilde F(0.11, 1.88, 2.88) &
\end{align}
and for $\Xi_c^0\to\Lambda$
\begin{align}
	f_1^{} & = \tilde F(0.28, 1.5, 2.32) \,, & f_2^{}  & = \tilde F(0.38, 1.35, 2.3) \,, &  \nonumber \\
	g_1^{} & = \tilde F(0.25, 1.18, 1.77) \,, & g_2^{} & = \tilde F(0.04, 1.71, 2.78) \,, &
\end{align}
but $f_3^{}=g_3^{}=0$ in the formalism of ref.\,\cite{Geng:2020gjh}.

\section{Predictions of the standard model\label{smpreds}}

Before dealing with the SM case, we consider the more general, effective couplings of invisible spin-1/2 Dirac fermion fields \,\texttt f and \,$\texttt f'$ to vector and axialvector $c\to u$ currents described by
\begin{align} \label{cuff'}
	{\cal L}_{\texttt{ff}'} & \,=\, -\overline u\gamma^\mu c~ \overline{\texttt f}
	\gamma_\mu^{}\big({\texttt C}_{\texttt{ff}'}^{\texttt V}+\gamma_5^{}{\texttt C}_{\texttt{ff}'}^{\texttt A}\big) {\texttt f}'
	- \overline u\gamma^\mu\gamma_5^{}c\, \overline{\texttt f}\gamma_\mu^{} \big(
	\tilde{\textsc c}_{\texttt{ff}'}^{\textsc v}+\gamma_5^{}\tilde{\textsc c}_{\texttt{ff}'}^{\textsc a}\big) {\texttt f}' \,, &
\end{align}
where the constants ${\texttt C}_{\texttt{ff}'}^{\texttt{V,A}}$ and $\tilde{\textsc c}_{\texttt{ff}'}^{\textsc{v,a}}$ may be complex.
It will induce $D^0\to\gamma\texttt f\bar{\texttt f}'$, \,${\mathbb D}\to{\mathbb P}{\texttt f}\bar{\texttt f}',{\mathbb V}\texttt f\bar{\texttt f}'$, and $\Lambda_c^+\to p\texttt f\bar{\texttt f}'$ if kinematically allowed.
The amplitudes for these decays can be derived after applying the hadronic matrix elements detailed in sections \ref{D0->SS'}-\ref{Lc->pSS'} to the quark bilinears in ${\cal L}_{\texttt{ff}'}$.
Permitting $m_{\texttt f}$ and $m_{\texttt f'}$ to be unequal, we then arrive at the (differential) rates
\begin{align} \label{BD2ff'}
	\Gamma_{D^0\to\texttt f\bar{\texttt f}'}^{} & \,=\, \frac{\lambda^{1/2}\big(m_{D^0}^2,m_{\texttt f}^2,m_{\texttt f'}^2\big)}{8\pi\,m_{D^0}^3} \Big[ |\tilde{\textsc c}_{\texttt{ff}'}^{\textsc v}|^2 \big( m_{D^0}^2-\widetilde\mu_+^2\big) \widetilde\mu_-^2
	+ |\tilde{\textsc c}_{\texttt{ff}'}^{\textsc a}|^2\big(m_{D^0}^2-\widetilde\mu_-^2\big)\widetilde\mu_+^2 \Big] f_D^2 \,, ~~
	\\ \raisebox{5ex}{}
	\frac{d\Gamma_{D^0\to\gamma\texttt f\bar{\texttt f}'}}{d\hat s} & \,=\, \frac{ \alpha_{\rm e}^{}\, \tilde\lambda_{\texttt{ff}'}^{1/2}\, \big(m_{D^0}^2-\hat s\big)\raisebox{2pt}{$^3$} }{192\pi^2\,m_{D^0}^5 \hat s^2} \!\! \begin{array}[t]{l} \displaystyle
		\Big[ \Big( |{\texttt C}_{\texttt{ff}'}^{\texttt V}|^2 F_V^2 + \big|\tilde{\textsc c}_{\texttt{ff}'}^{\textsc v}\big|\raisebox{1pt}{$^2$} F_A^2 \Big)
		(3\hat s-\widetilde s_+) \widetilde s_-
		\\ \,+\; \Big( |{\texttt C}_{\texttt{ff}'}^{\texttt A}|^2 F_V^2 + |\tilde{\textsc c}_{\texttt{ff}'}^{\textsc a}|^2 F_A^2 \Big)
		(3\hat s-\widetilde s_-) \widetilde s_+ \Big] \Big\} \,, \end{array}
	\nonumber \\ \raisebox{4ex}{}
	\frac{d\Gamma_{{\mathbb D}\to{\mathbb P}\texttt f\bar{\texttt f}'}}{d\hat s} & \,=\, \frac{ 4\,
		\tilde\lambda_{\mathbb{DP}}^{1/2}\, \tilde\lambda_{\texttt{ff}'}^{1/2}}{3(8\pi m_{\mathbb D}\,\hat s)^3}
	\!\! \begin{array}[t]{l} \displaystyle \Big\{ \Big[ |{\texttt C}_{\texttt{ff}'}^{\texttt V}|^2
		(3\hat s-\widetilde s_+) \widetilde s_- + |{\texttt C}_{\texttt{ff}'}^{\texttt A}|^2
		(3\hat s-\widetilde s_-) \widetilde s_+ \Big] \tilde\lambda_{\mathbb{DP}}\, F_+^2
		\\ \,+\; 3 \Big( |{\texttt C}_{\texttt{ff}'}^{\texttt V}|^2\, \widetilde\mu_-^2\, \widetilde s_+^{} + |{\texttt C}_{\texttt{ff}'}^{\texttt A}|^2\, \widetilde\mu_+^2\, \widetilde s_-^{} \Big) F_0^2\, {\texttt m}_+^2 {\texttt m}_-^2 \Big\} \,, \end{array}   \label{D2Pff'}
\end{align}
\begin{align} \label{D2Vff'}
	\frac{d\Gamma_{{\mathbb D}\to{\mathbb V}\texttt f\bar{\texttt f}'}}{d\hat s} \,=\, \frac{ 4\,
		\tilde\lambda_{\mathbb{DV}}^{3/2}\, \tilde\lambda_{\texttt{ff}'}^{1/2}}{(8\pi m_{\mathbb D}\,\hat s)^3}
	& \! \begin{array}[t]{l} \displaystyle \Bigg\{
		\Big[ |{\texttt C}_{\texttt{ff}'}^{\texttt V}|^2 (3\hat s-\widetilde s_+) \widetilde s_- + |{\texttt C}_{\texttt{ff}'}^{\texttt A}|^2 (3\hat s-\widetilde s_-) \widetilde s_+ \Big] \frac{2 \hat s V^2}{3\, \tilde{\texttt m}_+^2}
		\\ \displaystyle \,+\;
		\Bigg[ \frac{A_1^2\, \widetilde{\texttt m}_+^2}{6 m_{\mathbb V}^2} \Bigg( \frac{1}{2} + \frac{6 m_{\mathbb V}^2\hat s}{\tilde\lambda{}_{\mathbb{DV}}} \Bigg)
		+ \frac{\tilde\lambda_{\mathbb{DV}} A_2^2}{12\, \widetilde{\texttt m}_+^2 m_{\mathbb V}^2}
		+ \frac{\hat s-\widetilde{\texttt m}_+\widetilde{\texttt m}_-}{6 m_{\mathbb V}^2} A_1^{} A_2^{} \Bigg]
		\\ \displaystyle ~~\times
		\Big[ |\tilde{\textsc c}_{\texttt{ff}'}^{\textsc v}|^2 (3\hat s-\widetilde s_+) \widetilde s_- + |\tilde{\textsc c}_{\texttt{ff}'}^{\textsc a}|^2 (3\hat s-\widetilde s_-) \widetilde s_+ \Big]
		\\ \displaystyle \,+\;
		\Big( |\tilde{\textsc c}_{\texttt{ff}'}^{\textsc v}|^2 \widetilde\mu_-^2 \widetilde s_+^{}
		+ |\tilde{\textsc c}_{\texttt{ff}'}^{\textsc a}|^2 \widetilde\mu_+^2 \widetilde s_-^{} \Big) A_0^2
		\Bigg\} \,, \end{array}
	\nonumber \\
	\frac{d\Gamma_{\Lambda_c^+\to p\texttt f\bar{\texttt f}'}}{d\hat s} \,=\, \frac{ 4\, \tilde\lambda_{\Lambda_c p}^{1/2}\, \tilde\lambda_{\texttt{ff}'}^{1/2}}{3(8\pi m_{\Lambda_c}\,\hat s)^3} & \! \begin{array}[t]{l}
		\displaystyle \Big\{ \Big[ |{\texttt C}_{\texttt{ff}'}^{\texttt V}|^2 (3\hat s-\widetilde s_+) \widetilde s_- + |{\texttt C}_{\texttt{ff}'}^{\texttt A}|^2 (3\hat s-\widetilde s_-) \widetilde s_+ \Big] \big(2f_\perp^2\hat s+f_+^2{\texttt M}_+^2\big) \hat\sigma_-^{}
		\\ \displaystyle \,+\; 3 \Big( |{\texttt C}_{\texttt{ff}'}^{\texttt V}|^2\, \widetilde\mu_-^2\, \widetilde s_+^{}
		+ |{\texttt C}_{\texttt{ff}'}^{\texttt A}|^2\, \widetilde\mu_+^2\, \widetilde s_-^{} \Big) \hat\sigma_+^{} f_0^2\, {\texttt M}_-^2
		\\ \displaystyle \,+\; \Big[ |\tilde{\textsc c}_{\texttt{ff}'}^{\textsc v}|^2 (3\hat s-\widetilde s_+)
		\widetilde s_- + |\tilde{\textsc c}_{\texttt{ff}'}^{\textsc a}|^2 (3\hat s-\widetilde s_-^{})
		\widetilde s_+ \Big] \big(2 g_\perp^2\hat s+ g_+^2\, {\texttt M}_-^2\big) \hat\sigma_+^{}
		\\ \displaystyle \,+\; 3 \Big( |\tilde{\textsc c}_{\texttt{ff}'}^{\textsc v}|^2\, \widetilde\mu_-^2\,
		\widetilde s_+^{} + |\tilde{\textsc c}_{\texttt{ff}'}^{\textsc a}|^2\,\widetilde\mu_+^2\,\widetilde s_-^{} \Big) \hat\sigma_-^{} g_0^2\, {\texttt M}_+^2 \Big\} \,, \end{array}
\end{align}
where
\begin{align}
	\widetilde\mu_\pm^{} & = m_{\texttt f}^{}\pm m_{\texttt f'}^{} \,, &
	{\texttt m}_\pm^{} & = m_{\mathbb D}^{}\pm m_{\mathbb P}^{} \,, &
	{\texttt M}_\pm^{} & = m_{\Lambda_c}^{}\pm m_p^{} \,, &
	\nonumber \\
	\widetilde s_\pm^{} & = \hat s-\widetilde\mu_\pm^2 \,, &
	\widetilde{\texttt m}_\pm^{} & = m_{\mathbb D}^{}\pm m_{\mathbb V}^{} \,, &
	\hat\sigma_\pm^{} &= {\texttt M}_\pm^2 - \hat s \,. &
\end{align}

In the SM, the FCNC charmed-hadron decays with neutrinos in the final states get short-distance contributions arising from loop diagrams and brought about by the effective Hamiltonian
\begin{align} \label{sm-c2unn}
	{\cal H}_{c\to u\nu\bar\nu}^{\rm SM} & \,=\, \frac{\alpha_{\rm e}^{}G_{\rm F}^{}}{\sqrt2\,\pi\,\sin^2\theta_W}\, \raisebox{2pt}{\footnotesize$\displaystyle\sum_{\ell=e,\mu,\tau}$}~ \raisebox{2pt}{\footnotesize$\displaystyle\sum_{q=d,s,b}$} \hat\lambda_q ~ \overline{\!D}(r_q,r_\ell)~ \overline u\gamma^\eta P_L^{}c\, \overline{\nu_\ell^{}}\gamma_\eta^{}P_L^{}\nu_\ell^{} \,, &
\end{align}
where $G_{\rm F}$ denotes the Fermi constant, $\theta_W$ is the Weinberg angle, the factor $\hat\lambda_q=V_{uq}^*V_{cq}^{}$ comprises Cabibbo-Kobayashi-Maskawa (CKM) matrix elements, $r_f=m_f^2/m_W^2$, and the loop function~\cite{Inami:1980fz}
\begin{align}
	\overline{\!D}(x,y) & \,=\, \frac{x(4-y)^2}{8(1-y)^2}\, \frac{y \ln y}{x-y}
	+ \frac{x(4-x)^2}{8(1-x)^2}\, \frac{x\ln x}{y-x} + \frac{4-2x+x^2}{8(1-x)^2} x \ln x
	- \frac{4+2 x+5 y-2 x y}{8(1-x)(1-y)} x \,.
\end{align}
Accordingly, in the notation of eq.\,(\ref{cuff'}), each $\ell$ term in eq.\,(\ref{sm-c2unn}) yields
\begin{align}
	{\texttt C}_{\texttt{ff}'}^{\texttt V} & \,=\, -{\texttt C}_{\texttt{ff}'}^{\texttt A} \,=\, -\tilde{\textsc c}_{\texttt{ff}'}^{\textsc v} \,=\, \tilde{\textsc c}_{\texttt{ff}'}^{\textsc a} \,=\, \raisebox{2pt}{\footnotesize$\displaystyle\sum_{q=d,s,b}$}\, \frac{\alpha_{\rm e}^{}G_{\rm F}^{}\hat\lambda_q\, \overline{\!D}(r_q,r_\ell)}{4\sqrt2\, \pi\, \sin^2\theta_W} \,, &
\end{align}
with \,${\texttt f}={\texttt f}'=\nu_\ell$.
Incorporating this into eqs.\,\,(\ref{D2Pff'}) and (\ref{D2Vff'}), employing the form factors specified in the previous appendix and the values of the parameters in eq.\,(\ref{sm-c2unn}) and of the relevant hadron lifetimes and particle masses from ref.\,\cite{ParticleDataGroup:2022pth}, with $m_{\texttt f}=m_{\texttt f'}=0$, and adding the rates corresponding to the $\ell=e,\mu,\tau$ flavors of the neutrinos, we then obtain the predictions listed in eq.\,(\ref{Bsmsd}).  
The nonzero SM contribution to $D^0\to\slashed E$ is mainly from ${\cal B}(D^0\to\nu\bar\nu\nu\bar\nu)\sim3\times10^{-27}$~\cite{Bhattacharya:2018msv}.
The long-distance contributions are difficult to determine reliably, but estimates for a few modes produced results which could be somewhat bigger than their short-distance counterparts~\cite{Burdman:2001tf,Kamenik:2009kc,Colangelo:2021myn}, but not by several orders of magnitude.
It follows that the SM backgrounds to our charmed-hadron decays of interest can be safely ignored.

\bibliography{reference}

\end{document}